\documentclass[%
 prd,
 twocolumn,
 superscriptaddress,
 numerical,
 showpacs,
 amsmath,amssymb,
 aps,
nofootinbib,
longbibliography,
 floatfix
]{revtex4-1}

\usepackage{bm}
\usepackage{bbm}
\usepackage{slashed}
\usepackage{verbatim}
\usepackage{graphicx}
\usepackage{combelow} 
\usepackage{mathtools}
\usepackage[usenames,dvipsnames,svgnames,table]{xcolor}
\usepackage[colorlinks=True,linkcolor=blue,citecolor=blue,urlcolor=blue]{hyperref}
\let\oldciteauthor=\citeauthor
\def\citeauthor#1{\hypersetup{citecolor=black}\oldciteauthor{#1}}

\let\oldciten=\onlinecite
\def\onlinecite#1{\hypersetup{citecolor=blue}\oldciten{#1}}

\let\oldcite=\cite
\def\cite#1{\hypersetup{citecolor=blue}\oldcite{#1}}

\newcommand{\beqn}{\begin{eqnarray}}
\newcommand{\eeqn}{\end{eqnarray}}
\newcommand{\beqs}{\begin{subequations}}
\newcommand{\eeqs}{\end{subequations}\\[-2mm]\noindent}
\newcommand{\eq}[1]{(\ref{#1})}

\newcommand{\Z}{{\mathbb Z}}

\newcommand{\bs}{\boldsymbol}

\newcommand{\sumint}[1]{\sum \hskip -5mm\int\limits_{#1} \hskip 2mm}

\definecolor{purple}{rgb}{0.8,0,0.6}
\definecolor{PURPLE}{rgb}{0.8,0,0.6}
\definecolor{orange}{rgb}{1,0.55,0}
\definecolor{limegreen}{rgb}{0.2,0.8,0.2}

\newcommand{\om}[1]{{\overline{\mathcal{#1}}}}
\newcommand{\bk}{\mathbf{k}}
\newcommand{\bp}{\mathbf{p}}
\newcommand{\bx}{\mathbf{x}}

\allowdisplaybreaks
\begin{document}

\title{Rigidly-rotating scalar fields: Between real divergence and imaginary fractalization}

\author{Victor E. Ambru\cb{s}}
\affiliation{
Department of Physics, West University of Timi\cb{s}oara, Bd.~Vasile P\^arvan 4, Timi\cb{s}oara 300223, Romania}
\email{victor.ambrus@e-uvt.ro}

\author{Maxim N. Chernodub}
\affiliation{Institut Denis Poisson UMR 7013, Universit\'e de Tours, 37200 France}
\email{maxim.chernodub@univ-tours.fr}
\thanks{corresponding author.}

\begin{abstract}
The thermodynamics of rigidly rotating systems experience divergences when the system dimensions transverse to the rotation axis exceed the critical size imposed by the causality constraint. The rotation with imaginary angular frequency, suitable for numerical lattice simulations in Euclidean imaginary-time formalism, experiences fractalization of thermodynamics in the thermodynamic limit, when the system's pressure becomes a fractal function of the rotation frequency. Our work connects these two phenomena by studying how thermodynamics fractalizes as the system size grows. We examine an analytically-accessible system of rotating massless scalar matter on a one-di\-men\-si\-on\-al ring and the numerically-treatable case of rotation in the cylindrical geometry and show how the ninionic deformation of statistics emerges in these systems. We discuss a no-go theorem on analytical continuation between real- and imaginary-rotating theories. Finally, we compute the moment of inertia and shape deformation coefficients caused by the rotation of the relativistic bosonic gas. In all cases, we show that finite-mass effects are quantitative, leaving our conclusions qualitatively unchanged.
\end{abstract}

\maketitle

\section{Introduction}\label{sec_intro}

Effects of rotation on the state of physical bodies have been a subject of passionate interest throughout the decades. In metals, the uniform rotation acts on electrons via a centrifugal force that produces a slight but experimentally perceptible gradient of electric potential measured at $\sim 10^{2-3} \, \mathrm{Hz}$~\cite{Beams1968}. At the level of electronic spins,  one of the numerous examples of rotation-generated phenomena is the Barnett effect~\cite{Barnett1915} which -- with its celebrity reciprocal, the Einstein--de Haas effect~\cite{Einstein1915} -- relates the mechanical torque and magnetization in ferromagnets. The nuclear analog of the Barnett effect substantially affects the polarization of the protons (ions of hydrogen) in the water rotating with the frequency $\sim 10^{4} \, \mathrm{Hz}$~\cite{Arabgol2019}.

However, the fastest rotation of matter has been produced in noncentral collisions of relativistic heavy ions that create quark-gluon plasma in which the vorticity reaches the values $\sim 10^{22} \, \mathrm{Hz}$~\cite{STAR:2017ckg,Deng:2016gyh,Jiang:2016woz}. The fast rotation affects the local properties of quark-gluon plasma, leading to various spin polarization phenomena, allowing us to probe experimentally the interior of rapidly rotating plasma in terms of its local vortical structure~\cite{Becattini:2020ngo,Huang:2020dtn}. 

There are various theoretical shreds of evidence that fast rotation also affects the chiral~\cite{Chen:2015hfc,Jiang:2016wvv,Chernodub:2016kxh, Chernodub:2017ref,Wang:2018sur,Zhang:2020hha,Sadooghi:2021upd} and (de)confining transitions~\cite{Braguta:2020biu,Chen:2020ath,Chernodub:2020qah,Fujimoto:2021xix,Braguta:2021jgn,Golubtsova:2021agl,Chen:2022smf,Golubtsova2022,Zhao:2022uxc,Chernodub:2022veq,Braguta:2023yjn} of the quark-gluon plasma. Theoretical methods, however, prevailingly assume a rigid rotation that makes every physical point rotate about a fixed axis with the same angular velocity. While the rigid character of rotation substantially simplifies the analytical treatment of the problem~\cite{Ambrus:2014uqa,Ambrus:2015lfr}, the consensus on the thermodynamic properties of quark-gluon plasma, even in this simplest case, is still absent, thus opening a gap between numerical and various analytical calculations. Moreover, the latest first-principle simulation reveals the instability of the rigidly rotating gluon plasma below the ``supervortical'' critical temperature~\cite{Braguta:2023yjn}, indicating the complexity of rotation in strongly interacting systems.

First-principle information about the quark-gluon plasma comes from lattice simulation in the Euclidean imaginary time formalism where the real angular momentum $\Omega$ brings the sign problem~\cite{Yamamoto:2013zwa} which makes the numerical simulations impossible. This inconvenience can traditionally be overcome by turning the angular momentum into the complex plane and considering the purely imaginary rotation $\Omega_I = - i \Omega$ in full analogy with the baryon chemical potential~\cite{Yamamoto:2013zwa,Braguta:2020biu,Chernodub:2020qah,Braguta:2021jgn,Chen:2022smf,Chernodub:2022wsw}. 

The imaginary rotation differs, however, from the imaginary baryonic chemical potential: the analytical continuation to real rotation used in numerical lattice simulations has some unusual features including the emergence of (stable) ghost-like excitations~\cite{Chernodub:2020qah} characterized by ``ninionic'' deformation of statistics and the appearance of the fractal features of thermodynamics under imaginary rotation. The fractalization imposes a no-go theorem on an analytical continuation for rotating systems in the thermodynamic limit~\cite{Chernodub:2022qlz}. 

In our work, we discuss the effect of rotation on the thermodynamics of the simplest possible system represented by the massless scalar fields. This limit is convenient as it allows many of the calculations to be performed analytically. We explore the effect of finite mass in all considered setups and confirm that all features remain qualitatively identical to those in the massless case.

First, we briefly introduce the real and imaginary rotation in Sec.~\ref{sec_imaginary_rotation}. Then, in Sec.~\ref{sec_ring}, we analyze the interrelation between the fractal features, analytical continuation, and the causality constraint for the model formulated on a one-dimensional ring which can be treated analytically.

Section~\ref{sec_RKT} approaches real and imaginary rotation within the scope of the relativistic kinetic theory applied to the three-dimensional rotating gas. It also addresses the mechanical features of the rotating gas, including its moment of inertia and shape-deformation coefficients. This analysis is followed by Sec.~\ref{sec_KG}, where we pursue, for simplicity, a ``hybrid'' quantization approach based on the cylindrical waves with continuous momentum in a spatially unbounded region. 

We show the advantages of both discussed approaches in Sec.~\ref{sec_bounded}, where the rotating gas in the cylindrically-bounded region and the discrete quantization of the transverse modes is treated numerically in great detail. Furthermore, we reveal the analytical fractalization of the thermodynamics numerically in the three-dimensional rotating gas and explicitly show strong parallels with the fractalization of thermodynamics in the analytically-accessible one-dimensional ring. For simplicity, we implement the boundary using Dirichlet boundary conditions~\cite{Duffy:2002ss}, however it is reasonable to expect that similar results regarding fractalization can be expected when von Neumann or, more generally, Robin boundary conditions are considered~\cite{Romeo:2000wt,Romeo:2001dd,Ambrus:2019vkx}.

Our last section is devoted to conclusions and a summary of our results. Throughout the article, we work with the conventions $\hbar = c = k_B = 1$.

\section{Imaginary rotation and statistics}
\label{sec_imaginary_rotation}

\subsection{Real rotation and imaginary rotation}

Let us consider a quantum-mechanical system of bosonic particles rotating uniformly (rigidly, as a solid body) with the constant angular frequency $\Omega$ about the $z$ axis. For simplicity, we can assume that the system of particles is rotating inside a cylinder possessing reflective boundary conditions, such as Dirichlet boundary conditions \cite{Duffy:2002ss}. In the co-rotating frame, the free energy of the system takes the following form:
\beqn
\mathcal{F} = \frac{V}{\beta} \sumint{\alpha,m} 
\ln \left(1 {-} e^{- \beta (\omega_{\alpha,m} + m \Omega)} \right),
\label{eq_F_rot}
\eeqn
where $\beta = 1/T$ is the inverse temperature, $V$ is the volume of the system, $\omega_{\alpha,m}$ is the energy spectrum of the particles in the laboratory reference frame, and $\alpha$ is a collective notation of quantum numbers other than the projection of angular momentum, $m \equiv m_z \in \Z$. We work with zero-charge systems so that the chemical potential does not enter the free energy of the system~\eq{eq_F_rot}. We also ignore the zero-point contribution associated with the vacuum Casimir energy since it does not affect the thermodynamics of the system.

In order to determine the thermodynamic characteristics (for example, energy, pressure, entropy, angular momentum, moment of inertia, etc), it is sufficient to evaluate the statistical integral~\eq{eq_F_rot}. For bosonic particles, the contribution of each quantum level to the thermodynamic quantities is given by the Bose-Einstein distribution
\begin{align}
    n^{({\mathrm {bos}})}_{\omega} = \frac{1}{e^{\beta \omega} - 1}\,, \qquad \mbox{[bosonic statistics]},
    \label{eq_n_Bose}
\end{align}
where $\omega$ is the energy of the quantum level. In a rigidly rotating system, the statistical weight is determined by the energy in the co-rotating reference frame: 
\begin{align}
    \omega = {\tilde \omega}_{\alpha,m} \equiv \omega_{\alpha,m} - m \Omega\,,
    \label{eq_omega_varepsilon}
\end{align}
thus demonstrating explicitly how rotation with $\Omega \neq 0$ affects the statistical particle distribution. 

It is convenient to calculate the thermodynamic properties of rotating particles using the imaginary-time formalism in which the time coordinate is turned to a complex variable via the Wick transformation, $t \to \tau = i t$. The imaginary time $\tau$ is compactified to a circle of length $\beta = 1/T$ related to thermal equilibrium temperature $T$. The compactification imposes the matching conditions on the fields: all scalar fields $\phi$ are periodic functions along the thermal direction,
\beqn
\phi(\bx,\tau) = \phi\left(\bx, \tau + \beta\right)\,,
\label{eq_BC_Bosonic}
\eeqn
while all fermionic fields (not considered in this article) obey anti-periodic boundary conditions.
The Bose-Einstein statistical distribution~\eq{eq_n_Bose} for bosonic fields can be recovered automatically from the periodic boundary conditions~\eq{eq_BC_Bosonic}~\cite{Kapusta2006}.

The imaginary-time approach is intensively used in numerical lattice simulations of quantum field theories where the partition function is formulated in terms of a statistical integral in Euclidean spacetime~\cite{Rothe2012}. The lattice simulations are especially useful for obtaining information about non-perturbative effects that cannot be treated with standard perturbative methods~\cite{Rothe2012}.

However, the imaginary-time techniques cannot be directly applied to rotating systems because the action of the Euclidean theory becomes a complex quantity at a nonzero angular frequency, $\Omega \neq 0$, thus exhibiting the so-called ``sign problem''~\cite{Yamamoto:2013zwa}. The latter property does not allow us to treat the partition function of a rotating system as a statistical integral, bringing us to an inconvenient similarity with finite-density systems, where the (baryonic) chemical potential also makes the Euclidean action a complex quantity~\cite{Rothe2012}. The only practical way to avoid the sign problem for rotation is to consider the angular frequency as a purely imaginary variable:
\beqn
\Omega = i \Omega_I\,.
\label{eq_Omega_I}
\eeqn
The shift of the angular frequency to the complex plane~\eq{eq_Omega_I} restores the real-valuedness of the Euclidean action~\cite{Yamamoto:2013zwa,Braguta:2020biu}. Having calculated the desired quantities at a set of imaginary $\Omega_I$, one can then apply an analytical continuation to map the results back to the realistic case of real rotation~\cite{Braguta:2020biu,Braguta:2021jgn}. This prescription, applied to the angular frequency $\Omega$, follows a standard set of practices invoked to avoid the sign problem in simulations of finite-density systems~\cite{deForcrand:2009zkb}. 

In the context of the imaginary-time formalism, there are two methods how one can implement the imaginary rotation~\eq{eq_Omega_I}. The first approach, originally proposed in Ref.~\cite{Yamamoto:2013zwa} and adopted in various numerical Monte Carlo simulations of (quark-) gluon plasmas~\cite{Yamamoto:2013zwa,Braguta:2021jgn,Chernodub:2022veq,Braguta:2023yjn}, consists in (i) considering the system in a non-inertial co-rotating reference frame in Minkowski spacetime; (ii) turning the system, via a Wick transformation, to the curved Euclidean spacetime with a complex metric tensor; (iii) implementing the substitution~\eq{eq_Omega_I} which makes the metric tensor real-valued again; (iv) simulating the thermodynamics at a set of non-zero $\Omega_I$ with the standard periodic boundary conditions~\eq{eq_BC_Bosonic}; (v) fitting the obtained numerical results by a reasonable analytical function and, finally, (vi) making an analytical continuation of the lattice results to the real-valued frequency by setting 
\begin{align}
    \Omega^2_I \to - \Omega^2\,.
\label{eq_continuation}
\end{align}

The second approach implements the imaginary rotation in the imaginary-time formalism in a more straightforward way using the property that the imaginary frequency $\Omega_I$ corresponds, after all, to a uniform rotation of a subspace of a timeslice of the Euclidean spacetime about a certain fixed axis~\cite{Chernodub:2020qah,Chen:2022smf}. As the imaginary time variable $\tau$ advances for a full period from $\tau = 0$ to $\tau = \beta$, the system experiences a spatial rotation by the angle: 
\beqn
\chi = \beta \Omega_I
\equiv 2 \pi \nu\,, 
\qquad 
\nu = \frac{\beta \Omega_I}{2\pi}\,.
\label{eq_chi}
\eeqn
The turn of the space necessitates a modification of the standard bosonic boundary conditions~\eq{eq_BC_Bosonic} which should now incorporate a translation in imaginary time with the uniform rotation of the Euclidean spacetime.

\begin{figure}[!thb]
\centerline{\includegraphics[width=0.35\textwidth,clip=true]{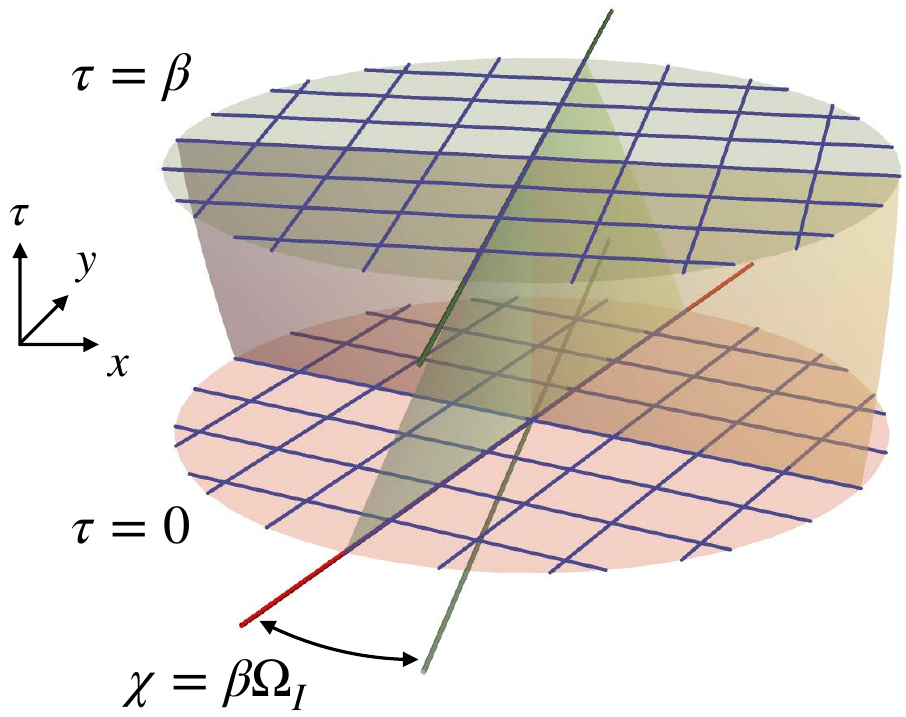}}
\caption{The rotwisted boundary conditions~\eq{eq_rotwisted} characterized by the statistical angle~\eq{eq_chi} produced by the imaginary angular velocity $\Omega_I$.}
\label{fig_rotwisting}
\end{figure}

Under the imaginary rotation, the bosonic wavefunction appears to satisfy the rotwisted boundary condition:
\beqn
\phi(\bx,\tau) = \phi\left({\hat R}_{{\bs \chi}} \bx,\tau + \beta\right)\,,
\label{eq_rotwisted}
\eeqn
where the $3 \times 3$ matrix 
\beqn
{\hat R}_{{\bs \chi}} = 
\begin{pmatrix}
    \phantom{-} \cos\chi & \sin\chi & 0 \\
    - \sin\chi & \cos\chi & 0 \\
    0 & 0 & 1
\end{pmatrix},
\label{eq_R}
\eeqn
written in Cartesian coordinates, corresponds to the global rotation of the whole spatial Euclidean subspace, $\bx \to \bx' = {\hat R}_{{\bs \chi}} \bx$, by the angle~\eq{eq_chi} around the $z$ axis. In the absence of rotation, the transformation~\eq{eq_R} becomes a unit matrix and the boundary condition~\eq{eq_rotwisted} reduces to the standard periodic condition for bosons~\eq{eq_BC_Bosonic}. The rotwisted boundary conditions, visualized in Fig.~\ref{fig_rotwisting}, have already been discussed in the context of the Euclidean lattice simulations of field theories~\cite{Chen:2022smf,Chernodub:2022veq}. 

The boundary conditions~\eq{eq_rotwisted} are obviously invariant under $2\pi$ shifts of $\chi$, or equivalently, shifts by one unit in $\nu$: 
\begin{align}
    \chi \to \chi + 2 \pi\,, \qquad
    \nu \to \nu + 1\,,
    \label{eq_chi_periodicity}
\end{align}
and, in the parity-unbroken systems, under the reversal of the rotation angle:
\begin{align}
    \chi \to - \chi\,, \qquad 
    \nu \to -\nu\,.
    \label{eq_chi_reflection}
\end{align}
The latter condition holds for a system of neutral particles that we consider.
The symmetry under clockwise and counterclockwise rotations~\eq{eq_chi_reflection} can be broken, for example, for charged particles subjected to a background magnetic field which leads, in particular, to a rotation diode effect in semiconductors~\cite{Chernodub:2021fpm}.

\subsection{Imaginary rotation and ninionic statistics}

The differences between the two implementations of the imaginary rotation are two-fold: one can either consider the curved Euclidean spacetime with corotating coordinates and ordinary boundary conditions~\eq{eq_BC_Bosonic} implemented along the compactified time (the first approach) or use Cartesian coordinates with the rotwisted boundary conditions~\eq{eq_rotwisted} following the second approach. In our article, we consider field theories subjected to imaginary rotation introduced via the rotwisted boundary condition. 

The boundary conditions imposed on the fields in the imaginary time direction have one-to-one correspondence with the statistical distribution of the particles. For example, the equal-time commutation relations for bosonic fields imply the periodic boundary conditions~\eq{eq_BC_Bosonic}, which lead to the Bose-Einstein distribution for bosons~\eq{eq_n_Bose}. Analogously, anti-commuting fermionic variables possess anti-periodic conditions in the compactified time direction which correspond to the Fermi-Dirac statistics~\cite{Kapusta2006}. Therefore, it is appropriate to ask which statistical distribution corresponds to the rotwisted boundary conditions~\eq{eq_rotwisted}? 

It turns out that the imaginary rotation deforms the statistical distribution of fermions and bosons, leading to a ``ninionic'' deformation which matches neither the bosonic nor the fermionic statistical distributions~\cite{Chernodub:2022qlz}. For example, for bosons, the ninionic deformation takes the following form:
\beqn
n^{({\mathrm{nin}})}_{\omega} (\xi) = \frac{e^{\beta \omega}  \cos \xi - 1}{1 - 2 e^{\beta \omega} \cos \xi + e^{2 \beta \omega}}\,, 
\label{eq_ninionic}
\eeqn
where $\omega \equiv \omega_{\alpha,m}$ is associated with the energy of the quantum state in the laboratory reference frame and $\xi = m \chi = 2\pi m \nu$ is the deformation parameter associated with the ``statistical angle'' $\chi$, Eq.~\eq{eq_chi}. The latter depends on the angular velocity $\Omega_I$ of the imaginary rotation in Euclidean spacetime. Notice that at the zero (modulo $2\pi$) statistical angle, the ninionic deformation~\eq{eq_ninionic} of the bosonic distribution~\eq{eq_n_Bose} disappears: $n^{(\mathrm{nin})}_{\omega} (\xi = 2 \pi k) = n^{(b)}_{\omega}$ with an integer $k \in {\mathbb Z}$.

The ninionic deformation~\eq{eq_ninionic} can be understood as the real part of the bosonic occupation number~\eq{eq_n_Bose},
\begin{align}
    n^{({\mathrm{nin}})}_{\omega} (\xi) = {\mathrm{Re}}\, n^{({\mathrm{bos}})}_{\omega + i \xi/\beta}\,,
    \label{eq_statistics_ninionic}
\end{align}
at an imaginary chemical potential $\mu = \xi/\beta$.

Given the unusual form of the ninionic deformation of the bosonic statistical distribution~\eq{eq_ninionic}, it is appropriate to ask how this deformation modifies the statistical properties of the thermal state? What are the consequences which are brought to the theory by the introduction of the new dimensionless parameter, the statistical angle~\eq{eq_chi}? The answer to this question, which depends on the volume of the rotating system, is one of the aims of our article. 

In respect of the causality, the rigid rotation with real-valued angular velocity is a well-defined notion only for transversely-bounded systems. On the contrary, the imaginary rotation does not impose any bounds on the size of the system due to the absence of the notion of the light cone in the Euclidean space (in other words, there is no causality constraint in the imaginary time formalism because it has no notion of real time). Therefore, the imaginary rotation does not lead to causality problems~\cite{Yamamoto:2013zwa} and can be formulated in the thermodynamic limit in the whole Euclidean space~\cite{Chen:2022smf}. The relation between imaginary and real rotation in terms of the analytical continuation is another aim of our paper.

\subsection{Ninionic statistics and fractal thermodynamics}

Sticking to an infinite-volume system, one can show, both in the scope of a classical interacting field theory~\cite{Chernodub:2022wsw} as well as in a free bosonic quantum field theory~\cite{Chernodub:2022qlz}, that the imaginary rotation characterized by the nonvanishing value of the statistical angle $\chi$ modifies the relation between the physical temperature $T \rightarrow T_I(\beta,\chi) \equiv T(\beta, i \beta \Omega_I)$ and the length of the compactified direction~$\beta$:\footnote{Throughout this paper, we shall use the subscript (or superscript) $I$ to indicate when an observable is computed under imaginary rotation.}
\beqn
T_I(\beta,\chi) = \frac{1}{\beta} f_{\mathsf{T}} \left( \frac{\chi}{2\pi} \right)\,,
\label{eq_T_rational_bosons}
\eeqn
where
\begin{align}
    f_{\mathsf{T}}(x) & =
\left\{
\begin{array}{rl}
    \frac{1}{q} &   \mathrm{if}\ x = \frac{p}{q} \in {\mathbb Q}, \ \mathrm{with}\ p,q\in {\mathbb N} \ \mathrm{coprimes},\\[2mm]
    0 &  \mathrm{if}\  x \notin {\mathbb Q}\,,
\end{array}
\right.
\label{eq_Thomae}
\end{align}
is the Thomae function. In other words, function~\eq{eq_Thomae} gives zero for all irrational numbers and equals to a nonzero number $1/q$ determined by the denominator $q$ of the rational argument $x = p/q \in {\mathbb Q}$ with two natural coprime numbers $p,q \in {\mathbb N}$.

\begin{figure}[!thb]
\centerline{\includegraphics[width=0.475\textwidth,clip=true]{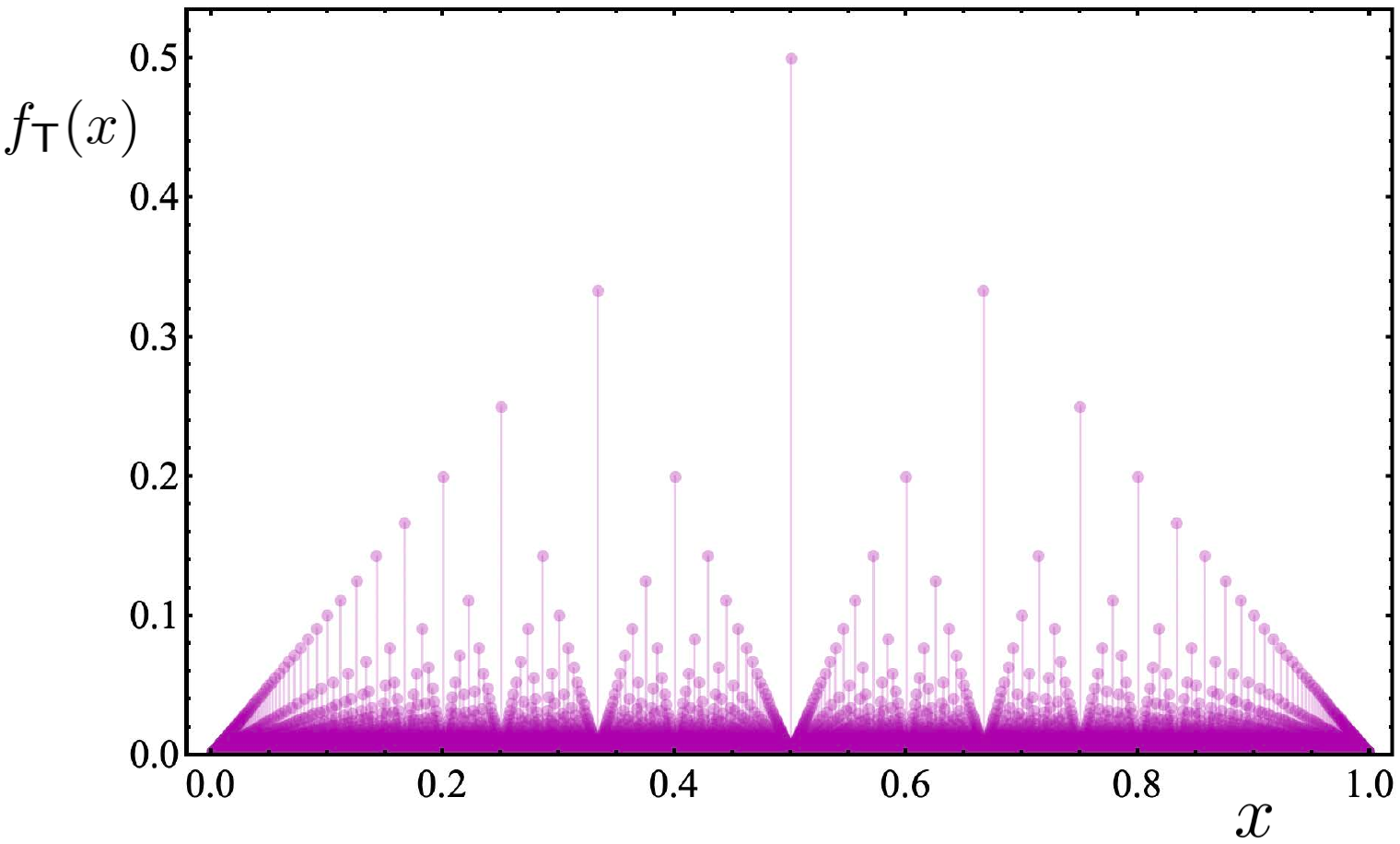}}
\caption{Thomae function~\eq{eq_Thomae}.}
\label{fig_Thomae}
\end{figure}

The Thomae function~\eq{eq_Thomae}, shown in Fig.~\ref{fig_Thomae}, is known also under other names, such as the raindrop function, the modified Dirichlet function, the popcorn function, etc. This function has the amazing counter-intuitive property stating that the function is discontinuous if its argument $x$ is rational, and it is continuous provided $x$ is irrational. The Thomae function emerges naturally in various physical systems such as one-dimensional disordered crystals, Hubbard model of particles on a ring, phyllotaxis problem, and fractional quantum Hall effect (we refer to Ref.~\cite{Flack:2023uop} for a comprehensive discussion). The Thomae function~\eq{eq_Thomae} possesses a nontrivial fractal structure~\cite{Flack:2023uop,Devaney1999,nechaev2017number} which equips the thermodynamics of imaginary rotation with fractal properties. The fractalization (and ``defractalization'') of thermodynamics of imaginary rotating systems will also be addressed in this paper.

Notice that the behavior of the physical temperature~\eq{eq_T_rational_bosons} as a function of the statistical angle~\eq{eq_chi} is determined solely by the denominator $q$ of the rational number $\chi/(2 \pi)$ and not by its numerator. Irrational (in units of $2\pi/\beta$) frequencies correspond to zero temperature~\eq{eq_T_rational_bosons}.

In the absence of the imaginary rotation, $\chi=0$, one gets the standard relation between temperature $T$ and the length of the imaginary time direction $\beta$, as expected:
\begin{align}
    T(\beta,0) = \frac{1}{\beta}\,.
    \label{eq_T_beta_0}
\end{align}

The ninionic deformation of bosonic statistics can be readily understood in the imaginary time formalism for free massless bosons in the thermodynamic limit (on an infinite spatial line) where the particles possess the linear energy dispersion, $\omega_k = |k|$. In this conformal system, the thermal pressure of bosons $P$ is equal to their energy density, $E \equiv P$, taking a well-known expression in the absence of imaginary rotation:
\begin{align}
    P_0 = \int_{- \infty}^{\infty} \frac{d k}{2 \pi}
    n^{(\mathrm{bos})}_{\omega_k} \, \omega_k = \frac{\pi}{6 \beta^2}\,, \qquad [\Omega_I = 0]\,.
    \label{eq_E_large_ring_0}
\end{align}
The temperature of the system is given by the inverse length $\beta$ of the compactified imaginary-time direction~\eq{eq_T_beta_0}. 

Looking ahead a little, one can also discuss the thermodynamics of the same system compactified into a ring of an infinitely large radius $R$ which is subjected to the imaginary rotation with the angular velocity $\Omega_I$. The pressure can be derived via the ninionic statistics~\eq{eq_statistics_ninionic}:  
\begin{align}
    P_I = \lim_{R \to \infty} \frac{1}{R} \sum_{m \in {\mathbb{Z}}}
    n^{(\mathrm{nin})}_{\omega_m} (\xi_m) \, 
    \omega_m = \frac{\pi}{6 \beta^2}
    f^2_{\mathsf{T}} \left( \frac{\beta \Omega_I}{2\pi} \right)\,,
    \label{eq_E_large_ring_Omega_I}
\end{align}
where the ninionic parameter $\xi_m = \chi m \equiv \beta \Omega_I m$ is expressed via the angular momentum $m$ and the statistical angle~\eq{eq_chi}. The energy spectrum in the statistical sum~\eq{eq_E_large_ring_Omega_I}, 
\begin{align}
    \omega_m = \frac{1}{R} |m|\,,
    \label{eq_lab_energy}
\end{align}
corresponds to the laboratory frame. In the thermodynamic limit, $R \to \infty$, the energy gaps of the discrete spectrum~\eq{eq_lab_energy} shrink, the variable $m/R$ becomes the continuum momentum $k$, and the sum in Eq.~\eq{eq_E_large_ring_Omega_I} reduces to an integral thus bridging the gap between the exotic~\eq{eq_E_large_ring_Omega_I} and standard~\eq{eq_E_large_ring_0} statistical sums in the thermodynamic limit. 

However, the presence of the imaginary rotation $\Omega_I$ makes the system nontrivial even in the thermodynamic limit. Indeed, the pressure of the imaginary rotating system~\eq{eq_E_large_ring_Omega_I} has the same expression as the pressure of the non-rotating one~\eq{eq_E_large_ring_0} with only one important difference that the temperature of the former~\eq{eq_T_rational_bosons} becomes a fractal function~\eq{eq_Thomae} of the imaginary angular frequency~$\Omega_I$. In the next section, we discuss the particularities of fractalization of thermodynamics by imaginary rotation working with an analytically-solvable example of a free massless particle confined to a one-dimensional ring.

\section{Real and imaginary rotations \\
on the ring}
\label{sec_ring}

\subsection{Classical (non-quantum) system} \label{sec_ring_RKT}

Let us consider an ensemble of bosonic massless particles, constrained to move on a ring of fixed radius $R$ with angle coordinates $\varphi$. Imposing that the system rotates with angular velocity $\Omega$, its free energy at inverse temperature $\beta$ reads
\begin{equation}
 \mathcal{F} = \frac{1}{\beta} \int_0^{2\pi} d\varphi \int_{-\infty}^\infty \frac{dk_\varphi}{2\pi} 
 \ln [1 -e^{-\beta(k^0 - \Omega J^z)}],
 \label{eq:ring_RKT_F}
\end{equation}
where $J^z = -k_\varphi$ represents the angular momentum of a particle spinning with azimuthal momentum $k^\varphi = -k_\varphi / R^2$, satisfying the dispersion relation $k^0 = |k_\varphi / R| = |R k^\varphi| > 0$. The above integral can be readily evaluated,
\begin{equation}
 \mathcal{F} = -\frac{\pi^2 R}{3\beta^2} \Gamma^2(R), \quad 
 \Gamma(R) = (1 - \Omega^2 R^2)^{-1/2}.
 \label{eq_ring_RKT_F_m0}
\end{equation}
Other thermodynamic quantities such as total entropy $\mathcal{S}$, thermodynamic pressure $\mathcal{P}$, and total angular momentum $\mathcal{M}$, can be obtained starting from the relation \cite{LL5}
\begin{equation}
 d\mathcal{F} = \beta^{-2} \mathcal{S} d\beta - \mathcal{P} dV - \mathcal{M} d\Omega,
 \label{eq:dF}
\end{equation}
where $V = 2\pi R$ is the ``volume'' of the ring. 
Denoting by an overhead bar volume-averaged quantities, $\om{A} \equiv \mathcal{A} / V$, we have 
\begin{align}
 \om{S} = \beta^2 \frac{\partial \om{F}}{\partial \beta}, \quad 
 \om{M} = -\frac{\partial \om{F}}{\partial \Omega}, \quad 
 \mathcal{P} = \frac{1}{2\pi} \frac{\partial \mathcal{F}}{\partial R},
 \label{eq_thermo_rel}
\end{align}    
while the average energy $\om{E}$ can be found from the Euler relation,
\begin{equation}
 \om{E} = \om{F} + \beta^{-1} \om{S} + \Omega \om{M}.
 \label{eq_Euler}
\end{equation}
Taking into account the thermodynamic relations in Eq.~\eqref{eq_thermo_rel}, the above relation can be recast as
\begin{equation}
 \om{E} = \left(\frac{\partial(\beta \om{F})}{\partial \beta}\right)_{\beta\Omega}.
 \label{eq_F_from_E}
\end{equation}

Starting from Eq.~\eqref{eq_ring_RKT_F_m0}, it is easy to derive
\begin{align}
 \om{F} &= -\frac{\pi \Gamma^2(R)}{6\beta^2}, &
 \mathcal{P} &= \frac{\pi \Gamma^2(R)}{6\beta^2} [2\Gamma^2(R) - 1], \nonumber\\
 \om{S} &= \frac{\pi \Gamma^2(R)}{3\beta}, & 
 \om{M} &= \frac{\pi \Omega R^2}{3\beta^2} \Gamma^4(R),
 \label{eq_ring_classical_thermo}
\end{align}
while $\om{E} = \mathcal{P}$ for this one-dimensional system. It can be checked that $\om{E}$ agrees with the energy density $E$ computed using the well-known formula
\begin{align}
 E &= \int_{-\infty}^\infty \frac{R dk^\varphi}{2\pi} \frac{k^0}{e^{\beta(k^0 - R^2 \Omega k^\varphi)} - 1} \nonumber\\
 &= \frac{\pi \Gamma^2(R)}{6\beta^2} [2\Gamma^2(R) - 1] \equiv \om{E}.
 \label{eq_ring_classical_E}
\end{align}

\begin{figure}
    \centering
    \includegraphics[width=.99\linewidth]{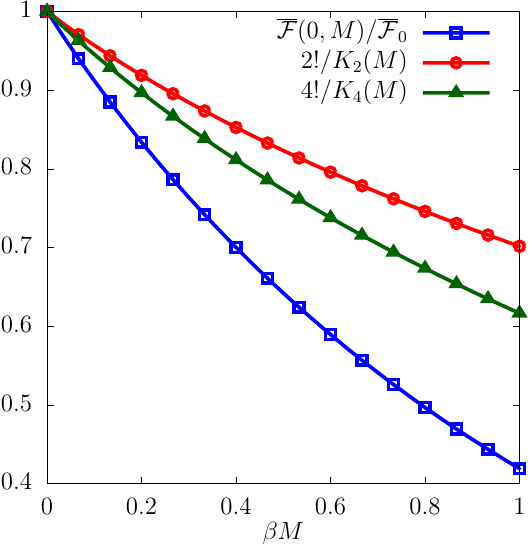}
    \caption{Effect of particle mass $M$ on free energy and shape coefficients $K_2$ and $K_4$, computed in relativistic kinetic theory. The blue line shows the ratio $\om{F}(0,M) / \om{F}_0$. with $\om{F}_0 = -P_0$ being the average free energy at vanishing mass [cf.~Eq.~\eqref{eq_E_large_ring_0}]. The red and green lines show the ratios $K_2(0) / K_2(M)$ and $K_4(0) / K_4(M)$, with $K_{2n}(0) = (2n)!$ corresponding to vanishing mass.}
    \label{fig:ring_RKT}
\end{figure}

Considering now the free energy at small values of the rotation parameter $\Omega$, we expand the free energy density in a power series in the velocity of a corotating particle 
\begin{equation}
 v_R = \Omega R\,,
\label{eq_v_R}
\end{equation}
at the system boundary $\rho = R$, following Ref.~\cite{Braguta:2023yjn}:
\begin{equation}
 \om{F} = \om{F}(0) \sum_{n = 0}^\infty \frac{v_R^{2n}}{(2n)!} K_{2n}, \quad 
 K_{2n} = \left.\frac{1}{\om{F}(0)} \frac{\partial^{2n} \om{F}}{\partial v_R^{2n}} \right|_{\Omega = 0},
 \label{eq_Kn_def}
\end{equation}
where $K_{2n}$ are $\Omega$-independent dimensionless coefficients, $\om{F}(0)$ is the free energy density in the absence of rotation, and we took into account that $\om{F}$ is an even function of $\Omega$. By construction, one gets $K_0 = 1$. Comparing Eqs.~\eqref{eq_Kn_def} and \eqref{eq_ring_classical_thermo}, it can be seen that 
\begin{equation}
 \om{F}(0) = -\frac{\pi}{6\beta^2}, \quad 
 K_{2n} = (2n)!, \quad 
 K_{2n+1} = 0.\label{eq_ring_K2n}
\end{equation}
Clearly, $\om{F}(0)$ and the $K_{2n}$ coefficients are independent of the system size.

The results for the coefficients $K_{2n}$ are also valid in a multicomponent non-interacting gas 
since they reflect the rotational response normalized per degree of freedom. The zero-rotation limit of the moment of inertia, $I_0$ of a one-component bosonic gas evaluates to
\begin{align}
 I_0 &\equiv \lim_{\Omega \rightarrow 0} I(\Omega) = -\lim_{\Omega \rightarrow 0} \frac{1}{\Omega} \frac{\partial \om{F}}{\partial \Omega} \nonumber \\
 & = -\om{F}(0) R^2 K_2 = 
 \frac{\pi R^2}{3\beta^2},
\end{align}
which follows from the definition $\om{M}(\Omega) = I(\Omega) \Omega$ of the moment of inertia $I(\Omega)$ in terms of the average angular momentum $\om{M}$, obtained via the thermodynamic relation~\eq{eq_thermo_rel}.

It is interesting to notice that recent first-principle simulations~\cite{Braguta:2023yjn} indicate that in the high-temperature limit, the rotating gluon gas possesses the dimensionless moment of inertia $K_2 = 2.23(39)$ consistent with our estimation~\eq{eq_ring_K2n}:
\begin{align}
    K_2 = 2\,, \qquad\ \text{[per one bosonic d.o.f.]}\,.
    \label{eq_K2}
\end{align}
This result is not unexpected since at sufficiently high temperatures, the gluon plasma becomes a weakly-interacting gas of gluons. It is worth mentioning that the value of $K_2$ receives corrections only if the quanta are massive or if interactions are taken into account.

The next non-zero coefficient in the series~\eq{eq_ring_K2n}, 
\begin{align}
    K_4 = 24\,, \qquad\ \text{[per one bosonic d.o.f.]}\,.
    \label{eq_K4}
\end{align}
corresponds to the correction to the free energy caused by the deformation of the rotating gas due to rotation. This correction also affects the moment of inertia, $I(\Omega) = I_0 + I_2 v_R^2/2 + \dots$, with the universal non-interacting coefficient $I_2/I_0 = 4$.

We now consider the effect of the particle mass on the $K_{2n}$ coefficients. To this end, we evaluate $\mathcal{F}$ from Eq.~\eqref{eq:ring_RKT_F} for the case when $k^0 = \sqrt{k_\varphi^2 / R^2 + M^2}$, with $M$ being the particle mass. Performing a Lorentz transformation to the local rest frame, we get
\begin{equation}
 \mathcal{F} = \frac{2R \gamma}{\beta} \int_0^\infty dk\, \ln(1 - e^{-\beta k^0 / \gamma}).
\end{equation}
Changing the integration variable to $z = k^0 / m$ and expanding the logarithm using $\ln (1 - x) = -\sum_{j = 1}^\infty x^j / j$, we arrive at 
\begin{equation}
 \mathcal{F}(\Omega,M) = -\frac{2R \gamma M}{\beta} \sum_{j = 1}^\infty \frac{1}{j} K_1 \left(\frac{j \beta M}{\gamma}\right).
 \label{eq_ring_RKT_Fexp}
\end{equation}
We can therefore evaluate the coefficients as 
\begin{align}
 \om{F}(0,M) &= -\frac{M}{\pi \beta} \sum_{j = 1}^\infty \frac{1}{j} K_1(j M \beta),\nonumber\\
 \overline{K}_2(M) &= -\frac{M^2}{\pi} \sum_{j = 1}^\infty K_2(j M \beta),\nonumber\\
 \overline{K}_4(M) &= -\frac{3M^2}{\pi} \sum_{j = 1}^\infty [j M \beta K_1(j M \beta) + 4 K_2(j M \beta)],
 \label{eq_ring_RKT_K2n}
\end{align}
where we introduced the notation $\overline{K}_{2n}(M) \equiv \om{F}(0, M) K_{2n}(M)$.
As shown in Fig.~\ref{fig:ring_RKT}, increasing $M$ decreases the average free energy $\om{F}(0,M)$, while the shape coefficients $K_2(M)$ and $K_4(M)$ increase with respect to their massless limits.

\subsection{Relativistic rotation, particle spectrum}\label{sec_ring_KG}

\begin{figure}[!thb]
\centerline{\includegraphics[width=0.3\textwidth,clip=true]{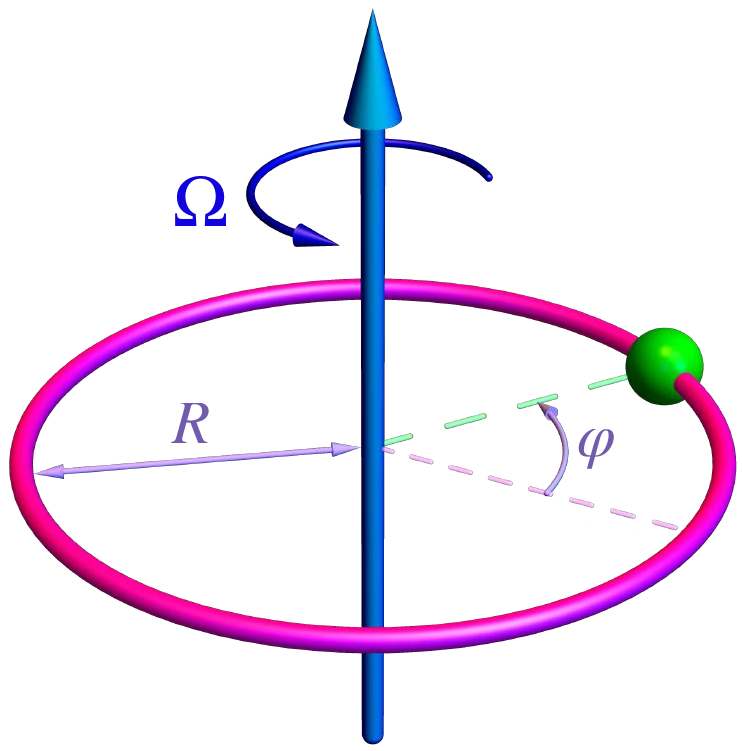}}
\caption{Illustration of a particle on a ring of the radius $R$ and the angular coordinate $\varphi$. The ring rotates with the angular velocity $\Omega$ counterclockwise.}
\label{fig_ring}
\end{figure}

In this section, we consider a free massless particle on a ring of a fixed radius $R$ with the angle coordinate $\varphi$ as shown in Fig.~\ref{fig_ring}. For a static ring, the particle wavefunction is described by the Klein-Gordon equation:
\begin{align}
    \left( \frac{\partial^2}{\partial t^2} - \frac{1}{R^2} \frac{\partial^2}{\partial \varphi^2} \right) \Phi(t,\varphi) = 0\,,
    \label{eq_Klein_ring_1}
\end{align}
which is formulated in the inertial, laboratory frame.

Let us consider the ring rotating with the constant angular velocity $\Omega$. The coordinates associated with the co-rotating reference frame (denoted by a tilde) are related to the laboratory coordinates as follows:
\beqn
t = {\tilde t}\,, \qquad {\tilde \varphi} = \varphi - \Omega t \ \ \mbox{mod} \ 2 \pi\,.
\eeqn
In the co-rotating frame, the Klein-Gordon equation~\eq{eq_Klein_ring_1} transforms into the following equation:
\begin{align}
    \left[ \left( \frac{\partial}{\partial \tilde t} - \Omega \frac{\partial}{\partial \tilde \varphi} \right)^2 - \frac{1}{R^2} \frac{\partial^2}{\partial \tilde \varphi^2}\right] \Phi(\tilde t,\tilde \varphi) = 0\,,
    \label{eq_Klein_ring_2}
\end{align}
which possesses the energy spectrum in the rotating reference frame:
\begin{align}
    {\tilde \omega}_m = \frac{1}{R} |m| - \Omega m\,, \qquad m \in \Z\,,
    \label{eq_varepsilon_m}
\end{align}
corresponding to the following eigenfunctions:
\begin{align}
    \Phi(\tilde t,\tilde \varphi)  = 
    \frac{1}{2\sqrt{\omega \pi R}}
    e^{ - i {\tilde \omega} \tilde t + i m \tilde \varphi}\,.
\end{align}
The energy spectrum~\eq{eq_varepsilon_m} is bounded from below provided the causality condition is satisfied:
\begin{align}
    R |\Omega| < 1\,.
    \label{eq_causality}
\end{align}
The thermodynamics of the system is determined in the rotating reference frame where all statistical distributions are set by the energy in the co-rotating frame ${\tilde \omega}_m$ rather than by its laboratory-frame counterpart~\eq{eq_lab_energy}.

\subsection{Free energy of rotating scalar field}\label{sec_ring_F}

We consider statistical mechanics of scalar particles in the rotating environment. The corresponding statistical sum,
\begin{align}
    {\mathcal Z} \equiv e^{- \beta \mathcal{F}}
    & = \prod_{m \in {\mathbb Z}} \sum_{n_m = 0}^\infty e^{- \beta (\omega_m - \Omega m) n_m} \nonumber \\
                 & = \prod_{m \in {\mathbb Z}} \left[ 1 - e^{- \beta (\omega_m - \Omega m)} \right]^{-1}\,,
    \label{eq_F_generic}
\end{align}
is formulated via the sum over states labeled by the angular momentum $m$ with the occupation number $n_m$ of system's levels that possess the total energy ${\tilde E}_{m,n_m} = {\tilde \omega}_m n_m$ in the rotating reference frame and the total angular momentum $L_{n,m_n} = m n_{m}$. In Eq.~\eq{eq_F_generic}, $\mathcal{F}$ stands for the free energy in the co-rotating reference frame. In the statistical sum, we do not take into account a $m=0$ contribution which corresponds to the zero-energy mode and contributes to the zero-point (Casimir) vacuum energy~\cite{Milton2001}. We concentrate on the thermal part of the free energy which possesses interesting fractal properties in the thermodynamic limit.

The thermodynamic free energy~\eq{eq_F_generic}, 
\begin{align}
\mathcal{F}(\Omega)  = \frac{1}{\beta} \sum_{m=1}^\infty \ln & \Bigl[ \Bigr(1 - e^{- \beta (1/R - \Omega)m} \Bigr) \nonumber \\
& \cdot \Bigl(1 - e^{- \beta (1/R + \Omega) m} \Bigr)  \Bigr]\,,
\label{eq_F_neutral}
\end{align}
can be evaluated explicitly:
\begin{align}
    \mathcal{F}(\Omega) = \frac{1}{12 R} + \frac{1}{\beta} 
    \ln & \biggl\{ \eta \left[ \frac{i \beta}{2\pi}  \left(\frac{1}{R} - \Omega\right)\right] \nonumber \\
               & \cdot \eta \left[ \frac{i \beta }{2\pi} \left(\frac{1}{R} + \Omega\right)\right]\biggr\},
\label{eq_F_Omega_2}
\end{align}
via the Dedekind $\eta$ function:
\beqn
    \eta(z) = e^{\frac{i \pi z}{12}} \prod_{n=1}^\infty \left( 1 - e^{2 \pi i n z} \right)\,.
\label{eq_Dedekind}
\eeqn

Notice that the Dedekind function~\eq{eq_Dedekind} is defined only in the upper complex plane ${\mathrm{Im}}\,z > 0$, which implies that the free energy~\eq{eq_F_Omega_2} is well-defined if and only if the causality condition~\eq{eq_causality} is satisfied. The causality condition is absent for the case of the imaginary rotation~\eq{eq_Omega_I}, when the angular frequency~$\Omega$ 
becomes a purely imaginary quantity. Indeed, according to the analytical properties of the Dedekind function~\eq{eq_Dedekind}, the free energy~\eq{eq_F_Omega_2} is a well-defined analytical function for any real value of the imaginary angular frequency~$\Omega_I$ at any radius of the ring~$R$. Consequently, the rigid imaginary rotation, contrary to rotation in Minkowski spacetime, can be formulated in the thermodynamic limit.

For convenience of our subsequent analysis, we consider all physical quantities in units of the inverse length $1/\beta$ of the imaginary time direction. We introduce the dimensionless length of the ring (the one-dimensional volume) $L$ and the frequency of rotation $\nu$, respectively:
\begin{align}
    L = \frac{2 \pi R}{\beta}\,, \qquad \nu = \frac{\beta \Omega_I}{2 \pi} \equiv \frac{\chi}{2\pi}\,.
\label{eq_l_nu}
\end{align}
The normalized frequency~$\nu$ corresponds to the normalized statistical angle~\eq{eq_chi}. 

The free energy density $\om{F}_I = \mathcal{F}_I / 2\pi R$ is given by
\begin{equation}
 \om{F}_I = \frac{\pi}{6\beta^2 L^2}
  + \frac{1}{\beta^2 L} \ln\left[\eta\left(\frac{i}{L}-\nu\right) \eta\left(\frac{i}{L}+\nu\right)\right].
  \label{eq_pressure_ring}
\end{equation}
After taking the thermodynamic limit $R \rightarrow \infty$ and in the absence of rotation, $\om{F}_I \rightarrow \om{F}_0 = -P_0 = -\pi / 6\beta^2$, as established by Eq.~\eqref{eq_E_large_ring_0}. Note that $\om{F}_0$ agrees with the classical expression $\om{F}(0)$ in Eq.~\eqref{eq_ring_K2n}.

The first term in Eq.~\eq{eq_pressure_ring} could be erroneously mistaken for the regularized zero-point (Casimir) energy contribution to the free energy. To show that this identification is not correct, let us consider the trivial case $\nu = 0$ which corresponds to a vanishing statistical angle, $\chi = 0$ (nontrivial angles will be considered shortly after). The low-temperature limit, $\beta \to \infty$, for a ring with a fixed radius $R$ corresponds to a vanishing parameter $L$. Using the following relation, valid for vanishingly-small positive~$L$,
\begin{align}
    \ln \eta \left( \frac{i}{L} \right) = - \frac{\pi}{12 L} + \dots,
\end{align}
(where the ellipsis denote subleading terms in the limit $L \to 0$), 
one gets from Eq.~\eq{eq_pressure_ring} that the normalized free energy vanishes in the low-temperature limit: 
\begin{align}
    \lim_{\beta \rightarrow \infty} \om{F}_I = 0.
\label{eq_f_bar_limit}
\end{align}

For the sake of convenience, we present here the expressions for the normalized Casimir free energy and the Casimir pressure,
respectively:
\begin{align}
    \om{F}_{\mathrm{Cas}} = -P_{\rm Cas} = \frac{1}{24 \pi R^2} = \frac{\pi}{6 \beta^2 L^2},
\label{eq_f_Cas}
\end{align}
which coincides with the first term in Eq.~\eqref{eq_pressure_ring}. The thermodynamic contribution~\eq{eq_pressure_ring} does not contain the zero-point energy since the latter should diverge in the $L \to 0$ limit~\eq{eq_f_Cas}, which is not the case~\eq{eq_f_bar_limit}. Therefore, Eq.~\eq{eq_pressure_ring} represents a purely thermodynamic contribution.

\subsection{Fractalization of thermodynamics}\label{sec_ring_fractal}

Using the property $\eta(-z^*) = [\eta(z)]^*$ valid for any complex number $z$ from the upper complex semi-plane, ${\mathrm{Im}}\, z > 0$, one gets for the thermodynamic part of the free energy density~\eq{eq_pressure_ring}, the following expression:
\begin{align}
    \om{F}_I = \frac{\pi}{6\beta^2 L^2} + \frac{2}{\beta^2 L} \ln \left|\eta \left(\nu + \frac{i}{L}
    \right)\right|\,,
    \label{eq_f_T}
\end{align}
where we used the notations in~\eq{eq_l_nu}.

The thermodynamic limit, $L \to \infty$, of the free energy on the ring~\eq{eq_f_T} can be deduced from the beautiful result of Ref.~\cite{nechaev2017number}, which relates the Dedekind $\eta$ function~\eq{eq_Dedekind} with the Thomae $f_{\mathsf T}$ function~\eq{eq_Thomae} as the following limit:
\begin{align}
    \lim_{\epsilon \to +0} \epsilon\, |\eta(x + i \epsilon)| = - \frac{\pi}{12} f_{\mathsf{T}}^2(x)\,.
    \label{eq_Nechaev}
\end{align}

Applying~\eq{eq_Nechaev} to the thermal part of the free energy density~\eq{eq_f_T}, we get that the (normalized) thermodynamic energy density ``fractalizes'' in the thermodynamic limit:
\begin{align}
    \lim_{L \to \infty}  \om{F}_I = -P_I =  - \frac{\pi}{6 \beta^2} f_{\mathsf{T}}^2(\nu).
    \label{eq_fbar_limit}
\end{align}
The non-analyticity of the Thomae function $f_{\mathsf T}$ has a fractal nature~\cite{nechaev2017number} implying the fractalization of thermodynamics under imaginary rotation~\cite{Chernodub:2022qlz}. The result in Eq.~\eq{eq_fbar_limit} implies that close to the thermodynamic limit, $L \gg 1$, the non-analytical fractal part of the free energy density dominates over the analytical term in the thermodynamic free energy~\eq{eq_f_T}. 
Interpreting the expression for $\om{F}$ in light of Eq.~\eqref{eq_E_large_ring_0}, we are led to define a rotation-dependent temperature $T_{\mathsf{T}}(\nu)$ via
\begin{align}
     T_{\mathsf T}(\nu) = \beta^{-1} f_{\mathsf T}(\nu),
     \label{eq_f_thermodynamic_limit}
\end{align}
which depends on the statistical parameter $\nu \equiv \beta \Omega_I/2\pi = p/q$ via the discontinuous Thomae function $f_{\mathsf T}$ as given in Eq.~\eq{eq_T_rational_bosons}. Notice that in the thermodynamic limit, the thermodynamics of the system is determined by the ninionic statistics~\eq{eq_ninionic}. This fact can be seen from the expression for pressure, given in Eq.~\eq{eq_fbar_limit}, which coincides with Eq.~\eq{eq_E_large_ring_Omega_I}.

The fractalization~\eq{eq_fbar_limit}, characterized by the non-analytical behaviour of free energy, is achieved only in the thermodynamic limit when the radius $R$ of the ring becomes infinitely large. At any finite $R$, all thermodynamic characteristics of the systems are analytical. Thus, it is instructive to see how thermodynamics acquires its fractal properties under imaginary rotation as the radius of the ring increases. 

\begin{figure}[!thb]
\centerline{\includegraphics[width=0.475\textwidth,clip=true]{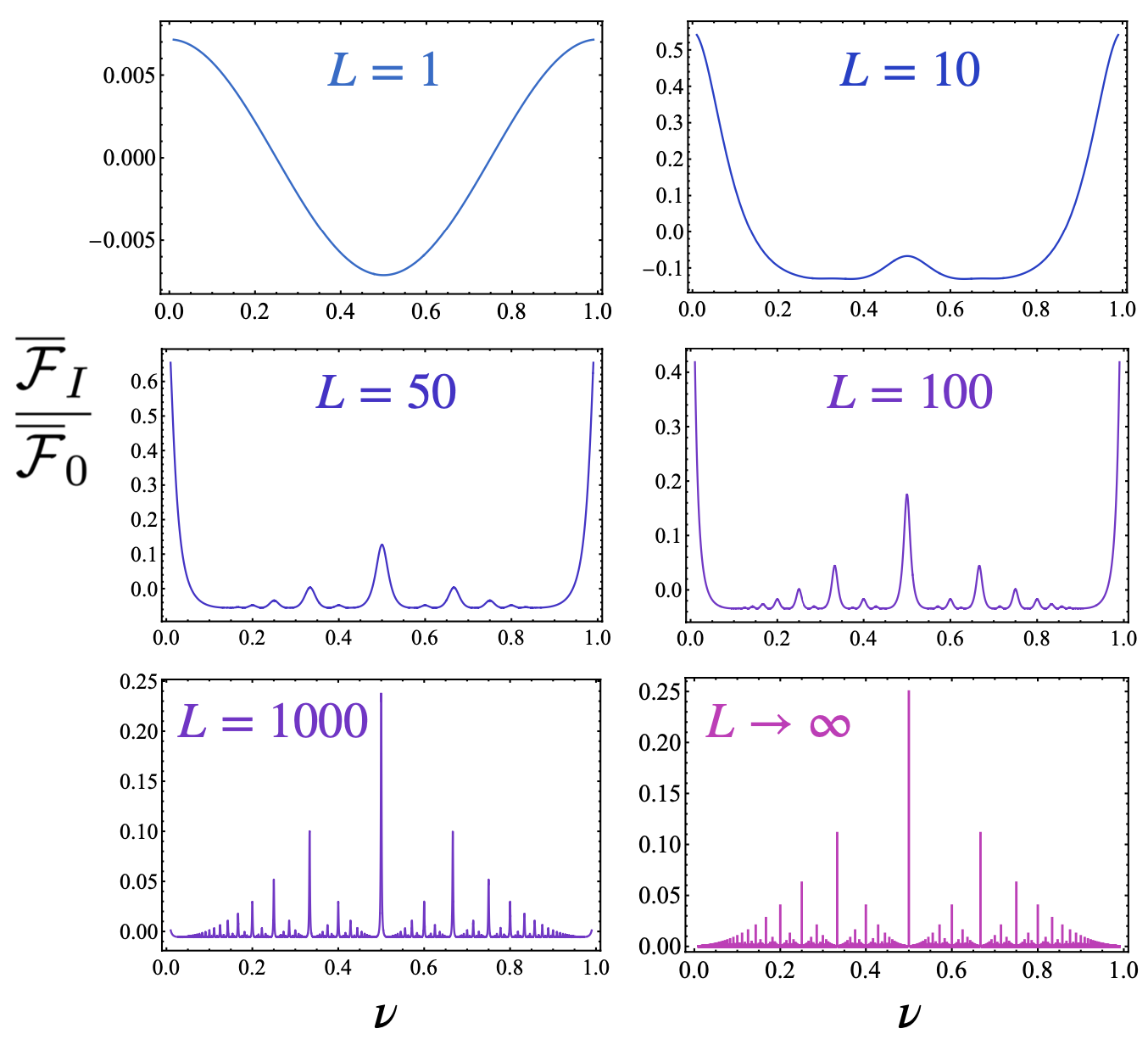}}
\caption{Fractalization of thermodynamics of scalar particles on the ring under the imaginary rotation $\Omega_I$: free energy density $-\om{F}_I$ Eq.~\eq{eq_pressure_ring}, shown in units of pressure~$P_0$ for an infinite ring $L \to \infty$ in the absence of rotation, Eq.~\eq{eq_E_large_ring_0}, as a function of the normalized statistical angle $\nu = \chi/(2\pi) \equiv \beta \Omega_I/(2\pi)$ for various (normalized) lengths $L$ of the ring~\eq{eq_l_nu}. The plots at finite values of $L$ are given for the analytical behaviour~\eq{eq_pressure_ring} in terms of the Dedekind $\eta$ function~\eq{eq_Dedekind}. The behaviour in the thermodynamic limit, $L \to \infty$, corresponds to the non-analytical fractal result~\eq{eq_fbar_limit} expressed via the Thomae function~\eq{eq_Thomae}. The behaviour near the points $\nu=0$ and $\nu=1$ is not shown to preserve a convenient vertical scale.}
\label{fig_ring_thermal}
\end{figure}

In Fig.~\ref{fig_ring_thermal} we show the average free energy of bosons~\eq{eq_pressure_ring} as a function of the normalized statistical angle $\nu$, Eq.~\eq{eq_l_nu}, at various (normalized) lengths $L$ of the ring. Pressure, which is a smooth analytical function of $\nu$ at small radius of the ring, develops a series of minima and maxima as the length of the ring increases. For large sizes $L \sim 10^3$, pressure of bosonic particles develops self-similar features. At  $L \sim 5 \times 10^3$, the pressure becomes almost indistinguishable from its limiting form ($L \to \infty$) given by Eq.~\eq{eq_E_large_ring_Omega_I} and governed by the fractal properties of the Thomae function~\eq{eq_Thomae}. In this limit, the thermodynamic pressure becomes a fractal dictated by the ninionic statistics~\eq{eq_ninionic}.

To see how the other thermodynamic quantities fractalize in the imaginary rotation case, it is convenient to rewrite the free energy in Eq.~\eqref{eq_F_neutral}, as follows.
Employing a series expansion of the Bose-Einstein factor, 
\begin{equation}
 \ln[1 - e^{-\frac{\beta m}{R} (1 \pm \Omega R)}] = -\sum_{j = 1}^\infty \frac{1}{j} e^{-\frac{j \beta m}{R} (1 \pm \Omega R)},
\end{equation}
we arrive at
\begin{equation}
 \mathcal{F} = -\frac{1}{\beta} \sum_{j = 1}^\infty \frac{1}{j} \left[\frac{1}{e^{\frac{j \beta}{R}(1 - \Omega R)} - 1} + \frac{1}{e^{\frac{j \beta}{R}(1 + \Omega R)} - 1}\right].
 \label{eq_ring_peasant_F}
\end{equation}
Switching now to imaginary rotation, $\Omega = i \Omega_I$, we have 
\begin{equation}
 \mathcal{F}_I = -\frac{1}{\beta} \sum_{j = 1}^\infty \frac{1}{j} 
 \frac{e^{-\pi j / L} \sinh(\pi j / L) - \sin^2(\pi j \nu)}
 {\sinh^2(\pi j /L) + \sin^2(\pi j \nu)}.
 \label{eq_ring_peasant_F_nu}
\end{equation}
The pressure and average angular momentum can be derived from Eq.~\eqref{eq:dF}:
\begin{align}
 \mathcal{P}_I &= \frac{\pi}{\beta^2 L^2} \sum_{j = 1}^\infty \frac{\sinh^2(\pi j / L) - \cosh(2\pi j / L) \sin^2(\pi j \nu)}{[\sinh^2(\pi j /L) + \sin^2(\pi j \nu)]^2}, \nonumber\\
 \om{M}_I &= \frac{1}{2\beta L} \sum_{j = 1}^\infty \frac{\sinh(2\pi j / L) \sin(\pi j \nu) \cos(\pi j \nu)}{[\sinh^2(\pi j /L) + \sin^2(\pi j \nu)]^2}, \label{eq_ring_peasant_thermo_nu}
\end{align}
where we introduced $\om{M}_I = -i \om{M} = \partial \om{F}_I / \partial \Omega_I$, while $\om{S}_I =\beta (\mathcal{P}_I - \om{F}_I) + 2\pi \nu \om{M}_I$. Using Eq.~\eqref{eq_Euler}, it can be shown that $\om{E}_I = \mathcal{P}_I$.

To reveal the fractal features of the expressions in Eqs.~\eqref{eq_ring_peasant_F_nu} and \eqref{eq_ring_peasant_thermo_nu}, we consider $\nu$ to be a rational number, represented by the irreducible fraction $\nu = p / q$. Then, the trigonometric function $\sin^2(\pi j \nu)$ and $\sin(\pi j \nu) \cos(\pi j \nu)$ have a periodicity with respect to $j \rightarrow j + q$. To take advantage of this periodicity, we write $j = q Q + j'$, with $0 \le Q < \infty$ and $1 \le j' \le q$, covering the entire $1 \le j < \infty$ summation range.
Out of the new range for $j'$, the value $j' = q$ is special, since then the trigonometric functions vanish:
\begin{equation}
 \left.\sin (\pi j \nu)\right|_{j = q(Q + 1)} = \sin [\pi p (Q + 1)] = 0.
\end{equation}
In what follows, we split our observables $\mathcal{A} \in \{\om{F}_I, \mathcal{P}_I, \om{M}_I, \om{S}_I, \om{E}_I\}$ as 
\begin{equation}
 \mathcal{A} = \mathcal{A}_q + \delta \mathcal{A}, \label{eq_ring_fractal_Adec}
\end{equation}
where $\mathcal{A}_q$ corresponds to the $j' = q$ contribution and $\delta \mathcal{A}$ collects all terms with $1 \le j' < q$. In particular, we have 
\begin{align}
 \om{F}^I_q &= -\frac{1}{\beta^2 L q} \sum_{Q = 1}^\infty \frac{1}{Q} \frac{\exp(-\pi q Q / L)}{\sinh(\pi q Q / L)} \rightarrow -\frac{\pi}{6\beta^2 q^2}, \nonumber\\
 \mathcal{P}^I_q &= \frac{\pi}{\beta^2 L^2} \sum_{Q = 1}^\infty \frac{1}{\sinh^2(\pi q Q / L)} \rightarrow \frac{\pi}{6\beta^2 q^2},\nonumber\\
 \om{M}^I_q &= 0,
 \label{eq_ring_fractal_peasant}
\end{align}
where the right arrow $\rightarrow$ implies taking the thermodynamic limit $L \rightarrow \infty$.
The above also imply that
\begin{equation}
 \om{S}^I_q = \beta(\mathcal{P}^I_q - \om{F}^I_q) \rightarrow \frac{\pi}{3\beta q^2}.
 \label{eq_ring_fractal_peasant_S}
\end{equation}
The presence of the denominator $q$ of the irreducible fraction $p/q = \nu$ in Eqs.~\eqref{eq_ring_fractal_peasant}--\eqref{eq_ring_fractal_peasant_S} is the hallmark of fractal thermodynamics and hence of loss of analyticity. 

The remainders $\delta \mathcal{A}$ in Eq.~\eqref{eq_ring_fractal_Adec} play the role of ``defractalizing'' the expectation values $\mathcal{A}$ for small systems (small $L$). At large $L$, it is not difficult to see that $\delta \mathcal{A}$ decays to $0$. For example, in the case of the free energy, we have
\begin{align}
 \delta \om{F}_I &\simeq \frac{1}{q \beta^2 L} \sum_{j = 1}^{q-1} \sum_{Q = 0}^\infty 
 \frac{\exp\left[-\frac{2 \pi q}{L}(Q + \frac{j}{q})\right]}{Q + \frac{j}{q}}.
\end{align}
The above sums are readily evaluated when considering the derivative of $\delta \om{F}_I L$ with respect to $L$:
\begin{align}
 \frac{\partial (\delta \om{F}_I L)}{\partial L} &= \frac{2\pi}{\beta^2 L^2} \frac{1 - e^{-\frac{2\pi}{L} (q - 1)}}{(1 - e^{-2\pi q / L})(e^{2\pi / L} - 1)} \nonumber\\
 &\simeq \frac{q-1}{q \beta^2 L} - \frac{\pi^2 (q-1)}{3\beta^2 L^3} + O(L^{-5}).
\end{align}
Integrating now the series expansion appearing on the second line of the above equation, we find the leading-order behaviour for $L\rightarrow \infty$:
\begin{equation}
 \delta\om{F}_I \simeq \frac{q-1}{q \beta^2 L} \ln L + O(L^{-1}).
\end{equation}

\begin{figure}
    \centering
    \includegraphics[width=.99\linewidth]{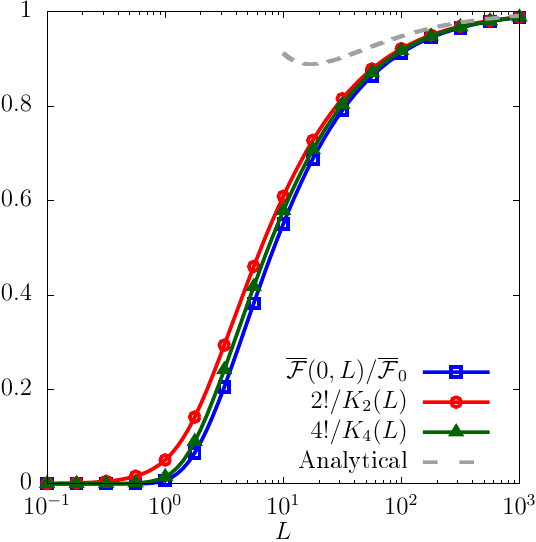}
    \caption{Effect of ring radius $R = \beta L / 2\pi$ on the average free energy $\om{F}(0,L)$ and on the coefficients $K_2$ and $K_4$, computed in the limit of vanishing rotation from Eq.~\eqref{eq:ring_QFT_K2n}. The quantity $\om{F}_0 = -P_0$ is given in Eq.~\eqref{eq_E_large_ring_0}. The dashed gray curve corresponds to the anayltical estimates in Eqs.~\eqref{eq:ring_QFT_Kbar2n_an} and \eqref{eq:ring_QFT_K2n_an}.}
    \label{fig:ring_QFT_M0}
\end{figure}

Finally, we evaluate the $K_{2n}$ coefficients defined in Eq.~\eqref{eq_Kn_def}, which in the present case reduce to
\begin{align}
 \om{F}(0) &= \frac{2}{L \beta^2} \sum_{m = 1}^\infty \ln (1 - e^{-2\pi m / L}), \nonumber\\
 \overline{K}_2 &= -\frac{2\pi^2}{L^3 \beta^2} \sum_{m = 1}^\infty \frac{m^2}{\sinh^2(\pi m / L)}, \nonumber\\
 \overline{K}_4 &= -\frac{4\pi^4}{L^5 \beta^2} \sum_{m = 1}^\infty \frac{m^4[2 + \cosh(2\pi m / L)]}{\sinh^4(\pi m / L)},
 \label{eq:ring_QFT_K2n}
\end{align}
where, as before, $\overline{K}_{2n} \equiv \om{F}(0) K_{2n}$. The above sums are performed numerically and the results can be seen in Fig.~\ref{fig:ring_QFT_M0}. Strikingly, the shape coefficients $K_{2n}(L)$ exceed their thermodynamic limit, $K_{2n}(L) > K_{2n}(\infty) = 2n!$, for any finite value of $L$.

For the purpose of evaluating the coefficients $\om{F}(0,L)$ and $K_{2n}(L)$ analytically, we rewrite $\om{F}$ from Eq.~\eqref{eq_ring_peasant_F_nu} as 
\begin{align}
 \om{F} &= -\frac{2}{\beta^2 L} \sum_{j = 1}^\infty \frac{e^{-2\pi j / L}}{j} \Delta \om{F}, \nonumber\\
 \Delta \om{F} &= \frac{(1 - e^{-2\pi j / L}) - 2\sin^2(\pi j \nu)}{(1 - e^{-2\pi j / L})^2 + 4 e^{-2\pi j /L} \sin^2(\pi j \nu)}.
\end{align}
The coefficients $K_{2n}$ can be computed by expanding $\Delta \om{F}$ in powers of $\nu$,
\begin{equation}
 \Delta\om{F} = \sum_{n = 0}^\infty \Delta \om{F}_{(2n)} \left(\frac{2\pi j}{L}\right)^{2n} \frac{(\nu L)^{2n}}{(2n)!},
\end{equation}
where we took into account that $\om{F}$ is an even function of $\nu$. Specifically, we have 
\begin{align}
 \Delta\om{F}_{(0)} &= \frac{1}{1 - e^{-2\pi j / L}},\quad
 \Delta\om{F}_{(2)} = -\frac{1 +e^{-2\pi j/L}}{(1-e^{-2\pi j/L})^3},\nonumber\\
 \Delta\om{F}_{(4)} &= \frac{1 + 11 e^{-2\pi j/L} + 11 e^{-4\pi j/L} + e^{-6\pi j/L}}{(1-e^{-2\pi j/L})^5}.
\end{align}
Taking into account that $(\nu L)^{2n} = (-1)^n v_R^{2n}$, we can identify 
\begin{equation}
 \overline{K}_{2n} = -\frac{2(-1)^n}{\beta^2 L} \sum_{j = 1}^\infty \frac{e^{-2\pi j/ L}}{j} 
 \left(\frac{2\pi j}{L} \right)^{2n} \Delta \om{F}_{(2n)}.
\end{equation}
We further expand the coefficients $\Delta\om{F}_{(2n)}$ in powers of $L^{-1}$:
\begin{equation}
 \Delta\om{F}_{(2n)} = \sum_{k = 0}^\infty \frac{\Delta\om{F}_{(2n,k)}}{k!} \left(\frac{2\pi j}{L}\right)^{k - 2n - 1}.
\end{equation}
In the above, we denoted 
\begin{equation}
 \Delta\om{F}_{(2n,k)} = \left. \frac{d^k}{dx^k}[x^{2n+1} \Delta\om{F}_{(2n)}]\right|_{x = 0},
\end{equation}
where the derivative is taken with respect to $x = 2\pi j / L$, and we took into account that $\Delta \om{F}_{(2n)} \sim x^{-2n-1}$ when $x \rightarrow 0$. We arrive at
\begin{equation}
 \overline{K}_{2n} = \frac{(-1)^{n+1}}{\pi \beta^2} \sum_{k = 0}^\infty \frac{\Delta\om{F}_{(2n,k)}}{k!} \left(\frac{2\pi}{L}\right)^k \sum_{j = 1}^\infty \frac{e^{-2\pi j / L}}{j^{2-k}}.
\end{equation}
The sum over $j$ can be taken in terms of the polylogarithm,
\begin{equation}
 \sum_{j = 1}^\infty \frac{e^{-2\pi j / L}}{j^n} = {\rm Li}_n(e^{-2\pi / L}),
\end{equation}
leading to
\begin{equation}
 \overline{K}_{2n} = \frac{(-1)^{n+1}}{\pi \beta^2} \sum_{k = 0}^\infty \frac{\Delta\om{F}_{(2n,k)}}{k!} \left(\frac{2\pi}{L}\right)^k {\rm Li}_{2 -k}(e^{-2\pi / L}).
\end{equation}
Using the series expansions 
\begin{align}
 {\rm Li}_2(e^{-2\pi / L}) &= \frac{\pi^2}{6} - \frac{2\pi}{L} \left(1 + \ln\frac{L}{2\pi}\right) + O(L^{-2}), \nonumber\\
 {\rm Li}_1(e^{-2\pi / L)} &= \ln \frac{L}{2\pi} + \frac{\pi}{L} + O(L^{-2}),\nonumber\\
 {\rm Li}_0(e^{-2\pi / L}) &= \frac{L}{2\pi} + O(L^0),\nonumber\\
 {\rm Li}_{-n}(e^{-2\pi / L}) &= \left(\frac{L}{2\pi}\right)^{n+1} n! + O(L^0),
\end{align}
it can be seen that the $k = 0$ term gives the leading-order, $L$-independent contribution to $\om{F}(0) K_{2n}$, while the $k = 0$ and $k = 1$ terms contribute to the next-to-leading-order $L^{-1} \ln L$ contribution. The coefficient of the $L^{-1}$ term receives contribution from all values of $k$:
\begin{multline}
 \overline{K}_{2n} = \frac{(-1)^{n+1}}{\pi \beta^2} \Bigg[\frac{\pi^2}{6} \Delta \om{F}_{(2n,0)} \\
 + \frac{2\pi}{L} \ln\frac{L}{2\pi} \left(\Delta\om{F}_{(2n,1)} - \Delta\om{F}_{(2n,0)}\right) \\
 -\frac{2\pi}{L}\Delta\om{F}_{(2n,0)} + \frac{2\pi}{L} \sum_{k = 0}^\infty \frac{\Delta\om{F}_{(2n,k+2)}}{(k+1)(k+2)}\Bigg].
\end{multline}
Alas, the sum over $k$ seems not to converge. This is easily visible in the case of $\om{F}(0)$, when 
\begin{equation}
 \Delta\om{F}_{0,k} = (-1)^k B_k,
\end{equation}
where $B_k$ represent the Bernoulli numbers, with $B_0 = 1$, $B_1 = -1/2$, $B_2 = 1/6$, $\dots$. Since $\sum_{k = 0}^\infty (-1)^k B_{k+2} / [(k+1)(k+2)]$ diverges, we refrain from computing the $L^{-1}$ term and list only the leading and next-to-leading terms, while warning the reader that the $L^{-1} \ln L$ term may also be flawed:
\begin{align}
 \om{F}(0) &= -\frac{\pi}{6\beta^2} + \frac{1}{L \beta^2} \ln \frac{L}{2\pi} + O(L^{-1}), \nonumber\\
 \overline{K}_2 &= -\frac{\pi}{3\beta^2} + O(L^{-1}), \nonumber\\
 \overline{K}_4 &= -\frac{4\pi}{\beta^2} + O(L^{-1}),
 \label{eq:ring_QFT_Kbar2n_an}
\end{align}
where we took into account that $\Delta \om{F}_{(2,0)} = \Delta \om{F}_{(2,1)} = -2$ and 
$\Delta \om{F}_{(4,0)} = \Delta \om{F}_{(4,1)} = 24$. Thus, we can extract $K_2$ and $K_4$ as 
\begin{equation}
 K_2 \simeq \frac{2}{1 - \frac{6}{\pi L} \ln \frac{L}{2\pi}}, \quad 
 K_4 \simeq \frac{24}{1 - \frac{6}{\pi L} \ln \frac{L}{2\pi}},
 \label{eq:ring_QFT_K2n_an}
\end{equation}
where we neglected $O(L^{-1})$ contributions.
A comparison between the above analytical estimates and the exact numerical evaluation of $\om{F}(0)$, $K_2$ and $K_4$ is shown in Fig.~\ref{fig:ring_QFT_M0}.

\subsection{Analytical continuation: the disk of analyticity}\label{sec_ring_analyticity}

Finally, let us discuss how the fractalization of thermodynamics in the thermodynamic limit leads to the absence of the analytical continuation from the imaginary angular frequencies to the real ones. In other words, we would like to see that the thermodynamic quantities obtained in an infinite volume limit at imaginary rotation cannot be directly connected to the thermodynamics of real rotation. Qualitatively, the validity of this statement can be deduced from both mathematical and physical arguments. 

Mathematically, it is clear that a non-analytical function cannot be analytically continued to an analytical domain because the result will depend on the prescription used for the continuation procedure. Moreover, at the imaginary-rotating side in the thermodynamic limit, the pressure $P$ cannot be expressed as a function of the imaginary velocity squared, $\Omega_I^2$, Eq.~\eq{eq_E_large_ring_Omega_I}, which renders inapplicable the continuation prescription to the real rotation summarized in Eq.~\eqref{eq_continuation}.

Physically, the causality condition~\eq{eq_causality} is incompatible with the continuation prescription~\eq{eq_continuation} in the thermodynamic limit, $R \to \infty$, for any finite $\Omega_I$. However, outside of the thermodynamic limit at any finite $R$, the analytical continuation does exist. Let us briefly discuss this point for the example of the ring. 

Since the length of the ring is always a positive number, $L>0$, the argument of the Dedekind $\eta$ function in the free energy density~\eq{eq_f_T} always belongs to the upper part of the complex plane, where the Dedekind function is an analytical and well-defined function. Therefore, the imaginary rotation is well-defined at any imaginary angular frequency~$\nu$, contrary to its real counterpart~\eq{eq_F_Omega_2}. 

It is convenient to write explicitly the normalized free energies $\beta^2 \om{F}$ for real ($\om{F}^{\,\mathrm{Re}}$) and imaginary ($\om{F}^{\, \mathrm{Im}}$) angular frequencies, respectively:
% \begin{align}
%     {\bar P}^{\,\mathrm{Re}}(x_R,y) & = -\frac{\pi y^2}{6} - y\, \ln \Bigl[ \eta \left(i x_R + i y \right) \eta \left(- i x_R + i y \right)\bigr],
% \label{eq_bar_f_real} \\
%     {\bar P}^{\, \mathrm{Im}}(x_I,y) & = -\frac{\pi y^2}{6} - y\, \ln \bigl[ \eta \left( x_I + i y \right) \eta\left( - x_I + i y \right)\bigr].
% \label{eq_bar_f_imaginary}
% \end{align}
\begin{align}
    \beta^2 \om{F}^{\,\mathrm{Re}}(x_R,y) & = \frac{\pi y^2}{6} + y\, \ln \Bigl[ \eta \left(i x_R + i y \right) \eta \left(- i x_R + i y \right)\bigr],
\label{eq_bar_f_real} \\
    \beta^2 \om{F}^{\, \mathrm{Im}}(x_I,y) & = \frac{\pi y^2}{6} + y\, \ln \bigl[ \eta \left( x_I + i y \right) \eta\left( - x_I + i y \right)\bigr].
\label{eq_bar_f_imaginary}
\end{align}
Here we defined the following real-valued quantities:
\begin{align}
    x_R =\frac{\beta \Omega}{2 \pi}, \qquad x_I =\frac{\beta \Omega_I}{2 \pi}, \qquad y = \frac{\beta}{2 \pi R}\,,
\end{align}
which represent the normalized angular frequencies for real and imaginary rotation ($x_R$ and $x_I$, respectively), and the inverse size of the ring~$y>0$. The analytical continuation can be formulated in terms of the relation between~\eq{eq_bar_f_real} and \eq{eq_bar_f_imaginary}.

Despite similarity of Eqs.~\eq{eq_bar_f_real} and \eq{eq_bar_f_imaginary}, these quantities have different properties. The free energy for real rotation \eq{eq_bar_f_real} is defined only in the strip $-y < x_R < y$ because the Dedekind eta function is defined only in the upper part of the complex plane (excluding the real axis). Physically, the same condition coincides with the causality requirement~\eq{eq_causality}. The pressure of the gas under imaginary rotation \eq{eq_bar_f_imaginary} is defined for any real-valued $x_I \in {\mathbb R}$. 

Equations~\eq{eq_bar_f_real} and \eq{eq_bar_f_imaginary} can be written in the following unified form:
\begin{align}
    %-{\bar P}(z,z_0) 
    \beta^2 \om{F}(z,z_0)
    & = - \frac{\pi z_0^2}{6} +   y\, \ln \Bigl[ \eta \left(i z + z_0 \right) \eta \left(- i z + z_0 \right)\bigr]\,,
    \label{eq_barf_z}
\end{align}
with 
\begin{align}
    z = x_R + i x_I \equiv \frac{\beta \left( \Omega + i \Omega_I \right)}{2\pi},
    \quad 
    z_0 = i y \equiv \frac{i}{L}.
    \label{eq_z_z0}
\end{align}
For any $y>0$, the analyticity properties of the free energy density at real rotation~\eq{eq_bar_f_real} imply that function~\eq{eq_barf_z} can be expanded in a series of powers of $z$ around the point $z = 0$ in the disk $|z| < |z_0|$ of radius $|z_0| =  y > 0$. In the original notations, the disk of analyticity can be defined as the generalization of the causality condition~\eq{eq_causality} in the plane of complex angular frequencies:
\begin{align}
    \left( \Omega^2 + \Omega_I^2 \right) R^2 < 1\,.
\label{eq_Omega_R2}
\end{align}

\begin{figure}[!thb]
\centerline{\includegraphics[width=0.45\textwidth,clip=true]{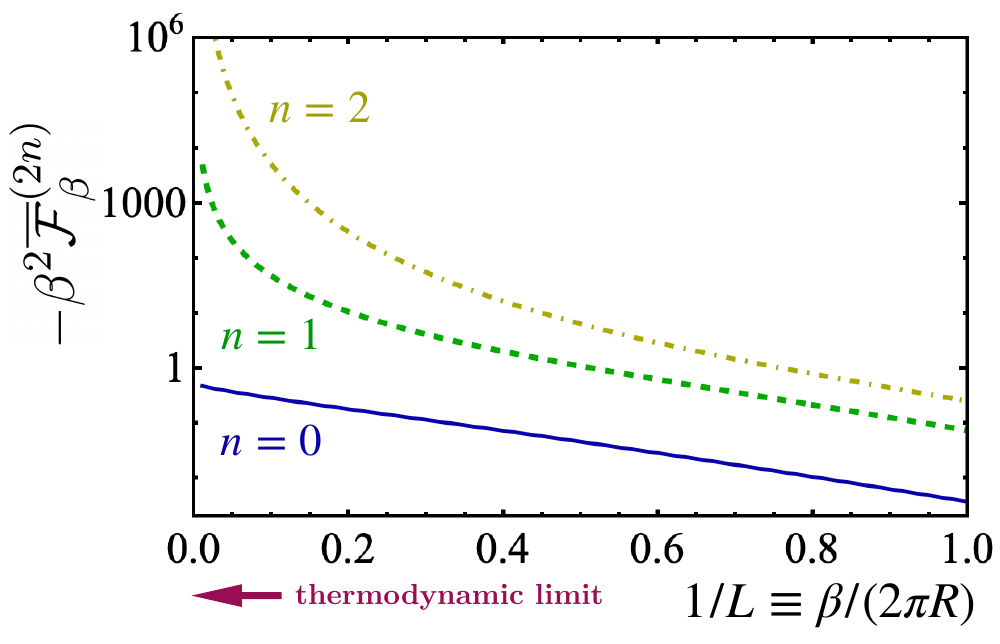}}
\caption{The first three nonzero coefficients in the series~\eq{eq_f_T_series} of the free energy of the ring as a function of the (inverse) normalized radius $1/L$. The direction of the thermodynamic limit is shown by the arrow.}
\label{fig_fi}
\end{figure}

The free energy $\om{F}$ can be written as the following series
\begin{align}
    \om{F}(z,z_0) = \sum_{n=0}^\infty 
    %{\bar P}^{(2 n)}_\beta(z_0) 
    \om{F}_\beta^{(2n)}
    z^{2n}\,, \quad |z| < |z_0|\,,
\label{eq_f_T_series}
\end{align}
with the first two coefficients in the explicit form:
% \begin{align}
%    {\bar P}^{(0)}_\beta(i y) & = -\frac{\pi y^2}{6} - y \log \bigl(\eta^2 (i y)\bigr) \,. \\
%    {\bar P}^{(2)}_\beta(i y) & = y \frac{\bigl(\eta''(i y)\bigr)^2 \eta (i y) - \bigl(\eta'(i y)\bigr)^2}{\eta^2 (i y)}\,.
% \end{align}
\begin{align}
   \beta^2 \om{F}^{(0)}_\beta(i y) & = \frac{\pi y^2}{6} + y \log \bigl(\eta^2 (i y)\bigr) \,. \\
   \beta^2 \om{F}^{(2)}_\beta(i y) & = -y \frac{\bigl(\eta''(i y)\bigr)^2 \eta (i y) - \bigl(\eta'(i y)\bigr)^2}{\eta^2 (i y)}\,.
\end{align}
One can check explicitly that the coefficients of the series~\eq{eq_f_T_series} diverge in the thermodynamic limit which is consistent with the shrinking radius of convergence~\eq{eq_Omega_R2} as $R \to \infty$. We show the first three non-zero coefficients $\beta^2 \om{F}^{(2n)}_\beta$ in Fig.~\ref{fig_fi}.

One can also rewrite the free energy~\eq{eq_f_T_series} in terms of physical variables:
% \begin{align}
%     {\bar P}_\beta({\mathit \Omega},R) = \sum_{n=0}^\infty P_{2n}(\beta, R) {\mathit \Omega}^{2n} ,
% \qquad 
% |{\mathrm{Re}}\, {\mathit \Omega}| R < 1\,,
% \label{eq_f_beta_series}
% \end{align}
\begin{align}
    \om{F}({\mathit \Omega},R) = \sum_{n=0}^\infty \om{F}_{2n} (\beta, R) {\mathit \Omega}^{2n} ,
\qquad 
|{\mathrm{Re}}\, {\mathit \Omega}| R < 1\,,
\label{eq_f_beta_series}
\end{align}
where ${\mathit \Omega} = \Omega + i \Omega_I$ and
% \begin{align}
% P_{2n}(\beta,R) \equiv \biggl(\frac{\beta}{2 \pi}\biggr)^{2 n} {\bar P}^{(2n)}_\beta \biggl(\frac{i \beta}{2 \pi R}\biggr)\,,
% \end{align}
\begin{align}
\om{F}_{2n}(\beta,R) \equiv \biggl(\frac{\beta}{2 \pi}\biggr)^{2 n} \om{F}_\beta^{(2n)} \biggl(\frac{i \beta}{2 \pi R}\biggr)\,,
\end{align}
and the radius of convergence in the complex $\Omega$ plane is determined by Eq.~\eq{eq_Omega_R2}: ${\mathit \Omega}_c = 1/R$. The radius shrinks to zero, ${\mathit \Omega}_c \to 0$ as $R \to \infty$, thus implying the absence of the direct analytical continuation between real and imaginary angular frequencies in the thermodynamic limit. Thus, thermodynamics of an infinite-volume system subjected to imaginary rotation is not directly connected to the thermodynamics of real rotation.

\subsection{How fractalization emerges as volume grows}\label{sec_ring_fractalization}

Figure~\ref{fig_ring_thermal} shows that at any fixed statistical angle $\chi = 2 \pi\nu$ (or, equivalently, at any imaginary frequency $\Omega_I$) and any finite radius $L$, the free energy is described by a smooth analytical function of $\chi$. For a rational normalized angle $\nu = p/q$ with coprime integer numbers $p$ and $q$ ($0 < p < q$), the pressure depends both on the numerator $p$ and the denominator $q$ (we remind that in these units, $\Omega_I = (2 \pi/\beta) p/q$). However, as the length $L$ of the ring increases, the pressure turns into a fractal, implying that it loses the sensitivity to the numerator $p$ and keeps only the dependence on the denominator $q$ that defines the imaginary frequency.

\begin{figure*}[!thb]
\centerline{\includegraphics[width=0.75\textwidth,clip=true]{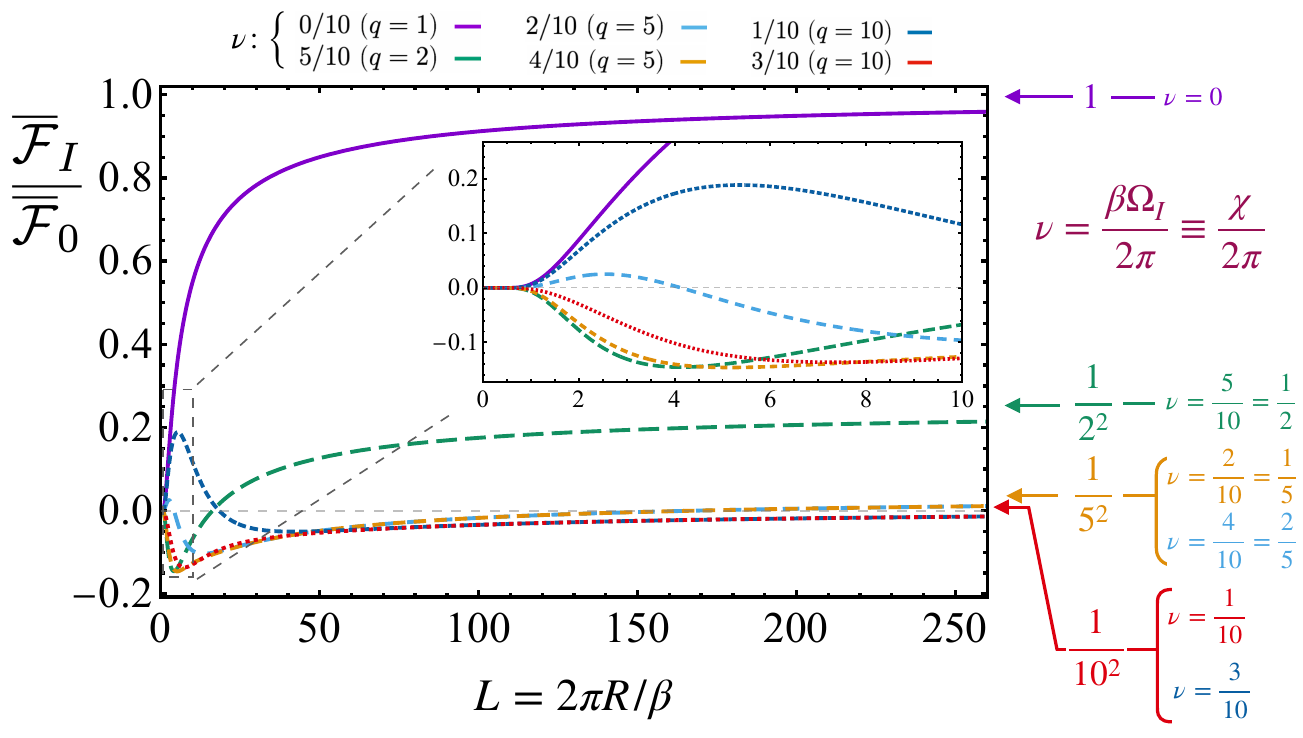}}
\caption{The thermal contribution to the free energy, $\om{F}_I \equiv \om{F}_I(L, \nu),$ as a function of the (normalized) length of the ring $L = 2 \pi R/\beta$ for various (normalized) statistical angles $\nu = \chi / (2 \pi)$ with the rational values $\nu = n/10$ at $n=0,1, \dots, 5$. The free energy is normalized to its value $\om{F}_0 = -P_0$, Eq.~\eq{eq_E_large_ring_0}, for a nonrotating ring, in the infinite-volume limit, $L \to \infty$. For rational~$\nu$, the free energy~\eq{eq_fbar_limit} in the infinite-volume limit takes fractal values (shown by the arrows) dictated by the Thomae function~\eq{eq_Thomae}. The inset shows the zoom in on the small-radius region.}
\label{fig_ring_pressure}
\end{figure*}

This curious fractalization transition is shown in Fig.~\ref{fig_ring_pressure} for the particular set of imaginary frequencies $\Omega_I \equiv 2\pi \nu/\beta = p \pi/(5 \beta)$ with $p = 0,1, \dots, 9$. Given the periodicity~\eq{eq_chi_periodicity} under $\nu \to \nu + 1$, as well the reflection symmetry~\eq{eq_chi_reflection} with respect to $\nu \to - \nu$, applied to the pressure, this particular choice leaves us with six distinct values of the normalized statistical angle: $\nu = 0/10, 1/10, \dots, 5/10$. 

At small and moderate ring sizes up to $L \simeq 4$, the free energy $\om{F} = \om{F}(L,\nu)$ depends on the normalized statistical angle $\nu$ monotonically, with $-\om{F}(L,\nu_a) < -\om{F}(L,\nu_b)$ for $1/2>$ $\nu_a > \nu_b$ in the mentioned set of values. In other words, in the analytical region, the thermodynamics of the system behaves analytically, exhibiting a dependence on the actual value of the rational-valued normalized statistical angle and not on its numerator or denominator separately.

As we have seen above, the transition to the fractal regime is associated with the loss of the analytical continuation from imaginary to real rotation. For purely imaginary rotation, the convergence segment~\eq{eq_Omega_R2} for the variable $\nu$ is defined by the condition:
\begin{align}
    |\nu| L < 1\,,
\label{eq_nu_l}
\end{align}
implying that for the largest value, $\nu = 1/2$, the non-analytical regime should come into play at the ring length $L=2$, while for the smallest nonzero value, $\nu = 1/10$, the critical length is larger, $L=10$. This region of the lengths -- shown in the inset of Fig.~\ref{fig_ring_pressure} -- is characterized by the breaking of the monotonic behavior of pressure on the statistical angle, which is a precursor of the fractal features observed at larger lengths of the ring. 

At higher values of $L$, the behavior of pressure on $\nu$ becomes more peculiar. To see this in detail, it is convenient to start from the non-rotating case, $\nu = 0/10 = 0$, and associate it to the pair $(p,q) = (1,1)$ since $\nu = 0/10 \equiv 0$ and $\nu = 1/1 \equiv 1$ correspond to the same static case related to each other by the translation symmetry, $\nu \to \nu + 1$. The $\nu = 0$ pressure, characterized by the denominator $q=1$, is shared both by real, $\Omega = 0$, and imaginary, $\Omega_I = 0$, static cases. In Fig.~\ref{fig_ring_thermal}, it provides us with a benchmark value for the pressure in the large-$L$ limit.

The values of the statistical angle $\nu = 1/10$ and $\nu = 3/10$ correspond to rotations with different imaginary angular frequencies $\Omega_I = \pi/(5 \beta)$ and $3 \pi/(5 \beta)$, respectively, but they share the same denominator $q=10$. According to the fractalization property~\eq{eq_T_rational_bosons}, both these cases -- which are characterized by the pairs of coprimes $(p,q) = (1,10)$ and $(p,q) = (3,10)$, respectively -- should correspond, in the thermodynamic limit, to the pressure of free bosonic gas at the same temperature $T = 1/(10 \beta)$ which is ten times smaller than the temperature in the non-rotating $\Omega_I=0$ ($\nu = 0$) case. The pressure for $\nu = 1/10$ and $\nu = 3/10$ is, consequently, $1/q^2 \equiv 1/100$ of the gas pressure in the absence of imaginary rotation. The described features are clearly seen in Fig.~\ref{fig_ring_thermal}: the $\nu=1/10$ and $\nu = 3/10$ pressures, very different at low $L \sim 1$, start to approach each other at $L \sim 10$, converging into a single curve already at $L \sim 50$. This asymptotic behaviour has fractal features as the thermodynamics of the gas is sensitive only to the denominator of the rational (properly normalized) angular frequency.

The cases $\nu = 2/10 \equiv 1/5$ and $\nu = 4/10 \equiv 2/5$ correspond to the coprime pairs $(p,q) = (1,5)$ and $(p,q) = (2,5)$ that share the same denominator $q=5$. The pressure for these imaginary angular frequencies collapse to a single line even earlier, at $L \sim 7$, as it can be seen from the inset of Fig.~\ref{fig_ring_pressure}. In both cases, pressure approaches the result for a free bosonic gas with temperature $T = 1/(5 \beta)$ which is $q^2 = 25$ times smaller than the pressure of the non-rotating gas. 

Finally, the normalized statistical angle $\nu = 5/10 \equiv 1/2$ gives the denominator $q=2$, temperature $T = 1/(2\beta)$ and a gas pressure which is $q^2 = 4$ times smaller than the one of the non-rotating gas. 

The monotonic analytical behaviour of the free energy $\om{F}_I(\nu) \equiv \om{F}_I(\nu, L)$, seen at small lengths of the ring $L$, 
% \begin{align}
%     & [\text{small $L$ (analytical)}]: \\[1mm]
%     & P(0) {>}  P \Big(\frac{1}{10}\Big)  {>}  P\Big(\frac{2}{10}\Big) {>} P\Big(\frac{3}{10}\Big) {>} P\Big(\frac{4}{10}\Big) {>} P\Big(\frac{5}{10}\Big), \nonumber
% \end{align}
\begin{align}
 & [\text{small $L$ (analytical)}]: \\[1mm]
 & \om{F}_I(0) {<}  \om{F}_I (\tfrac{1}{10})  {<}  \om{F}_I(\tfrac{2}{10}) {<} \om{F}_I(\tfrac{3}{10}) {<} \om{F}_I(\tfrac{4}{10}) {<} \om{F}_I(\tfrac{5}{10}), \nonumber
\end{align}
is completely lost for large $L$ giving us the fractal non-analytical hierarchy: 
% \begin{align}
%     & [\text{large $L$ (fractal)}]: \\[1mm]
%     & P(0) {>} P \Big(\frac{5}{10}\Big)  {>} P\Big(\frac{2}{10}\Big) =  P\Big(\frac{4}{10}\Big)   {>} P\Big(\frac{1}{10}\Big) =  P\Big(\frac{3}{10}\Big), \nonumber
% \end{align}
\begin{align}
 & [\text{large $L$ (fractal)}]: \\[1mm]
 & \om{F}_I(0) {<} 
 \om{F}_I (\tfrac{5}{10})  {<} 
 \om{F}_I(\tfrac{2}{10}) = \om{F}_I(\tfrac{4}{10})   {<} 
 \om{F}_I (\tfrac{1}{10}) =  \om{F}_I(\tfrac{3}{10}), \nonumber
\end{align}
as it is clearly seen in Fig.~\ref{fig_ring_thermal}.

\subsection{Negative thermodynamic pressure of ninions}\label{sec_ring_negativeP}

Apart from the fractal features of the thermodynamic limit -- already anticipated from the analytical approach discussed earlier -- the pressure at finite volumes $L \sim 1 \dots 10$ appears to possess an unexpected feature. Namely, there are regions of the statistical angle $\chi$ where the thermal contribution to pressure is negative, as it is clearly seen in Fig.~\ref{fig_ring_thermal}. In this sense, the ``ninions'' -- the auxiliary particles which are associated with the ninionic deformation of the standard statistical distribution~\eq{eq_ninionic} --  provide us with the similar phenomenon as the Casimir effect with one important difference: the negative ``ninionic'' pressure is produced by thermal, and not quantum, fluctuations. As temperature rises, the negative pressure rises as well. 

The effect of the negative pressure appears in the analytical region~\eq{eq_nu_l} as it is seen in Fig.~\ref{fig_ring_pressure} and especially in the inset of this figure. This unusual behavior is an exotic property of ninions which is not associated with the fractal statistics.

\subsection{Effect of finite particle mass}\label{sec_ring_massive}

\begin{figure}
    \centering
    \includegraphics[width=.99\linewidth]{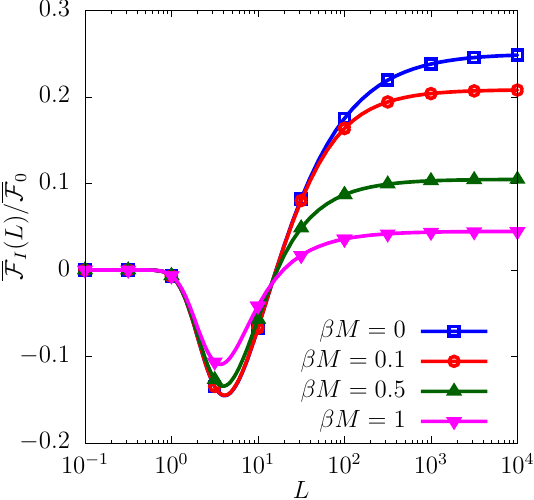}
    \caption{Evidence of fractalization at finite mass $M$ (various values of $\beta M$ are shown in the legend), for the rotation parameter $\nu = p/q = 1/2$, at the level of the ratio $\om{F}_I(L) / \om{F}_0$ between the free energy in Eq.~\eqref{eq_F_massive} and that of a boson gas, $\om{F}_0 = -P_0 = -\pi / 6\beta^2$ [see Eq.~\eqref{eq_E_large_ring_0}].}
    \label{fig:ring_QFT_fractalization_M}
\end{figure}

We now consider the effect of finite mass $M$ on the fractalization properties discussed in Subsec.~\ref{sec_ring_fractal}. Due to the relativistic dispersion relation $\omega_m = \sqrt{M^2 + m^2 / R^2}$, 
the resummation of the free energy as in Eq.~\eqref{eq_ring_peasant_F} is no longer possible. Writing $\mathcal{F}_I$ for imaginary rotation parameter $\Omega = i \Omega_I$ directly from Eq.~\eqref{eq_F_neutral} gives
\begin{align}
\mathcal{F}_I  = \frac{1}{\beta} \sum_{m=1}^\infty \ln \Bigl[ 1 - 2e^{-\beta \omega_m} \cos(2\pi m \nu) + e^{-2 \beta \omega_m}\Bigr].
\label{eq_F_massive}
\end{align}
Due to its slow convergence, the above sum is not amenable to asymptotic analysis. Instead, we rely on brute force to evaluate Eq.~\eqref{eq_F_massive}. It can be seen in Fig.~\ref{fig:ring_QFT_fractalization_M} that the plateau (fractal) value is reached for any values of $\beta M$. For convenience, the plot reports the ratio $\om{F}_I(L) / \om{F}_0$, where $\om{F}_0 = -P_0 = -\pi / \beta^2$ represents the free energy of a boson gas [see Eq.~\eqref{eq_E_large_ring_0}].

\begin{figure}
    \centering
    \begin{tabular}{c}
    \includegraphics[width=.99\linewidth]{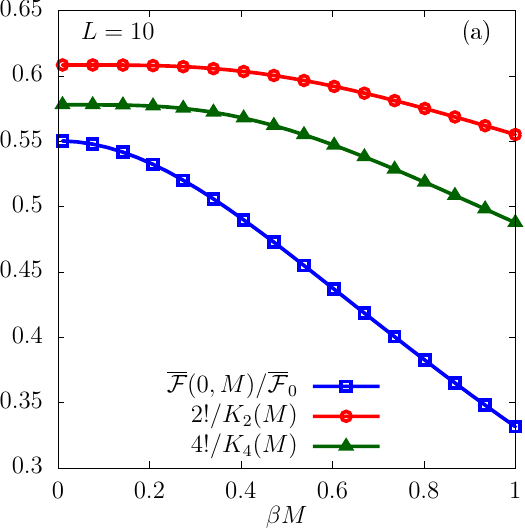} \\ 
    \includegraphics[width=.99\linewidth]{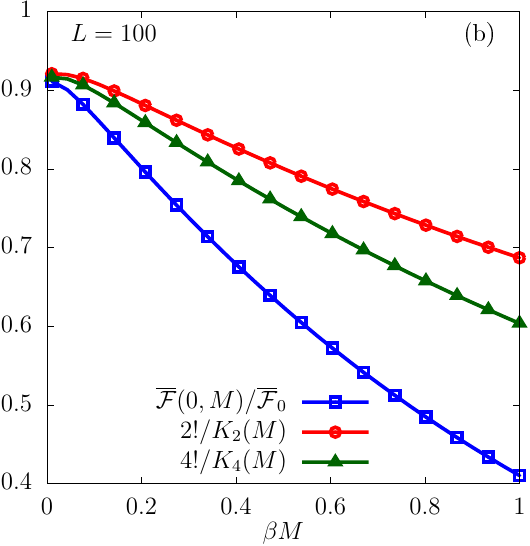}
    \end{tabular}
    \caption{Mass dependence of the free energy density $\om{F}(0,M)$, normalized with respect to the classical value $\om{F}_0=  -\pi / 6\beta^2$, as well as of the shape coefficients $K_2$ and $K_4$. The system size is set to (a) $L = 10$ and (b) $L = 100$.}
    \label{fig:ring_QFT_K2n_M}
\end{figure}

Finally, we analyze the effect of the particle mass $M$ onto the free energy in the absence of rotation, $\om{F}(0,M)$, and on the coefficients $K_{2n}$. The relevant expressions are
\begin{align}
 \om{F}(0, M) &= \frac{2}{\beta^2 L} \sum_{m=1}^\infty \ln (1 - e^{-\beta \omega_m}),\nonumber\\
 \overline{K}_2 &= -\frac{2\beta}{R^2} \sum_{m = 1}^\infty \frac{m^2 e^{-\beta \omega_m}}{(1 - e^{-\beta \omega_m})^2}, \nonumber\\
 \overline{K}_4 &= -\frac{2\beta^3}{R^4} \sum_{m = 1}^\infty \frac{m^4 e^{-\beta \omega_m}}{(1 - e^{-\beta \omega_m})^2} \left[1 + \frac{6 e^{-\beta \omega_m}}{(1 - e^{-\beta \omega_m})^2}\right].
\end{align}
Figure~\ref{fig:ring_QFT_K2n_M} indicates a generally decreasing trend of the free energy density $\om{F}(0,M)$, computed for a static system ($\nu = 0$) at finite mass, with respect to its massless, thermodynamic limit value, $\om{F}_0 = -\pi / 6\beta^2$. The shape coefficients $K_{2n}(M$) increase with $M$ with respect to their classical, massless values, $K_{2n} = (2n)!$. The results are consistent with those observed in the classical analysis shown in Fig.~\ref{fig:ring_RKT} and one can expect to reach agreement in the thermodynamic limit, when $L \rightarrow \infty$.

\section{Rigidly-rotating Bose-Einstein distribution}\label{sec:RKT}
\label{sec_RKT}

We now move on and consider a (3+1)d rigidly-rotating system comprised of uncharged, massless boson particles. In this section, we consider such a system from the perspective of relativistic kinetic theory, which is introduced briefly in Subsect.~\ref{sec:RKT:BE}. In Subsect.~\ref{sec:RKT:thermo}, we discuss the thermodynamic properties of a rigidly-rotating system with real rotation parameter $\Omega$, whose properties for slow rotation are discussed in Sec.~\ref{sec_slow_rotation_shape_0}. Finally, Subsec.~\ref{sec:RKT:imaginary} is dedicated to the case of imaginary rotation.

\subsection{Relativistic kinetic theory} \label{sec:RKT:BE}

Although throughout this article we consider non-interacting bosonic systems, it is instructive to discuss, for a brief moment, an interacting model. This approach will allow us to 
elucidate thermal distributions in thermodynamic equilibrium and shed some light on the physical nature of imaginary rotating systems.

In relativistic kinetic theory, the system dynamics are described using the relativistic Boltzmann equation \cite{Groot.1980,Cercignani.2002,Rezzolla.2013}:
\begin{equation}
 k^\mu \partial_\mu f_\bk = C[f],\label{eq_RKT_boltz}
\end{equation}
where $f_\bk \equiv f_\bk(x)$ is the one-particle distribution function  and $k^\mu = (k^0, \bk)$ is the on-shell momentum satisfying $k^2 = 0$. The macroscopic properties of the system can be described using the energy-momentum tensor, 
\begin{equation}
 T^{\mu\nu} = \int dK \, k^\mu k^\nu f_\bk,
\end{equation}
where $dK = g d^3k / [(2\pi)^3 k^0]$ is the Lorentz-invariant integration measure and the degeneracy factor of a single neutral scalar field considered in this paper is $g = 1$. 
The conservation law $\partial_\mu T^{\mu\nu} = 0$ demands that $k^\mu$ be a collision invariant, i.e.
\begin{equation}
 \int dK C[f] \, k^\mu  = 0.
\end{equation}

The prototypical collision term is that corresponding to $2$-to-$2$ scattering processes \cite{Groot.1980,Denicol.2012},
\begin{multline}
 C_{2\rightarrow 2}[f] = \frac{1}{2} \int dK' dP dP' W_{\bk \bk' \rightarrow \bp \bp'} \\\times 
 (f_\bp f_{\bp'} \tilde{f}_{\bk} \tilde{f}_{\bk'} - f_\bk f_{\bk'} \tilde{f}_\bp \tilde{f}_{\bp'}),\label{eq_RKT_C}
\end{multline}
where $\tilde{f}_\bk = 1 + f_\bk$ is the Bose enhancement factor and the Lorentz-invariant transition rate $W_{\bk \bk' \rightarrow \bp \bp'}$ can be written in terms of the quantum-mechanical differential cross section $d\sigma / d\Omega$ as
\begin{equation}
 W_{\bk \bk' \rightarrow \bp \bp'} = s \frac{d\sigma(s, \Theta_s)}{d\Omega_s} (2\pi)^6 \delta^{4}(k + k' - p -p'), \label{eq_RKT_W}
\end{equation}
with $s = (k + k')^2$ and $\Theta_s$  being the center of mass squared energy and emission angle, respectively, with
\begin{equation}
 \cos \Theta_s = \frac{(k - k') \cdot (p - p')}{(k -k')^2}.
\end{equation}

According to the H theorem, the collision term $C[f]$ drives the system towards local thermal equilibrium, described for the case of a free bosonic gas by the Bose-Einstein distribution:
\begin{equation}
 f_\bk^{\rm BE} = \frac{1}{\exp[u_\mu (x) k^\mu / T(x)] - 1}\,,
 \label{eq_RKT_fBE}
\end{equation}
where $T(x)$ is the local temperature and $u^\mu(x)$ is the local four-velocity. It is easy to check that $C[f] = 0$ when the gas is in thermal equilibrium, i.e. $f_* = f_*^{\rm BE}$ for $* \in \{\bk,\bk',\bp,\bp'\}$. In global thermal equilibrium, $f_\bk = f_\bk^{\rm BE}$ at each space-time point and Eq.~\eqref{eq_RKT_boltz} becomes:
\begin{equation}
 k^\mu k^\nu \partial_\mu \beta_\nu(x) = 0,
\end{equation}
where $\beta^\mu(x) = u^\mu(x) / T(x)$ is the temperature four-vector. Thus, in global equilibrium, $\beta^\mu$ satisfies the Killing equation, $\partial_\mu \beta_\nu + \partial_\nu \beta_\mu = 0$. In this paper, we seek the solution that corresponds to rigid rotation:
\begin{equation}
 \beta^\mu(\Omega) \partial_\mu = \beta (\partial_t - y \Omega \partial_x + x \Omega \partial_y) = \beta (\partial_t + \Omega \partial_\varphi),
\end{equation}
where $\Omega$ is the angular velocity and $\beta = 1/T_0$ is a constant corresponding to the inverse temperature on the rotation axis (where $x = y = 0$).  The equilibrium distribution \eqref{eq_RKT_fBE} thus reads:
\begin{equation}
 f_\bk^{\rm BE}(\Omega) = \frac{1}{ \exp[ \beta(k^0 + \Omega k_\varphi)] - 1}\,,
 \label{eq_RKT_fBE_Omega}
\end{equation}
where 
\begin{align}
    k_\varphi = -\rho^2 k^\varphi\,, 
    \qquad
    k^\varphi = \rho^{-2} (-y k^x + x k^y)\,, 
\label{eq_k_varphi}
\end{align}
with $\rho = \sqrt{x^2 + y^2}$ being the distance from the point at $(x,y,z)$ to the rotation axis.

\subsection{Thermodynamics of rigid rotation} \label{sec:RKT:thermo}

We now discuss the properties of the rigidly-rotating global equilibrium state. From the relation $\beta^\mu \beta_\mu = 1/T^2(x)$, we can identify the local temperature as \cite{Cercignani.2002}
\begin{equation}
 T(\rho) = \beta^{-1} \gamma(\rho), \quad 
 \gamma(\rho) = (1 - \rho^2 \Omega^2)^{-1/2},
    \label{eq_TE_gamma}
\end{equation}
where $\gamma(\rho)$ is the Lorentz factor of a co-rotating particle at a distance $\rho$ from the rotation axis. Relation~\eq{eq_TE_gamma} corresponds to the Tolman-Ehrenfest law~\cite{Tolman:1930ona,Tolman:1930zza}, which relates local temperature to the metric in a static gravitational field, in the curvilinear background of co-rotating reference frame.
Similarly, the local four-velocity reads
\begin{equation}
 u^\mu \partial_\mu = \gamma(\partial_t + \Omega \partial_\varphi).
\end{equation}
Both the Lorentz factor and the local temperature $T(\rho) = \gamma(\rho) / \beta$ diverge on the light cylinder, as $\rho \Omega \rightarrow 1$.

The energy-momentum tensor $T^{\mu\nu}(x)$ can be obtained via 
\begin{equation}
 T^{\mu\nu} = \int dK \, k^\mu k^\nu f_\bk = (E + P) u^\mu u^\nu - P g^{\mu\nu}.
\end{equation}
In the case of massless particles, the energy $E = 3P$ is expressed via the local pressure $P$, which reads
\begin{equation}
 P(\rho) = \frac{1}{3} \int dK\, (k\cdot u)^2 f^{\rm BE}_\bk = \frac{\pi^2 \gamma^4(\rho)}{90 \beta^4}.
 \label{eq_P_rho}
\end{equation}

We now consider the thermodynamic limit of our rigidly-rotating system. Identifying $F(\rho) = -P$ as the local free-energy density, we consider the average free energy $\om{F} = \mathcal{F} / V$ in a cylindrical volume $V = \pi R^2 L_z$ of height $L_z$ and radius $R$, centered on the rotation axis. The mean free-energy density reads
\begin{equation}
 \om{F}(\Omega, R) = -\frac{2}{R^2} \int_0^R d\rho\, \rho\, P(\rho)
 = -\frac{\pi^2}{90\beta^4} \gamma^2(R).\label{eq_RKT_Fbar}
\end{equation}
The same result can be obtained starting from the expression for the grand potential $\mathcal{F}$ of relativistic bosons in rotation \cite{LL5}, cf.~Eq.~\eqref{eq:ring_RKT_F},
\begin{align}
 \mathcal{F} &= \frac{1}{\beta} \int_V d^3x \int \frac{d^3k}{(2\pi)^3} \ln [1 - e^{-\beta(k^0 - \Omega J^z)}] \nonumber\\
 &= \frac{1}{2\beta} \int_V d^3x \int \frac{d^3k}{(2\pi)^3} \nonumber\\
 & \times \ln [1 - 2e^{-\beta k^0} \cosh(\beta \Omega k_\varphi) + e^{-2\beta k^0}], 
 \label{eq_RKT_Fcl_cosh}
\end{align}
where $J^z = -k_\varphi$ is the $z$ component of the particle's angular momentum.
Other thermodynamic quantities can be obtained starting from Eq.~\eqref{eq:dF}.
Taking into account that $V = \pi R^2 L_z$, it can be seen that the radial and vertical directions are not equivalent. Therefore, we replace the term $\mathcal{P} dV$ by
\begin{equation}
 \mathcal{P} dV \rightarrow 2\pi R L_z \mathcal{P}_R dR + \pi R^2 \mathcal{P}_z dL_z,
\end{equation}
with the hydrostatic pressure obtained as the weighted average $\mathcal{P} = (2\mathcal{P}_R + \mathcal{P}_z) / 3$. Thus, the thermodynamic pressures are given by
\begin{equation}
 \mathcal{P}_R = -\frac{1}{2R} \frac{\partial(\om{F} R^2)}{\partial R}, \qquad \mathcal{P}_z = -\om{F}. \label{eq_P_from_F}
\end{equation}
Similarly, the average entropy $\om{S}$, angular momentum $\om{M}$, and energy density $\om{E}$ are given by Eqs.~\eqref{eq_thermo_rel} and \eqref{eq_Euler}, respectively. 
We now evaluate the above quantities using the classical expression in Eq.~\eqref{eq_RKT_Fbar}:
\begin{align}
 \mathcal{P}_R &= \frac{\pi^2}{90\beta^4} \gamma^4(R), &
 \om{E} &= \frac{\pi^2}{90\beta^4} [2\gamma^4(R) + \gamma^2(R)], \nonumber\\
 \om{S} &= \frac{2\pi^2}{45\beta^3} \gamma^2(R), &
 \om{M} &= \frac{\pi^2 R^2 \Omega}{45\beta^4} \gamma^4(R).
 \label{eq_RKT_thermo}
\end{align}
It can be checked that $\om{E}$ and $\om{F}$ satisfy Eq.~\eqref{eq_F_from_E}. Furthermore, $\om{E}$ is compatible with the classical relation [cf.~\eqref{eq_ring_classical_E}]:
\begin{equation}
 \om{E} = \frac{2}{R^2} \int_0^R d\rho\, \rho\, T^{tt},
 \label{eq_average_E}
\end{equation}
with 
\begin{align}
    T^{tt} = P(\rho) [4\gamma^2(\rho) - 1] = \frac{\pi^2}{90\beta^4} \frac{3 + \rho^2\Omega^2}{(1 - \rho^2\Omega^2)^3}\,,
    \label{eq_Ttt}
\end{align}
where we took into account the explicit expressions for the local hydrostatic pressure~\eq{eq_P_rho} and the Lorentz factor~\eq{eq_TE_gamma}.

\subsection{Slow rotation: moment of inertia and shape}
\label{sec_slow_rotation_shape_0}

We now consider the coefficients $\om{F}(0)$ and $K_{2n}$ introduced in Eq.~\eqref{eq_Kn_def}. Comparing $\om{F}$ in Eq.~\eqref{eq_RKT_Fbar} for the $(3+1)$d system and in Eq.~\eqref{eq_ring_classical_thermo} for the ($1+1$)d ring, it can be seen that the $\Omega$ dependence is through the same $\gamma^2(R)$ factor. Therefore, the coefficients $K_{2n}$ are identical to those in Eq.~\eqref{eq_ring_K2n},
\begin{equation}
 K_{2n} = (2n)!, \quad K_{2n+1} = 0. \label{eq_RKT_K2n}
\end{equation}
The average free energy in the absence of rotation evaluates to
\begin{equation}
 \om{F}(0) = -\frac{\pi^2}{90\beta^4}. \label{eq_RKT_F0}
\end{equation}
The moment of inertai in the zero-rotation limit $I_0$ evaluates to 
\begin{align}
 I_0 = \frac{\pi^2 R^2}{45 \beta^4}.
\end{align}
The coefficient $I_2$ in the series $I(\Omega) = I_0 + I_2 v_R^2/2 + \dots$ is trivially 
\begin{equation}
 I_2 = 4 I_0 = \frac{4\pi^2 R^2}{45 \beta^4}.
\end{equation}
The positiveness of $I_2>0$ implies that the rotating matter tends to increase its angular momentum with an increase in angular frequency. This property signals the change in the shape of rotating system leading to a spatial redistribution of energy as a result of the rotation, which can already be seen from Eq.~\eq{eq_Ttt}: rotation tends to increase the contributions to the energy density~\eq{eq_average_E} coming from the outer regions as compared to those coming from the inner ones. The physical situation is somewhat similar -- neglecting viscosity effects -- to water rotating in a glass: its moment of inertia increases with rotation because the distribution of mass within the glass changes, with the water particles moving away from the axis of rotation, increasing the distance of each mass element from the axis, and, hence, the moment of inertia becomes larger.

Finite-size corrections, related to the finite transverse size of the system and, consequently, to quantization of the transverse modes, will be discussed below in Subsects.~\ref{sec_KG_slow} and \ref{sec_bounded_slow}.

\subsection{Imaginary rotation} \label{sec:RKT:imaginary}

We now turn to the case of imaginary rotation. Setting $\Omega = i \Omega_I$ with real $\Omega_I$ is not possible directly in $f_{\rm BE}$, because that would lead to a complex-valued distribution function. Instead, we can consider the properties of the system described by the distribution 
\begin{align}
 f^I_\bk & = \frac{1}{2} \bigl[f_\bk^{\rm BE}(i \Omega_I) + f_\bk^{\rm BE}(-i \Omega_I)\bigr] \nonumber\\
 &= \frac{e^{\beta k} \cos(\beta \Omega_I k_\varphi) - 1}
 {e^{2\beta k} - 2 e^{\beta k} \cos(\beta \Omega_I k_\varphi) + 1},
 \label{eq_RKT_fBE_iOmega}
\end{align}
which is nothing but a form of the ninionic deformation of the Bose-Einstein statistics~\eq{eq_ninionic}.
Since $\beta^\mu(i \Omega_I) \partial_\mu = \beta (\partial_t + i \Omega_I \partial_\varphi)$ still satisfies the Killing equation, the left-hand side of the Boltzmann equation \eqref{eq_RKT_boltz} vanishes. Somewhat unsurprisingly, the collision term on the right-hand side of the same equation does not vanish. This can be seen by considering the small-$\Omega$ expansion of $f_\bk^{\rm BE}(\Omega)$ introduced in Eq.~\eqref{eq_RKT_fBE_Omega}:
\begin{multline}
 f_\bk^{\rm BE}(\Omega) = f_\bk^{0}\Big[1 - \tilde{f}_\bk^{0} \beta \Omega k_\varphi 
 \\ + \frac{\beta^2 \Omega^2 k_\varphi^2}{2!} \tilde{f}_\bk^{0} (f_{\bk}^{0} + \tilde{f}_\bk^{0}) + O(\Omega^3)\Big],
\end{multline}
where $f_\bk^0 \equiv f_\bk^{\rm BE}(\Omega = 0)$ and $\tilde{f}_\bk^{0} = 1 + f_\bk^0$.
Considering now $\Omega \rightarrow \pm i \Omega_I$ and taking the average as described in Eq.~\eqref{eq_RKT_fBE_iOmega} gives
\begin{align}
 f^I_\bk &= f_\bk^{0} \left[1 - \frac{\beta^2 \Omega_I^2 k_\varphi^2}{2!} \tilde{f}_\bk^{0} (f_{\bk}^{0} + \tilde{f}_\bk^{0}) + O(\Omega_I^4) \right],\nonumber\\
 \tilde{f}^I_\bk &= \tilde{f}_\bk^{0} \left[1 - \frac{\beta^2 \Omega_I^2 k_\varphi^2}{2!} f_\bk^{0} (f_{\bk}^{0} + \tilde{f}_\bk^{0}) + O(\Omega_I^4) \right].
\end{align}
Taking this substitution back into the collision term \eqref{eq_RKT_C} shows that 
\begin{multline}
 \frac{f^I_\bp f^I_{\bp'} \tilde{f}^I_{\bk} \tilde{f}^I_{\bk'} -
 f^I_\bk f^I_{\bk'} \tilde{f}^I_\bp \tilde{f}^I_{\bp'}}{f_\bp^{0} f_{\bp'}^{0} \tilde{f}_\bk^{0} \tilde{f}_{\bk'}^{0}} = -\frac{\beta^2 \Omega_I^2}{2} \\
 \times \left[p_\varphi^2 (f_\bp^{0} + \tilde{f}_\bp^{0}) + 
 p^{\prime 2}_\varphi (f_{\bp'}^{0} + \tilde{f}_{\bp'}^{0}) \right.\\
 \left. - k_\varphi^2 (f_\bk^{0} + \tilde{f}_\bk^{0}) -
 k^{\prime 2}_\varphi (f_{\bk'}^{0} + \tilde{f}_{\bk'}^{0})\right]
 + O(\Omega^4).
 \label{eq_collission_nonzero}
\end{multline}
It can be seen that in general, $C[f]$ does not vanish when $f_\bk = f_\bk^I$, hinting that thermal equilibration will generically reduce the magnitude of $\Omega_I^2$. 

Keeping in mind that imaginary-rotation states are not in actual thermal equilibrium -- in a sense that their deformed distribution~\eq{eq_RKT_fBE_iOmega} does not have the equilibrium Bose-Einstein form~\eq{eq_n_Bose} and that the collision integral~\eq{eq_collission_nonzero} does not vanish -- we can still derive the macroscopic energy-momentum tensor, which becomes now diagonal:
\begin{subequations}\label{eq_RKT_im}
\begin{equation}
 T^{\mu\nu}_I = {\rm diag}(E_I, P_{\rho;I}, \rho^{-2} P_{\varphi;I}, P_{z;I}),\label{eq_RKT_im_Tmunu}
\end{equation}
with 
\begin{align}
 E_I &= \frac{\pi^2}{90 \beta^4} \gamma_I^4 (4\gamma_I^2 - 1), \label{eq_RKT_im_E}\\
 P_{\rho;I} &= P_{z;I} = \frac{\pi^2}{90 \beta^4} \gamma_I^4,\label{eq_RKT_im_Pr}\\
 P_{\varphi;I} &= \frac{\pi^2}{90 \beta^4} \gamma_I^4 (4\gamma_I^2 - 3), \label{eq_RKT_im_Pp}
\end{align}
\end{subequations}
where $\gamma_I(\rho)$ is reminiscent of the Lorentz factor of corotating particles,
\begin{equation}
 \gamma_I = \frac{1}{\sqrt{1 + \rho^2 \Omega_I^2}}.
 \label{eq_RKT_im_gamma}
\end{equation}

Notice that the Euclidean version of the quantum Tolman-Ehrenfest effect gives a different Lorentz factor~\cite{Chernodub:2022veq}:
\begin{align}
 \gamma^{\mathrm{TE}}_I = \frac{1}{\sqrt{1 + \rho^2 \beta^{-2} [\Omega_I \beta]^2_{2\pi}}},
 \label{eq_TE_im_gamma}    
\end{align}
where $[x]_{2\pi} = x + 2 \pi k \in (-\pi,\pi]$, with $k \in {\mathbb Z}$, is invariant under the $2\pi$ symmetry enforced by the natural periodicity of the imaginary rotation~\eq{eq_chi_periodicity}. The apparent non-compliance of the kinetic Euclidean Lorentz factor~\eq{eq_RKT_im_gamma} with the periodicity requirement~\eq{eq_chi_periodicity} can be traced back to the continuous nature of the angular component $k_\varphi$ of the momentum~\eq{eq_k_varphi}.

From a thermodynamic point of view, the structure of the energy-momentum tensor reveals an underlying equilibrium (perfect fluid) contribution, $T^{\mu\nu}_{{\rm pf};I} = {\rm diag}(E_I, P_I, \rho^{-2} P_I, P_I)$, with hydrostatic pressure $P_I = E_I / 3$, and a shear-stress tensor $\pi^{\mu\nu}_I = T^{\mu\nu}_I - T^{\mu\nu}_{{\rm pf};I}$ with components
\begin{equation}
 \pi^{\mu\nu}_I = \frac{2\pi^2 \gamma_I^4}{135\beta^4}(1 - \gamma^2_I) \times 
 {\rm diag}(0, 1, -2\rho^{-2}, 1).\label{eq_RKT_im_pimunu}
\end{equation}
It is instructive to note that, on the rotation axis, $E_I = E = \pi^2 / (30\beta^4)$ is independent of $\Omega_I$, while at $\rho = \sqrt{3} / |\Omega_I|$, the energy density reaches $0$. At larger distances, $E_I$ decreases to a minimum (negative) value $-\pi^2 / (9720\beta^4)$ (reached at $\rho = \sqrt{5} / |\Omega_I|$) and afterwards increases asymptotically towards its limit $0$. In this large-$\rho$ limit, Eq.~\eqref{eq_RKT_im} shows that $T^{\mu\nu}_I$ behaves as follows:
\begin{equation}
 T^{\mu\nu}_I \simeq \frac{\pi^2}{90 \beta^4} \gamma_I^4 {\rm diag}(-1, 1, -3\rho^{-2}, 1),
 \label{eq_RKT_im_Tmunu_far}
\end{equation}
with $\gamma_I \sim (\rho |\Omega_I|)^{-1}$. Thus, far away from the rotation axis, the azimuthal pressure becomes negative and three times larger in magnitude than the energy density, while the radial and vertical pressures remain positive, each being equal in magnitude to the energy density.

We now consider the large-volume limit of our system. The average energy inside a cylinder of radius $R$ is simply 
\begin{equation}
 \om{E}_I = \frac{\pi^2}{90\beta^4} [2 \gamma_I^4(R) + \gamma_I^2(R)],
\end{equation}
which agrees with the expression in Eq.~\eqref{eq_RKT_thermo} under the substitution $\gamma(R) \rightarrow \gamma_I(R)$. Substituting $P_I = E_I / 3$ in Eq.~\eqref{eq_RKT_Fbar} will clearly give a different result for the average free energy than $\om{F}$ in Eq.~\eqref{eq_RKT_Fbar}. To achieve agreement up to the substitution $\gamma(R) \rightarrow \gamma_I(R)$, we must replace $P(\rho)$ by $P_{\rho;I}(\rho)$. This choice is supported also by the 
more fundamental expression for the free energy obtained by setting $\Omega \rightarrow i \Omega_I$ in the third line of Eq.~\eqref{eq_RKT_Fcl_cosh}, i.e.
\begin{align}
 \mathcal{F}_I \, &= \int_V d^3x \int\frac{d^3k}{2\beta(2\pi)^3} \ln[1 - 2e^{-\beta k} \cos(\beta \Omega_I k_\varphi) + e^{2\beta k} ] \nonumber\\
 &= -\frac{\pi^2 V}{90\beta^4} \gamma_I^2(R),
 \label{eq_RKT_Fim}
\end{align}
which is consistent with the expression in Eq.~\eqref{eq_F_from_E}.
Applying the thermodynamic relations in Eqs.~\eqref{eq_thermo_rel} and \eqref{eq_P_from_F} gives expressions for quantities analogue to the system pressure, entropy, and angular momentum:
\begin{align}
 \mathcal{P}_{R;I} &= \frac{\pi^2}{90\beta^4} \gamma_I^4(R), &
 \mathcal{P}_{z;I} &= \mathcal{P}_{\varphi;I} = \frac{\pi^2}{90\beta^4} \gamma_I^2(R), \nonumber\\
 \om{S}_I &= \frac{2\pi^2}{45\beta^3} \gamma_I^2(R), &
 \om{M}_I &= -\frac{\pi^2 R^2 \Omega_I}{45\beta^4} \gamma_I^4(R),
 \label{eq_RKT_im_thermo}
\end{align}
with $\om{M}_I = -i \om{M} = \partial \om{F}_I / \partial \Omega_I$ [see Eq.~\eqref{eq_ring_peasant_thermo_nu}]. The above quantities are compatible with the Euler relation \eqref{eq_Euler},
\begin{equation}
 \om{E}^{\rm im}_{\rm cl} = \om{F}^{\rm im}_{\rm cl} + \beta^{-1} \om{S}^{\rm im}_{\rm cl} + \Omega_I \om{M}^{\rm im}_{\rm cl},
 \label{eq_RKT_im_Euler}
\end{equation}
formulated now for a system under imaginary rotation.

\subsection{Effect of finite mass} \label{sec:RKT:massive}

It is interesting to check how a finite particle mass affects the properties of the rigidly-rotating boson gas. Starting from Eq.~\eqref{eq_RKT_Fcl_cosh}, one may perform a Lorentz transformation in the $d^3k$ integral to the local rest frame, such that 
\begin{equation}
 \mathcal{F} = \frac{1}{\beta} \int_V d^3x\, \gamma(\rho) \int \frac{d^3k}{(2\pi)^3} \ln [1 - e^{-\beta k^0 /\gamma(\rho)}]. 
\end{equation}
Writing $d^3x = \rho d\rho d\varphi dz$ and $d^3k = k^2 dk d\Omega_{\mathbf{k}}$, the $\varphi$, $z$ and $\Omega_{\mathbf{k}}$ integrals can be performed automatically, leading to 
\begin{equation}
 \om{F} = \frac{1}{\pi^2 \beta R^2} \int_0^R d\rho\, \rho \gamma(\rho) \int_0^\infty dk\, k^2 \ln(1 - e^{-\beta k^0 / \gamma(\rho)}).
\end{equation}
Expanding the logarithm as in Eq.~\eqref{eq_ring_RKT_Fexp}, and introducing $y = 1/\gamma(\rho)$, with $dy = -\rho \gamma(\rho) \Omega^2 d\rho$, leads to
\begin{multline}
 \om{F} = -\frac{1}{\pi^2 \beta^2 R^2 \Omega^2} \sum_{j = 1}^\infty \frac{1}{j^2} \int_M^\infty dk^0  k \\\times \left(e^{-j \beta k^0 / \gamma(R)} - e^{-j \beta k^0}\right),
\end{multline}
where we changed the momentum integration variable using $k dk = k^0 dk^0$. Changing again the integration variable to $z = k^0 / M$ and using $\int_1^\infty dz\, e^{-a z} = K_1(a) / a$, we arrive at
\begin{equation}
 \om{F} = -\frac{M}{\pi^2 \beta^3 R^2 \Omega^2} \sum_{j = 1}^\infty \frac{1}{j^3} \left[K_1\left(\frac{j \beta M}{\gamma(R)}\right) \gamma(R) - K_1(j \beta M)\right].
\end{equation}

\begin{figure}
    \centering
    \includegraphics[width=.9\columnwidth]{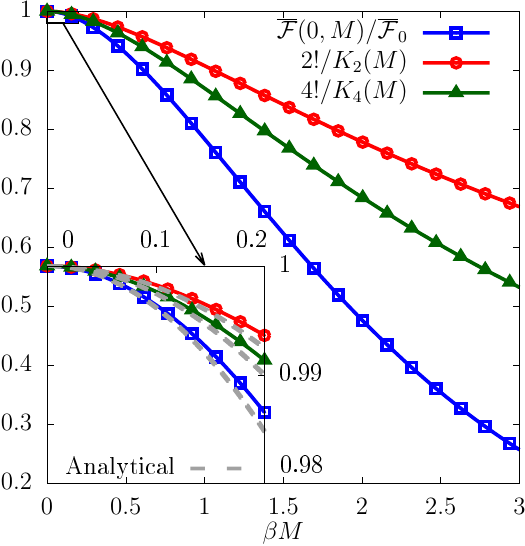} 
    \caption{Effect of particle mass $M$ on free energy and shape coefficients $K_2$ and $K_4$, respectively, computed in relativistic kinetic theory for the (3+1)d system (see Fig.~\ref{fig:ring_RKT} for explanation about notation). The inset magnifies the $0 \le \beta M \le 0.2$ region to demonstrate the domain of applicability of the asymptotic formulas in Eqs.~\eqref{eq:unb_RKT_Fbar_K2bar_K4bar_M} and \eqref{eq:unb_RKT_K2K4_M}}.
    \label{fig:unb_RKT_M}
\end{figure}

We are now in a position to evaluate the average free energy in the absence of rotation, $\om{F}(0,M)$, as well as the shape coefficients $K_{2n}(M) = d^{2n} \om{F}(\Omega,M) / d\Omega^{2n} \rfloor_{\Omega = 0}$:
\begin{align}
    \om{F}(0,M) &= -\frac{M^2}{2\pi^2 \beta^2} \sum_{j = 1}^\infty \frac{1}{j^2} K_2(j M \beta),\nonumber\\
    \overline{K}_2 &= 2\om{F}(0,M) - \frac{M^3}{4\pi^2 \beta} \sum_{j = 1}^\infty \frac{1}{j} K_1(j M \beta),\nonumber\\
    \overline{K}_4 &= 12 \overline{K}_2 - \frac{M^4}{2\pi^2} \sum_{j= 1}^\infty K_2(j M \beta).
    \label{eq:unb_RKT_Fbar_shape_M_exact}
\end{align}
At small $M$, it is possible to derive the asymptotic expansions
\begin{align}
 \om{F}(0,M) &\simeq \om{F}_0 \left(1 - \frac{15}{4 \pi^2} \beta^2 M^2\right),\nonumber\\
 \overline{K}_2 &\simeq \om{F}_0 \times 2! \left(1 - \frac{15}{8\pi^2} \beta^2 M^2\right),\nonumber\\
 \overline{K}_4 &\simeq \om{F}_0 \times 4! \left(1 - \frac{5}{4\pi^2} \beta^2 M^2\right),
 \label{eq:unb_RKT_Fbar_K2bar_K4bar_M}
\end{align}
where $\overline{K}_{2n} \equiv \om{F}(0,M) K_{2n}$ and
$\om{F}_0 = -\pi^2 / 90\beta^4$ is the free energy density for a massless bosonic gas. The coefficients $K_2$ and $K_4$ have therefore the following asymptotic behaviour:
\begin{align}
    K_2 &\simeq 2!\left(1 + \frac{15}{8\pi^2} \beta^2 M^2\right), & 
    K_4 &\simeq 4!\left(1 + \frac{5}{2\pi^2} \beta^2 M^2\right).
    \label{eq:unb_RKT_K2K4_M}
\end{align}
The above estimates are validated in Fig.~\ref{fig:unb_RKT_M} by comparison to the exact formulas in Eq.~\eqref{eq:unb_RKT_Fbar_shape_M_exact}. The results obtained in this section for the (3+1)d system are in good qualitative agreement with those obtained for the $(1+1)$d ring (cf. Figs.~\ref{fig:ring_RKT} and \ref{fig:ring_QFT_K2n_M}).

\section{Imaginary rotation of the scalar field vs. real rotation}
\label{sec_KG}

\subsection{Mode solutions} \label{sec_KG_modes}

We consider a real, massless scalar field $\hat{\phi}$. The decomposition of the field operator $\hat{\phi}(x)$ reads as follows:
\begin{equation}
 \hat{\phi}(x) = \sum_j [\hat{a}_j f_j(x) + \hat{a}^\dagger_j f^*_j(x)],
 \label{eq_phi}
\end{equation}
where $f_j(x)$ represent a complete basis of orthonormal mode solutions of the Klein Gordon equation, 
\begin{align}
    (\Box + m^2) f_j =0\,. 
\end{align}
These modes are taken as eigenfunctions of the Hamiltonian $H = i \partial_t$, momentum component $P^z = -i \partial_z$, and angular momentum component $J^z = -i \partial_\varphi$:
\begin{equation}
 f_j = \frac{1}{2\pi \sqrt{2\omega_j}} e^{-i \omega_j t + ik_j z + i m_j \varphi} J_{m_j} (q_j \rho),
 \label{eq_fj_rotation}
\end{equation}
with $q_j = \sqrt{\omega_j^2 - k_j^2}$. The one-particle operators $\hat{a}_j$ satisfy the canonical commutation relations,
\begin{equation}
 [\hat{a}_j, \hat{a}_{j'}] = 0, \quad 
 [\hat{a}_j, \hat{a}^\dagger_{j'}] = \delta(j,j'),
\end{equation}
where $\delta(j,j') = \omega_j^{-1} \delta(\omega_j - \omega_{j'}) \delta(k_j - k_{j'}) \delta_{m_j, m_{j'}}$. The sum over the quantum numbers is abbreviated as
\begin{equation}
 \sum_j \rightarrow \sum_{m_j = -\infty}^\infty \int_0^\infty d\omega_j \omega_j\, \int_{-\omega_j}^{\omega_j} dk_j.
 \label{eq_measure_sum_integral}
\end{equation}
In this and subsequent subsections, we pursue, for simplicity, a ``hybrid'' quantization approach: we work in the basis of the cylindrical waves~\eq{eq_fj_rotation} with continuous transverse momentum $q_j$ which is typical for the unbounded system while restricting the integral~\eq{eq_measure_sum_integral} over the longitudinal momentum, $|k_j| \leqslant \omega_j$, to preserve the hermiticity of the Hamiltonian. This set of modes is not suitable to describe rigid real rotation, since in that case, the system must be enclosed inside a boundary in order to preserve causality \cite{Duffy:2002ss}, as will be discussed in Sec.~\ref{sec_bounded}. In this section, we will focus primarily on the study of rigid rotation with imaginary angular velocity, for which no causality issues arise and the set of eigenmodes presented above is perfectly applicable.

\subsection{Thermal states}\label{sec_KG_tevs}

The statistical operator for a thermal state at inverse temperature $\beta$ which rotates with angular velocity $\Omega$ is 
\begin{equation}
 \hat{\rho}(\beta,\Omega) = e^{-\beta (:\widehat{H}: - \Omega :\hat{J}^z:)},
\end{equation}
where we took the normal-ordered operators
\begin{equation}
    :\widehat{H}: = \sum_j \omega_j \hat{a}^\dagger_j \hat{a}_j, \qquad 
    :\hat{J}^z: = \sum_j m_j \hat{a}^\dagger_j \hat{a}_j.
\end{equation}

Using the commutation relations
\begin{equation}
 [\widehat{H}, \hat{a}^\dagger_j] = \omega_j \hat{a}^\dagger_j, \quad 
 [\hat{J}^z, \hat{a}^\dagger_j] = m_j \hat{a}^\dagger_j,
 \label{eq_comm}
\end{equation}
it is not difficult to establish that
\begin{equation}
    \hat{\rho} \hat{a}^\dagger_j \hat{\rho}^{-1} = e^{-\beta \tilde{\omega}_j} \hat{a}^\dagger_j,
    \label{eq_rhoa}
\end{equation}
where $\tilde{\omega}_j \equiv \tilde{\omega}_j(\Omega) = \omega_j - \Omega m_j$ represents the co-rotating energy~\eq{eq_omega_varepsilon}.

The thermal expectation value (t.e.v.) of an arbitrary operator $\widehat{A}(x)$ is 
\begin{equation}
 A(x) \equiv \langle \widehat{A}(x) \rangle = 
 \mathcal{Z}^{-1} {\rm Tr} [\hat{\rho} \widehat{A}(x)],
\end{equation}
where $\mathcal{Z} = {\rm Tr}(\hat{\rho})$ is the partition function. Using Eq.~\eqref{eq_rhoa}, the t.e.v. of the product of two one-particle operators can be seen to satisfy
\begin{equation}
 \langle \hat{a}^\dagger_j \hat{a}_{j'} \rangle = 
 e^{-\beta \tilde{\omega}_j} \langle \hat{a}_{j'} \hat{a}^\dagger_j \rangle.
\end{equation}
Using the commutation relations in Eq.~\eqref{eq_comm}, we establish
\begin{equation}
 \langle \hat{a}^\dagger_j \hat{a}_{j'} \rangle = \frac{\delta(j, j')}{e^{\beta \tilde{\omega}_j} - 1}.\label{eq_tev_aadagger}
\end{equation}

Introducing the functions
\begin{multline}
 G_{abc} = \sum_j \frac{2{\rm Re}(f_j^* f_j)}{e^{\beta \tilde{\omega}_j} - 1} \omega_j^a q_j^b m_j^c \\
 = \sum_{m = -\infty}^\infty \int_0^\infty \frac{d\omega}{e^{\beta \tilde{\omega}} - 1} 
 \int_{0}^\omega \frac{dk}{2\pi^2} \omega^a q^b m^c J^2_m(q\rho),
 \label{eq_kg:Gabc}
\end{multline}
the scalar condensate\footnote{For brevity, we use the term ``condensate'' although the expectation value of the ordered operator $\hat{\phi}^2$ does not imply the presence of a nonvanishing coherent condensate $\langle \phi \rangle$.} becomes
\begin{equation}
 \phi^2 \equiv \langle :\hat{\phi}^2:\rangle = G_{000}. \label{eq_tev_phi2}
\end{equation}
Considering now the conformal energy-momentum tensor, defined classically as \cite{Christensen:1976vb}
\begin{equation}
 T_{\mu\nu} = \frac{2}{3} \nabla_\mu \phi \nabla_\nu \phi - \frac{1}{3} \phi \nabla_\mu \nabla_\nu \phi - \frac{1}{6} g_{\mu\nu} 
 (\nabla \phi)^2,
\end{equation}
its expectation value $T^{\mu\nu} = \langle :\widehat{T}^{\mu\nu}:\rangle$ can be expressed as
\begin{subequations}\label{eq_tev_SET}
\begin{align}
 T^{tt} &= G_{200} + \frac{1}{12\rho^2} G^{(2)}_{000}, \label{eq_tev_SET_Ttt}\\
 T^{\rho\rho} &= G_{020} - \frac{1}{\rho^2} G_{002} + 
 \frac{1}{4\rho^2}G^{(2)}_{000} + \frac{1}{6\rho^2} G^{(1)}_{000},\\
 T^{\varphi\varphi} &= \frac{1}{\rho^4} G_{002} - \frac{1}{12\rho^4} G^{(2)}_{000} - \frac{1}{6\rho^4} G^{(1)}_{000}, \\
 T^{zz} &= G_{200} - G_{020} - \frac{1}{12\rho^2} G^{(2)}_{000}, \\
 T^{t\varphi} &= \frac{1}{\rho^2} G_{101},
\end{align}
\end{subequations}
where we introduced the notations:
\begin{equation}
 G^{(1)}_{000} = \rho \frac{dG_{000}}{d\rho}, \qquad 
 G^{(2)}_{000} = \rho \frac{d}{d\rho} \rho \frac{d G_{000}}{d\rho},
 \label{eq_dG000}
\end{equation}
while all other components vanish.

Turning back to the definition of the functions $G_{abc}$ in Eq.~\eqref{eq_kg:Gabc}, we immediately notice divergences associated with the Bose-Einstein factor $[e^{\beta \tilde{\omega}} - 1]^{-1}$. For each value of $m$ such that $\Omega m > 0$, there will be a value of $\omega$ where this factor diverges. The only notable exception is the rotation axis, where $J^2_m(q\rho) = \delta_{m0}$ and $[e^{\beta \omega} - 1]^{-1}$ has the usual Bose-Einstein infrared divergence when $\omega = 0$. Thus, we are led to conclude that thermal rigidly-rotating states of the scalar field are ill-defined at each point outside the rotation axis due to long wavelength (super-horizon) modes, for which $\omega \le \Omega m$ \cite{Davies:1996ks}. We will come back to this issue in Sec.~\ref{sec_bounded} when we will discuss the Klein-Gordon field enclosed inside a cylindrical boundary.

\subsection{Evaluation for imaginary rotation}\label{sec_KG_OmegaI}

We now seek to construct states which undergo imaginary rotation, $\Omega = i \Omega_I$, where $\Omega_I \in \mathbb{R}$. As also noted in Sec.~\ref{sec:RKT:imaginary}, 
a state under imaginary rotation leads to complex expectation values of physical observables. This problem can be alleviated by considering the hermiticized version of $\hat{\rho}$, namely
\begin{equation}
 \hat{\rho}(\beta, \Omega) \rightarrow \frac{1}{2}[\hat{\rho}(\beta,\Omega) + \hat{\rho}^\dagger(\beta, \Omega)],
 \label{eq_rho_herm}
\end{equation}
which is equivalent to averaging over the results obtained for positive and negative values of $\Omega_I$.
Under the above hermitization, the t.e.v. in Eq.~\eqref{eq_tev_aadagger} becomes 
\begin{equation}
 \langle \hat{a}^\dagger_j \hat{a}_{j'} \rangle_\beta = \frac{e^{\beta \omega_j} \cos(\beta \Omega_I m_j) - 1}{e^{2\beta \omega_j} - 2 e^{\beta \omega_j} \cos(\beta \Omega_I m_j) + 1} \delta(j, j'),
 \label{eq_ninions}
\end{equation}
which is similar to the relativistic kinetic theory distribution $f_\bk^{\rm im}$ in Eq.~\eqref{eq_RKT_fBE_iOmega} under the substitution $k_\varphi \rightarrow -m_j$. The thermodynamic state corresponding to the t.e.v.~\eq{eq_ninions} is characterized by ninionic statistics~\eq{eq_ninionic}. In what follows, we perform the calculations considering averages using the statistical operator $\hat{\rho}(\beta, i\Omega_I)$, keeping in mind that the final result is obtained by taking the real part.

In order to analyse the functions $G_{abc}^I$, obtained by replacing $\Omega \rightarrow i \Omega_I$ in Eq.~\eqref{eq_kg:Gabc}, the Bose-Einstein factor can be expanded in a power series, as follows:
\begin{equation}
 \frac{1}{e^{\beta\tilde{\omega}} - 1} = \sum_{j =1}^\infty e^{-j\beta \tilde{\omega}},
 \label{eq_kg:sumj_BE}
\end{equation}
where $\tilde{\omega} = \omega - i \Omega_I m$ has a positive real part, $\omega > 0$.
Writing 
\begin{equation}
 G_{abc}^I = \sum_{j = 1}^\infty G^{j;I}_{abc},\label{eq_kg:sumj_Gabc}
\end{equation}
it can be seen that the power of $m$ can be accounted for by taking derivatives with respect to the rotation parameter:
\begin{equation}
 G^{j;I}_{abc} = \left(-\frac{i}{j \beta}\right)^c \frac{d^c G^{j;I}_{ab0}}{d\Omega^c_I}.
 \label{eq_Gabc_from_dOmega}
\end{equation}
On the other hand, the sum over $m$ can be performed in $G_{ab0}^{j;I}$ using the summation theorem for Bessel functions [see Eq.~(8.531.1) in Ref.~\cite{gradshteyn2015table}; and Eq.~(10.23.7) in Ref.~\cite{olver10}]:
\begin{equation}
 \sum_{m = -\infty}^\infty e^{-i m x} J_m^2(z) = J_0\left(2z \sin\frac{x}{2}\right),
\end{equation}
leading to
\begin{multline}
 G_{ab0}^{j;I} = 
 \int_0^\infty d\omega \, e^{-j \beta \omega} \omega^a
 \int_0^\omega \frac{dk}{2\pi^2} q^b J_0\left(2q\rho \sin \frac{j \beta \Omega_I}{2}\right)\\
 = \int_0^\infty \frac{dx e^{-x} x^{a+b+1}}{2\pi^2 (j\beta)^{a+b+2}}
 \int_0^{\pi/2} d\theta (\cos\theta)^{b+1} J_0(\alpha_j x \cos\theta),
 \label{eq_Gabc_aux}
\end{multline}
where $x = j \beta \omega$ and $\theta$ is defined by $(k,q) = \omega (\sin\theta,\cos\theta)$, while $\alpha_j$ is given by
\begin{equation}
 \alpha_j = \frac{2\rho}{j\beta} \sin \frac{j \beta \Omega_I}{2} = \frac{l}{\pi j} \sin(\pi j \nu),
 \label{eq_alphaj}
\end{equation}
with
\begin{subequations}\label{eq_notations}
\begin{align}
l & = \frac{2\pi \rho}{\beta}, \qquad \qquad\ L = \frac{2\pi R}{\beta}, \\
\nu & \equiv \nu_I = \frac{\beta \Omega_I}{2\pi}, \qquad
\nu_R = \frac{\beta \Omega}{2\pi},
\end{align}
\end{subequations}
where, for convenience, we reproduced Eq.~\eq{eq_l_nu} and introduced other notations to be used later (notice that $0 \leqslant l \leqslant L$).

In order to perform the integral with respect to $\theta$ in Eq.~\eqref{eq_Gabc_aux}, we replace the Bessel function $J_0(x)$ by its series expansion,
\begin{equation}
    J_0(x) = \sum_{k=0}^\infty \frac{(-1)^k x^{2k}}{4^k (k!)^2}\,.
\end{equation}
The integral with respect to $\theta$ can be performed now term by term using the relation (valid for ${\mathrm{Re}}\,\gamma > -1$)
\begin{align}
    \int_0^{\pi/2} \cos^\gamma \theta d\theta = \frac{\sqrt{\pi}\Gamma \left( \frac{1+\gamma}{2} \right) }{2 \Gamma \left(1+ \frac{\gamma}{2} \right)}\,, 
    \label{eq_cos:int}
\end{align}
Using the following identities for the gamma functions,
\begin{align}
    \Gamma(n + 1) {\biggl|}_{n \in {\mathbb N}} & = n!\,, \\
    \Gamma \left(\frac{1}{2} + n \right) {\biggl|}_{n \in {\mathbb N}} & = \sqrt{\pi} \frac{(2n)!}{4^n n!}\,,
\end{align}
we arrive at
\begin{subequations}
\begin{align}
    \int_0^{\pi / 2} d\theta\, \cos\theta J_0(\alpha_j x \cos\theta) & = \sum_{k=0}^\infty \frac{(-\alpha_j^2 x^2)^k}{(2k+1)!}\,,\label{eq:int_J0_b0}\\
    \int_0^{\pi / 2} d\theta\, \cos^3\theta J_0(\alpha_j x \cos\theta) & = \sum_{k=1}^\infty \frac{(2k)^2 (-\alpha_j^2 x^2)^{k-1}}{(2k+1)!}\,,\label{eq:int_J0_b2}
\end{align}
\end{subequations}
corresponding to the cases $b = 0$ and $2$ in Eq.~\eqref{eq_Gabc_aux}. The summation can be trivially performed, 
\begin{subequations}\label{eq_mysery}
\begin{align}
 \int_0^{\pi / 2} d\theta\, \cos\theta J_0(\alpha_j x \cos\theta) &= \frac{\sin(\alpha_j x)}{\alpha_j x},
 \label{eq_mysery:1}\\
 \int_0^{\pi / 2} d\theta\, \cos^3\theta J_0(\alpha_j x \cos\theta) &= 
 \frac{\cos(\alpha_j x)}{\alpha_j^2 x^2} \nonumber\\
 & \hspace{-35pt} + (\alpha_j^2 x^2 - 1) \frac{\sin(\alpha_j x)}{\alpha_j^3 x^3},
 \label{eq_mysery:2}
\end{align}
\end{subequations}
finally arriving at
\begin{align}
 G^{j;I}_{n00} &= \frac{1}{2\pi^2 \alpha_j (j\beta)^{n+2}}
 \int_0^\infty dx e^{-x} x^n \sin(x \alpha_j),\nonumber\\
 G^{j;I}_{n20} &= \frac{1}{2\pi^2 \alpha_j^3 (j\beta)^{n+4}}
 \int_0^\infty dx e^{-x} x^n [x \alpha_j \cos(x \alpha_j) \nonumber\\
 &\hspace{80pt} + (x^2 \alpha_j^2 - 1) \sin(x\alpha_j)].
\end{align}

Employing the identity
\begin{align}
    \int_0^\infty dx\, e^{-x + i \alpha_j x} x^n = \frac{n!}{(1 - i \alpha)^{n+1}}\,, 
\end{align}
or equivalently,
\begin{align}
 \int_0^\infty dx\, e^{-x} x^n \sin(\alpha_j x) &= n!\, {\rm Im}\left(\frac{1 + i \alpha_j}{1 + \alpha_j^2}\right)^{n+1},\\
 \int_0^\infty dx\, e^{-x} x^n \cos(\alpha_j x) &= n!\, {\rm Re} \, \left(\frac{1 + i\alpha_j}{1+ \alpha^2_j}\right)^{n+1},
\end{align}
with ${\rm Re}(z) = (z + z^*) / 2$ and ${\rm Im}(z) = (z - z^*) / 2i$ being the real and imaginary parts of a complex number $z$, we obtain
\begin{subequations}
\begin{align}
 G^{j;I}_{n00} &= \frac{n!}{2\pi^2 \alpha_j (j \beta)^{n+2}} {\rm Im}\left(\frac{1 + i \alpha_j}{1 + \alpha^2_j}\right)^{n+1}, \label{eq_Gn00} \\
 G^{j;I}_{n20} &= \frac{1}{2\pi^2 \alpha_j^3 (j \beta)^{n+4}} \Bigg[
 \alpha_j^2 (n+2)! {\rm Im}\left(\frac{1 + i \alpha_j}{1 + \alpha_j^2}\right)^{n+3} \nonumber\\
 &\hspace{-15pt} + \alpha_j (n+1)!\, {\rm Re}\left(\frac{1 + i \alpha_j}{1 + \alpha_j^2}\right)^{n+2} \hspace{-15pt}
 -n!\, {\rm Im}\left(\frac{1 + i \alpha_j}{1 + \alpha_j^2}\right)^{n+1} \Bigg].
\end{align}
Specifically, for $\phi^2$ and $T^{\mu\nu}$, we require:
\begin{align}
 G^{j;I}_{000} &= \frac{1}{2\pi^2 (j\beta)^2(1 + \alpha_j^2)},\label{eq_G000}\\
 G^{j;I}_{100} &= \frac{1}{\pi^2 (j\beta)^3(1 + \alpha_j^2)^2},\label{eq_G100}\\
 G^{j;I}_{200} &= \frac{3 - \alpha_j^2}{\pi^2 (j\beta)^4(1 + \alpha_j^2)^3},\label{eq_G200}\\
 G^{j;I}_{020} &= \frac{2(1 - \alpha_j^2)}{\pi^2 (j\beta)^4(1 + \alpha_j^2)^3}.
 \label{eq_G020}
\end{align}
Furthermore, $G^{j;I}_{001}$, $G^{j;I}_{101}$, and $G^{j;I}_{002}$ can be obtained by means of Eq.~\eqref{eq_Gabc_from_dOmega}:
\begin{align}
 G^{j;I}_{001} &= \frac{i \rho^2 \sin(j \beta \Omega_I)}{\pi^2 (j\beta)^4(1 + \alpha_j^2)^2},\label{eq_G001}\\
 G^{j;I}_{101} &= \frac{4i \rho^2 \sin(j \beta \Omega_I)}{\pi^2 (j\beta)^5(1 + \alpha_j^2)^3},\label{eq_G101}\\
 G^{j;I}_{002} &= \frac{\rho^2 \cos(j \beta \Omega_I)}{\pi^2 (j\beta)^4(1 + \alpha_j^2)^2} - 
 \frac{4\rho^4 \sin^2(j \beta \Omega_I)}{\pi^2 (j\beta)^6(1 + \alpha_j^2)^3}.
 \label{eq_G002}
\end{align}
Finally, the functions $G^{(1); j;I}_{000}$ and $G^{(2);j;I}_{000}$ corresponding to the expressions in Eq.~\eqref{eq_dG000} are
\begin{align}
 G^{(1);j;I}_{000} &= 2\alpha_j^2 \frac{dG^{j}_{000}}{d\alpha_j^2} = -\frac{\alpha_j^2}{\pi^2 j^2 \beta^2 (1 + \alpha_j^2)^2},\label{eq_G(1)000}\\
 G^{(2);j;I}_{000} &= 2\alpha_j^2 \frac{dG^{(1);j}_{000}}{d\alpha_j^2} = \frac{2\alpha_j^2 (\alpha_j^2 - 1)}{\pi^2 j^2 \beta^2 (1 + \alpha_j^2)^3}.\label{eq_G(2)000}
\end{align}
\end{subequations}

Substituting Eq.~\eqref{eq_G000} into Eq.~\eqref{eq_tev_phi2} allows $\phi^2$ to be expressed as
\begin{subequations}\label{eq_sumj}
\begin{align}
 \phi^2_I &= \frac{1}{2\pi^2 \beta^2}
 \sum_{j = 1}^\infty \frac{1}{j^2} \frac{1}{1+\alpha_j^2},\label{eq_sumj_phi2} 
\end{align}
while the non-vanishing components of $T^{\mu\nu}$, given in Eq.~\eqref{eq_tev_SET}, reduce to
\begin{align}
 T^{tt}_I &= \sum_{j = 1}^\infty \frac{3-\alpha_j^2 + \frac{2}{3}(\alpha_j^2 - 1) \sin^2(\pi j \nu)}{\pi^2 \beta^4 j^4 (1+\alpha_j^2)^3},  \label{eq_sumj_Ttt}\\
 T^{\rho\rho}_I &= \sum_{j = 1}^\infty \frac{3 - 2\sin^2(\pi j \nu)}{3\pi^2 \beta^4 j^4 (1 + \alpha_j^2)^2}, \label{eq_sumj_Trr}\\
 T^{\varphi\varphi}_I &= \sum_{j = 1}^\infty \frac{[3 - 2\sin^2(\pi j \nu)](1 - 3\alpha_j^2)}{3\pi^2 \rho^2 \beta^4 j^4 (1 + \alpha_j^2)^3}, \label{eq_sumj_Tpp} \\
 T^{zz}_I &= \sum_{j = 1}^\infty \frac{3 + 3 \alpha_j^2 + 2(1 - \alpha_j^2) \sin^2(\pi j \nu)}{3\pi^2 \beta^4 j^4 (1 + \alpha_j^2)^3},\label{eq_sumj_Tzz} \\
 T^{t\varphi}_I &= \sum_{j = 1}^\infty \frac{4i \sin(2\pi j \nu)}{\pi^2 (j\beta)^5(1 + \alpha_j^2)^3}.
 \label{eq_sumj_Ttp}
\end{align}
\end{subequations}

The above results show that the diagonal components of $T^{\mu\nu}$ are real-valued and even with respect to $\nu = \beta \Omega_I / 2\pi$, while $T^{t\varphi}_I$ is imaginary and odd with respect to $\nu \rightarrow -\nu$. Under the hermitization~\eqref{eq_rho_herm}, it is clear that $T^{t\varphi}_I$ vanishes and $T^{\mu\nu}_I$ remains diagonal. As in the case of the classical relativistic kinetic theory (RKT) analysis in Sec.~\ref{sec:RKT:imaginary}, the resulting state is not isotropic. Identifying as in the classical case $E_I = T_I^{tt}$ and the perfect-fluid contribution $T^{\mu\nu}_{{\rm pf};I} = {\rm diag}(E_I, P_I, \rho^{-2} P_I, P_I)$ with $P_I = E_I/3$, the quantum shear-stress tensor $\pi_I^{\mu\nu} = T_I^{\mu\nu} - T^{\mu\nu}_{{\rm pf};I} = {\rm diag}(0, \pi_I^{\rho\rho}, \pi_I^{\varphi\varphi}, \pi_I^{zz})$ can be obtained as
\begin{align}
 \pi_I^{\rho\rho} &= \frac{4}{9} \sum_{j = 1}^\infty \frac{3 \alpha_j^2 - (2\alpha_j^2 + 1) \sin^2(\pi j \nu)}{\pi^2 \beta^4 j^4 (1 + \alpha_j^2)^3}, \nonumber\\
 \pi_I^{\varphi\varphi} &= -\frac{4}{9} \sum_{j = 1}^\infty \frac{6 \alpha_j^2 - (4\alpha_j^2 - 1) \sin^2(\pi j \nu)}{\pi^2 \rho^2 \beta^4 j^4 (1 + \alpha_j^2)^3}, \nonumber\\
 \pi_I^{zz} &= \frac{4}{9} \sum_{j = 1}^\infty \frac{3 \alpha_j^2 - 2(\alpha_j^2 - 1) \sin^2(\pi j \nu)}{\pi^2 \beta^4 j^4 (1 + \alpha_j^2)^3}. 
\end{align}
At large distances from the rotation axis, $\alpha_j \rightarrow \infty$, implying that
\begin{multline}
 T_I^{\mu\nu} = {\rm diag} (-1,1,-3\rho^{-2}, 1)  \\\times 
 \sum_{j = 1}^\infty \frac{3 - 2 \sin^2(\pi j \nu)}{3\pi^2 \beta^4 j^4(1 + \alpha_j^2)^2}.
\end{multline}
The structure of the above result is similar to that obtained in Eq.~\eqref{eq_RKT_im_Tmunu_far}, with the important difference that the quantum field-theoretical (QFT) $T_I^{\mu\nu}$ depends on $\nu$ through the harmonic function $\sin(\pi j \nu)$. This property implies that $T_I^{\mu\nu}$ obtained in QFT is periodic with respect to $\nu$, with period $\Delta \nu = 1$, in agreement with the symmetry~\eq{eq_chi_periodicity} expected on general grounds. This periodic behaviour is fundamentally different from that observed in RKT (Subsect.~\ref{sec:RKT:imaginary}), where no periodicity in $\nu$ can be seen.

\subsection{Values on the rotation axis: no analytical connection between real and imaginary rotations}\label{sec_KG_axis}

\begin{figure}
\centering
\begin{tabular}{c}
\includegraphics[width=.98\linewidth]{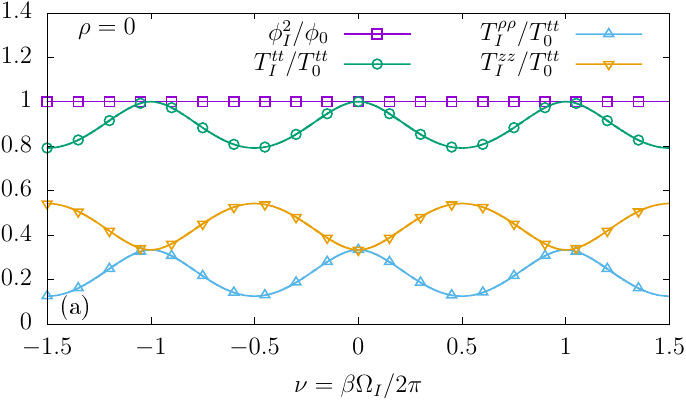} \\
\includegraphics[width=.98\linewidth]{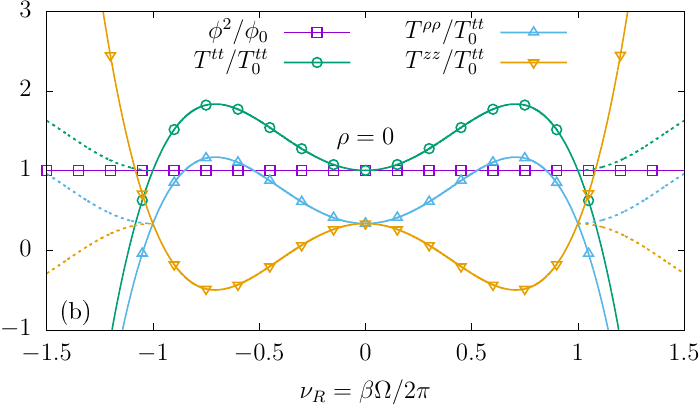} 
\end{tabular}
\caption{The condensate $\phi^2$ and the components of the energy-momentum tensor $T^{\mu\nu}$ on the axis of rotation of the cylinder, $\rho=0$, under (a) imaginary rotation with $\nu \equiv \nu_I = \beta \Omega_I / 2\pi$ and (b) real rotation with $\nu_R = \beta \Omega / 2\pi$, normalized with respect to their values in the absence of rotation. All quantities under imaginary rotation are periodic ($\nu \to \nu + 1$) in agreement with Eq.~\eq{eq_chi_periodicity}. The dashed lines extending in the region $|\nu_R| > 1$ indicate the expected behaviour if the components of $T_{\mu\nu}$ were periodic with respect to $\nu_R$.
\label{fig_axis}
}
\end{figure}

On the rotation axis, we have $\alpha_j = 0$. Using the relations $\sum_{j=1}^\infty j^{-2} = \pi^2/6$ and $\sum_{j = 1}^\infty j^{-4} = \pi^4/90$, we find that the expectation value of $\phi^2$ is not affected by the imaginary rotation while the energy-momentum tensor acquires a nontrivial dependence on the imaginary angular frequency:
\begin{subequations}\label{eq_rho0}
\begin{align}
 \left.\phi_I^2\right|_{\rho=0} &= \phi^2_0, \label{eq_rho0_phi2}\\
 \left.T^{tt}_I\right|_{\rho = 0} &= T_0^{tt} \left(1 - \frac{10 \{\nu\}^2}{3} + \frac{20\{\nu\}^3}{3} - \frac{10\{\nu\}^4}{3}\right),
 \label{eq_rho0_Ttt}\\
 \left.T^{\rho\rho}_I\right|_{\rho = 0} &= T_0^{tt} \left(\frac{1}{3} - \frac{10 \{\nu\}^2}{3} + \frac{20\{\nu\}^3}{3} - \frac{10\{\nu\}^4}{3}\right),
 \label{eq_rho0_Trr}\\
 \left.T^{zz}_I\right|_{\rho = 0} &= T_0^{tt} \left(\frac{1}{3} + \frac{10 \{\nu\}^2}{3} - \frac{20\{\nu\}^3}{3} + \frac{10\{\nu\}^4}{3}\right),
 \label{eq_rho0_Tzz}\\
 \left.T^{t\varphi}_I\right|_{\rho = 0} &= \pm \frac{4i\pi^3\{\nu\}}{45\beta^5} \bigl(1 - 10 \{\nu\}^2 + 15\{\nu\}^3 - 6\{\nu\}^4\bigr),
 \label{eq_rho0_Ttp}
\end{align}
\end{subequations}
and $\left.\rho^2 T_I^{\varphi\varphi}\right|_{\rho = 0} = \left. T_I^{\rho\rho}\right|_{\rho = 0}$. In the above, 
\begin{equation}
 \phi_0^2 = \frac{1}{12 \beta^2}, \qquad 
 T^{tt}_0 = \frac{\pi^2}{30 \beta^4},
 \label{eq_0}
\end{equation}
represent the expectation values in the absence of rotation ($\Omega_I = 0$). The notation $\{\nu\} = \nu - \lfloor \nu \rfloor$ represents the fractional part of $\nu$ ($0 \le \{\nu \} < 1$), while $\lfloor \nu \rfloor$ represents its integer part. The $\pm$ sign in the expression for $\left. T^{t\varphi}_I\right|_{\rho = 0}$ corresponds to the sign of $\nu$. 
Figure~\ref{fig_axis}(a) confirms that $T^{\mu\nu}_I$ is periodic with respect to the imaginary rotation parameter $\nu$, as implied by the presence of $\{\nu\}$. The energy density $T_I^{tt}$ and the radial pressure $T_I^{\rho\rho}$ are decreased by the imaginary rotation, while the azimuthal and vertical pressures $T_I^{\varphi\varphi} = T_I^{zz}$ are increased.

An alternative way of characterizing $T_I^{\mu\nu}$ on the rotation axis is by using the Bernoulli polynomial 
\begin{align}
 B_n(x) &= \sum_{k = 0}^n \binom{n}{k} B_{n-k} x^k \nonumber\\
 &= -\frac{n!}{(2\pi i)^n} \left[{\rm Li}_n(e^{2\pi i x}) + (-1)^n {\rm Li}_n(e^{-2\pi i x})\right],
\end{align} 
with ${\rm Li}_s(x) = \sum_{k = 1}^\infty x^k / k^s$ being the polylogarithm and
$B_n \equiv B_n(0)$ being the Bernoulli numbers:
\begin{align}
    B_n = \sum_{k = 0}^n \sum_{v = 0}^k (-1)^v \binom{k}{v} \frac{v^n}{k + 1}\,.
\end{align}
In terms of the Bernoulli polynomials, the components of the energy-momentum read
\begin{subequations}
\begin{align}
 \left.T_I^{tt}\right|_{\rho = 0} &= T^{tt}_0 \left[\frac{8}{9} - \frac{10}{3} B_4(\{\nu\})\right], \\
 \left.T_I^{\rho\rho}\right|_{\rho = 0} &= T^{tt}_0 \left[\frac{2}{9} - \frac{10}{3} B_4(\{\nu\})\right], \\
 \left.T_I^{zz}\right|_{\rho = 0} &= T^{tt}_0 \left[\frac{4}{9} + \frac{10}{3} B_4(\{\nu\})\right], \\
 \left.T_I^{t\varphi}\right|_{\rho = 0} &= -\frac{8i \pi^3}{15 \beta^5} B_5(\{\nu\}),
\end{align}
\end{subequations}
as well as $\left.\rho^2 T_I^{\varphi\varphi}\right|_{\rho = 0} = \left. T_I^{\rho\rho}\right|_{\rho = 0}$. These are the results on the axis of rotation for an infinite-volume system subjected to the imaginary rotation.

We now compare the results in Eq.~\eqref{eq_rho0} to those derived on the basis of real rotation in Refs.~\cite{Ambrus:2014itg,Becattini:2015nva,Ambrus:2017opa,Becattini:2020qol} using a perturbative approach for slow rotation, reproduced below for definiteness:
\begin{subequations}\label{eq_real_rho0}
\begin{align}
 \left.\phi^2\right|_{\rho=0} &= \phi^2_0, \label{eq_real_rho0_phi2}\\
 \left.T^{tt}\right|_{\rho = 0} &= T_0^{tt} \left(1 + \frac{10 \nu_R^2}{3} - \frac{10\nu_R^4}{3}\right),
 \label{eq_real_rho0_Ttt}\\
 \left.T^{\rho\rho}\right|_{\rho = 0} &= T_0^{tt} \left(\frac{1}{3} + \frac{10 \nu_R^2}{3} - \frac{10\nu_R^4}{3}\right),
 \label{eq_real_rho0_Trr}\\
 \left.T^{zz}\right|_{\rho = 0} &= T_0^{tt} \left(\frac{1}{3} - \frac{10 \nu_R^2}{3} + \frac{10\nu_R^4}{3}\right),
 \label{eq_real_rho0_Tzz}\\
 \left.T^{t\varphi}\right|_{\rho = 0} &= \frac{4\pi^3\nu_R}{45\beta^5} \left(1 + 10 \nu_R^2 - 6\nu_R^4\right),
\end{align}
\end{subequations}
and $\left.\rho^2 T^{\varphi\varphi}\right|_{\rho = 0} = \left. T^{\rho\rho}\right|_{\rho = 0}$, with $\nu_R$ defined in Eq.~\eq{eq_notations}. 

The above results can be derived from Eq.~\eqref{eq_rho0} using the following replacements: 
\begin{equation}\label{eq_no_analytical}
 \{\nu\}^2 \rightarrow -\nu_R^2, \qquad 
 \{\nu\}^3 \rightarrow 0, \qquad 
 \{\nu\}^4 \rightarrow \nu_R^4.
\end{equation}
It is remarkable to observe that the diagonal components of $T^{\mu\nu}$ satisfy 
\begin{align}
T^{\mu\nu}(\nu_R = \pm 1)  = T^{\mu\nu}(\nu_R = 0) \qquad \text{for} \ \mu = \nu\,,    
\end{align}
however contrary to the same quantities evaluated under imaginary rotations, they do not exhibit periodicity with respect to $\nu_R$. Figure~\ref{fig_axis}(b) shows that when $|\nu_R| > 1$, $T^{zz}$ increases dramatically, while $T^{tt}$ and $T^{\rho\rho} = \rho^2 T^{\varphi\varphi}$ eventually become negative. The dashed lines extending in the region $|\nu_R| > 1$ indicate the expected behaviour if $T^{\mu\nu}$ were periodic with respect to $\nu_R$.

Before ending this subsection, we remark that the presence of odd powers of $\nu$ in the expressions for $T^{\mu\nu}_I$ is unexpected and seemingly unsupported by the formulas in Eq.~\eqref{eq_sumj}. For example, in the case of $T^{tt}_I$ given by Eq.~\eqref{eq_sumj_Ttt}, a Taylor expansion of the summand around $\nu = 0$ fails to capture the $\nu^3$ term revealed in Eq.~\eqref{eq_rho0_Ttt}. Moreover, since $\nu$ always appears multiplied by the summation variable $j$, such an approach can reliably produce only the first two terms, proportional to $j^{-4} \nu^0$ and $j^{-2} \nu^2$. The third term proportional to $j^0 \nu^4$ cannot be computed due to the divergence of the sum over $j$. We are thus led to believe that the $\nu^3$ term appearing in Eq.~\eqref{eq_rho0} is related to an inherent non-analytic behavior of $T^{\mu\nu}_I$ with respect to the rotation parameter $\nu$. We remark that the results quoted in Eqs.~\eqref{eq_real_rho0} for the case of real rotation were obtained also using a Taylor series approach and may therefore omit similar non-analytical $\nu_R^3$-like terms.

\subsection{High temperature expansion} \label{sec_KG_highT}

Let us now consider the large temperature expansion, when $\beta \rightarrow 0$. Since $\beta$ comes multiplied by $j$ under the summation sign in Eqs.~\eqref{eq_sumj}, higher-order terms with respect to $\beta$ come with higher powers of $j$. Since the summation over $j$ and the power series with respect to $j \beta$ in general do not commute, this procedure allows only the coefficients of the $\beta^{-4}$ and $\beta^{-2}$ terms to be extracted. The results are
\begin{gather}
 \phi_I^2 = \frac{\gamma_I^2}{12\beta^2}, \quad 
 T_I^{tt} = \frac{\pi^2 \gamma_I^4}{90\beta^4} (4\gamma_I^2 - 1) - \frac{\Omega_I^2 \gamma_I^6}{36\beta^2} (6\gamma_I^2 - 5), \nonumber\\ 
 T_I^{\rho\rho} = \frac{\pi^2 \gamma_I^4}{90\beta^4} - \frac{\Omega_I^2 \gamma_I^6}{36\beta^2}, \qquad 
 T_I^{zz} = \frac{\pi^2 \gamma_I^4}{90\beta^4} + \frac{\Omega_I^2 \gamma_I^6}{36\beta^2}, \nonumber\\
 \rho^2 T_I^{\varphi\varphi} = \frac{\pi^2 \gamma_I^4}{90\beta^4} (4\gamma_I^2 - 3) - \frac{\Omega_I^2 \gamma_I^6}{36\beta^2} (6\gamma_I^2 - 5),\nonumber\\
 T_I^{t\varphi} = i \Omega_I \left[\frac{2\pi^2 \gamma_I^6}{45\beta^4} - \frac{\Omega_I^2 \gamma_I^6}{18\beta^2} (3\gamma_I^2 - 1)\right].
 \label{eq_highT}
\end{gather}
As discussed in the previous subsection, the $\nu^3$ terms revealed in Eq.~\eqref{eq_rho0} are not captured by the perturbative series expansion approach. Nevertheless, the results reported in Eq.~\eqref{eq_highT} are fully consistent with previously derived results, see Eq.~(4.2.51) of Ref.~\cite{Ambrus:2014itg}; Eqs.~(A.21,A.22) of Ref.~\cite{Ambrus:2017opa}; and Eqs.~(7.19,7.22,7.23) of Ref.~\cite{Becattini:2020qol}.

\subsection{Emergence of fractal structure}\label{sec_KG_fractal}

\begin{figure*}
    \centering
    \includegraphics[width=1.9\columnwidth]{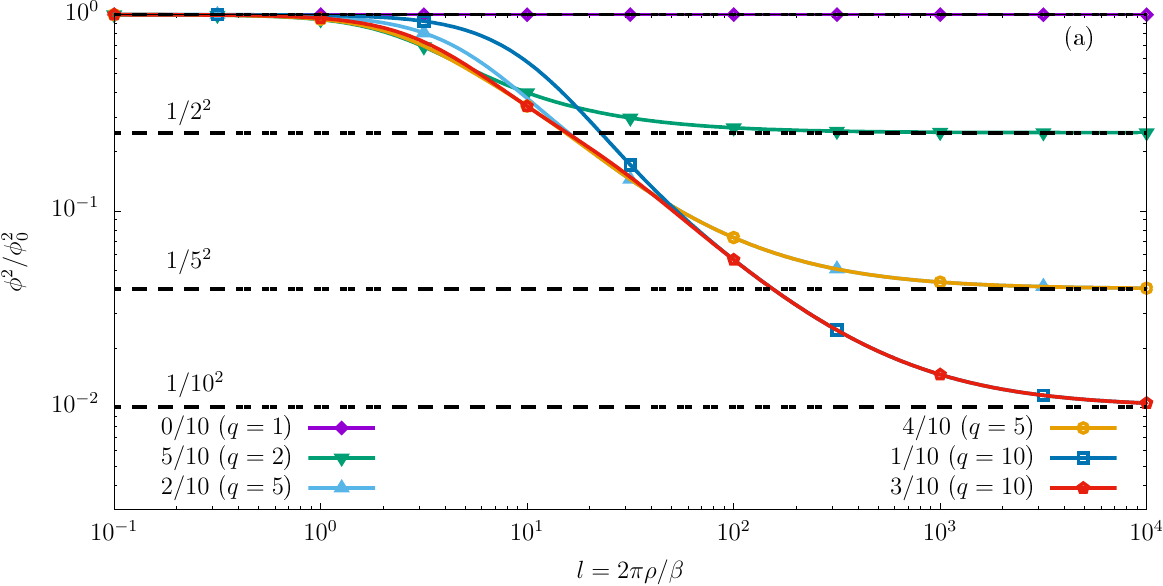} \\[2mm]
    \includegraphics[width=1.9\columnwidth]{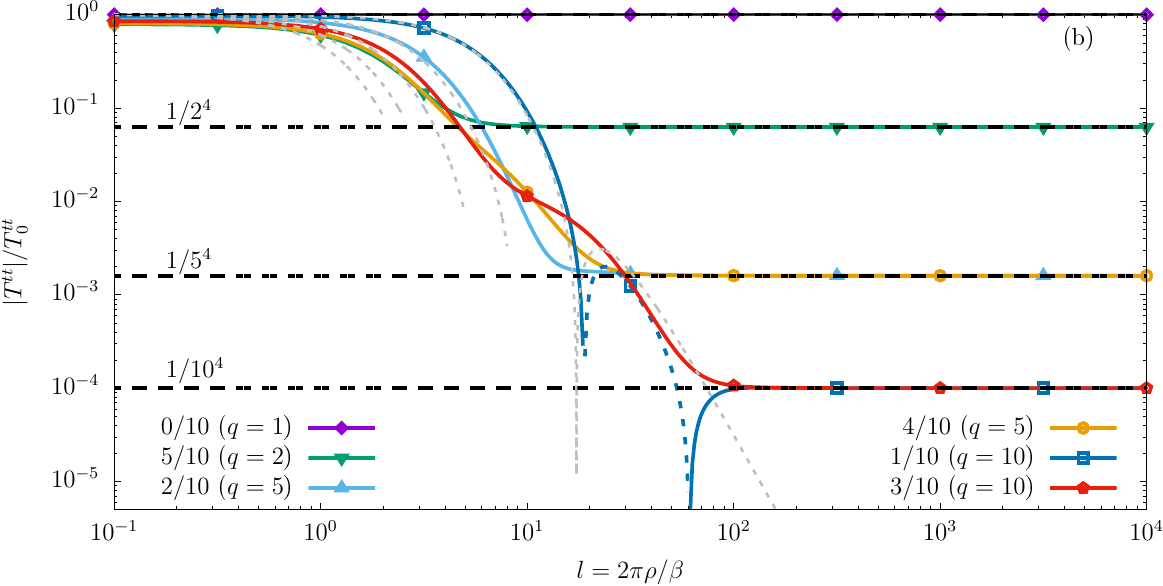}
    \caption{Fractalization of thermodynamics with increasing volume: The thermal expectation values of (a) $\phi^2$ and (b) $T^{tt}$ under imaginary rotation normalized with respect to their values in the absence of rotation ($\phi^2_0 = 1 / 12\beta^2$ and $T^{tt}_0 = \pi^2 / 30\beta^4$) as functions of dimensionless distance $l = \rho / (2\pi\beta)$ from the rotation axis of the cylinder. The lines and points show results for rational values of $\nu = \beta \Omega_I / 2\pi$ of the form $r / 10$, with $0 \le r \le 5$, corresponding to irreducible fractions $p/q$ with $q = 1$, $2$, $5$ and $10$, which are identical to the imaginary frequencies used for the rotating ring in Fig.~\ref{fig_ring_pressure}. The horizontal dashed black lines represent the expected large-distance plateau given by (a) $1/q^2$ and (b) $1/q^4$, which signal the fractal behaviour of thermodynamics. The gray dotted lines represent the relativistic kinetic theory prediction in Eq.~\eqref{eq_RKT_im_E}. A small segment of the result for $T^{tt}$ when $p/q=1/10$ corresponds to negative values and is represented with dashed lines. The values are obtained using Eq.~\eqref{eq_pq}. }
    \label{fig_Ttt_vs_rho}
\end{figure*}

We now consider the case when $\nu = p / q$ is a rational number, where $p/q$ is an irreducible fraction, as discussed in Sec.~\ref{sec_ring_fractal}. Writing $j = Q q + j'$, with $0 \le Q < \infty$ and $1 \le j' \le q$, the trigonometric functions taking as argument $j \pi \nu = \pi p Q + j' \pi p/q$ depend only on $j'$. Specifically, Eq.~\eqref{eq_sumj} becomes
\begin{subequations}\label{eq_sumjp_aux}
\begin{align}
 \phi_I^2 &= \frac{1}{2\pi^2 \beta^2 q^2} \sum_{j = 1}^{q} \sum_{Q = 0}^\infty 
 \frac{1}{(Q + \frac{j}{q})^2 + x_j^2},
 \label{eq_phi2_sumjp_aux} \\
 T_I^{tt} &= \frac{1}{3\pi^2 \beta^4 q^4} \sum_{j = 1}^q \sum_{Q = 0}^\infty \frac{(9 - 2s_j^2)(Q + \frac{j}{q})^2 - (3 - 2s_j^2) x_j^2}{[(Q + \frac{j}{q})^2 + x_j^2]^3},
 \label{eq_Ttt_sumjp_aux} \\
 T_I^{\rho\rho} &= \frac{1}{3\pi^2 \beta^4 q^4} \sum_{j = 1}^q \sum_{Q = 0}^\infty \frac{3 - 2s_j^2}{[(Q + \frac{j}{q})^2 + x_j^2]^2},
 \label{eq_Trr_sumjp_aux} \\
 T_I^{\varphi\varphi} &= \frac{1}{3\pi^2 \rho^2 \beta^4 q^4} \sum_{j = 1}^q \sum_{Q = 0}^\infty \frac{(3 - 2s_j^2) [(Q +\frac{j}{q})^2 - 3x_j^2]}{[(Q + \frac{j}{q})^2 + x_j^2]^3}, 
 \label{eq_Tpp_sumjp_aux} \\
 T_I^{zz} &= \frac{1}{3\pi^2 \beta^4 q^4} \sum_{j = 1}^q \sum_{Q = 0}^\infty \frac{(3 + 2s_j^2)(Q + \frac{j}{q})^2 + (3 - 2s_j^2) x_j^2}{[(Q + \frac{j}{q})^2 + x_j^2]^3},
 \label{eq_Tzz_sumjp_aux} \\ 
 T_I^{\varphi t} &= \frac{4i}{\pi^2 \beta^5 q^5} \sum_{j = 1}^q \sum_{Q = 0}^\infty \frac{2 s_j c_j (Q + \frac{j}{q})}{[(Q + \frac{j}{q})^2 + x_j^2]^3},
 \label{eq_Tpt_sumjp_aux}
\end{align}
\end{subequations}
where we introduced the notation
\begin{equation}
 x_j = \frac{l s_j}{\pi q}, \quad 
 s_j = \sin\left(\frac{\pi j p}{q}\right), \quad 
 c_j = \cos\left(\frac{\pi j p}{q}\right),
 \label{eq_xj_sj_cj_def}
\end{equation}
while $j'$ was relabeled as $j$ for convenience.

The sum over $Q$ introduced by the procedure shown in Eq.~\eqref{eq_sumjp_aux} can be performed using 
\begin{subequations}
\begin{align}
 \sum_{Q = 0}^\infty \frac{1}{(Q + \frac{j}{q})^2 + x_j^2} &= \frac{1}{x_j} {\rm Im} \, \psi_j,\label{eq_sum_dig1}\\
 \sum_{Q = 0}^\infty \frac{1}{[(Q + \frac{j}{q})^2 + x_j^2]^2} &= \frac{{\rm Im} \, \psi_j}{2x_j^3} - \frac{{\rm Re}\psi'_j}{2x_j^2},\label{eq_sum_dig2}\\
 \sum_{Q = 0}^\infty \frac{1}{[(Q + \frac{j}{q})^2 + x_j^2]^3} &= \frac{3{\rm Im} \, \psi_j}{8x_j^5} - \frac{3{\rm Re} \, \psi'_j}{8x_j^4}  - \frac{{\rm Im} \, \psi''_j}{8x_j^3}, \label{eq_sum_dig3}\\
 \sum_{Q = 0}^\infty \frac{Q + \frac{j}{q}}{[(Q+\frac{j}{q})^2 + x_j^2]^3} &= -\frac{{\rm Im} \, \psi'_j}{8x_j^3} + \frac{{\rm Re} \, \psi''_j}{8x_j^2}, \label{eq_sum_dig1p}
\end{align}
\end{subequations}
where 
\begin{equation}
 \psi_j \equiv \psi\left(\frac{j}{q} + ix_j\right),
 \label{eq_psij_def}
\end{equation}
$\psi(x) = \Gamma'(x) / \Gamma(x)$ is the digamma function and the primes denote differentiation with respect to the argument, e.g. $\psi''(x) = d^2 \psi(x) / dx^2$. Also, ${\rm Im}$ and ${\rm Re}$ denote the real and imaginary parts of their arguments, respectively: ${\rm Im}\, \psi_j = \frac{1}{2i} (\psi_j - \psi_j^*)$ and ${\rm Re}\, \psi_j' = \frac{1}{2} (\psi_j' + \psi_j'{}^*)$, with $\psi_j^* = \psi(\frac{j}{q} - ix_j)$.

When considering the summation over $j$ appearing in Eq.~\eqref{eq_sumjp_aux}, a special case corresponds to $j = q$, when the sine term $s_j = \sin(\pi j p / q)$ cancels. In this case, we employ
\begin{equation}
 \sum_{Q = 0}^\infty \frac{1}{(Q + 1)^2} = \frac{\pi^2}{6}, 
 \qquad 
 \sum_{Q = 0}^\infty \frac{1}{(Q + 1)^4} = \frac{\pi^4}{90}.
\end{equation}
Since $s_j = 0$ implies also $x_j = 0$, Eqs.~\eqref{eq_sumjp_aux} show that the $j = q$ contribution becomes $l$-independent. This allows all expectation values to be split as
\begin{equation}
 \phi_I^2 = \phi_q^2 + \frac{\delta \phi_I^2}{2\pi^2 q^2 \beta^2}, \quad 
 T_I^{\mu\nu} = T^{\mu\nu}_q + \frac{\delta T^{\mu\nu}_I}{2\pi^2 q^4 \beta^4},
 \label{eq_fractal_aux}
\end{equation}
where the first terms are coordinate-independent and correspond to a bosonic gas at rest with inverse temperature $q \beta$:
\begin{align}
 \phi^2_q &\equiv 
 \phi^2_0(q\beta) = \frac{1}{12q^2 \beta^2}, \nonumber\\
 T^{\mu\nu}_q &\equiv 
 T^{\mu\nu}_0(q\beta) = \frac{\pi^2}{30 q^4 \beta^4} {\rm diag}\left(1, \frac{1}{3}, \frac{1}{3} \rho^{-2}, \frac{1}{3}\right).
 \label{eq_phi20_Ttt0}
\end{align}
The factor $q$ represents the denominator of the irreducible fraction $\nu = p/ q$. More importantly, because these terms are independent of the transverse distance to the rotation axis given by $l$, they become dominant at large distances from the rotation axis, giving rise to fractal structures in the thermodynamic (infinite volume) limit. 
It is noteworthy that the fractal terms are completely absent in the relativistic kinetic theory analysis in Sec.~\ref{sec:RKT:imaginary} and thus represent a purely quantum effect.

The second terms in Eq.~\eqref{eq_fractal_aux} ``defractalize'' the result close to the rotation axis and are computed via:
\begin{subequations}\label{eq_pq}
\begin{align}
 \delta \phi_I^2 &= {\rm Im} \, \sum_{j = 1}^{q-1} \frac{\psi_j}{x_j},\label{eq_pq_phi2}\\
 \delta T_I^{tt} &= {\rm Im} \, \sum_{j = 1}^{q-1}\left[ \frac{s_j^2}{3} \left(\frac{\psi_j}{x_j^3} - \frac{i \psi'_j}{x_j^2} - \frac{\psi''_j}{x_j} \right) + \frac{\psi''_j}{x_j} \right],\label{eq_pq_Ttt} \\
 \delta T_I^{\rho\rho} &= {\rm Im} \, \sum_{j = 1}^{q-1}\left(1 - \frac{2s_j^2}{3}\right) \left(\frac{\psi_j}{x_j^3} - \frac{i \psi'_j}{x_j^2} \right),
 \label{eq_pq_Trr}\\
 \rho^2 \delta T_I^{\varphi\varphi} &=  {\rm Im} \, \sum_{j = 1}^{q-1}\left(1 - \frac{2s_j^2}{3}\right) \left(-\frac{2\psi_j}{x_j^3} + \frac{2i \psi'_j}{x_j^2} + \frac{\psi_j''}{x_j} \right),
 \label{eq_pq_Tpp}\\
 \delta T_I^{zz} &= {\rm Im} \, \sum_{j = 1}^{q-1}\left[\left(1 - \frac{s_j^2}{3}\right) \left(\frac{\psi_j}{x_j^3} - \frac{i \psi'_j}{x_j^2} \right) + \frac{s_j^2 \psi''_j}{3x_j}\right],
 \label{eq_pq_Tzz}\\
 \delta T_I^{t\varphi} &= -\frac{i}{q \beta} {\rm Im}\sum_{j= 1}^{q-1} \sin\frac{2\pi j p}{q} \left(\frac{\psi'_j}{x_j^3} - \frac{i \psi_j''}{x_j^2}\right).\label{eq_pq_Ttphi}
\end{align}
\end{subequations}
where $\psi_j$, $x_j$ and $s_j$ were introduced in Eqs.~\eqref{eq_xj_sj_cj_def} and \eqref{eq_psij_def}.

\begin{figure*}
\centering
\includegraphics[width=.98\linewidth]{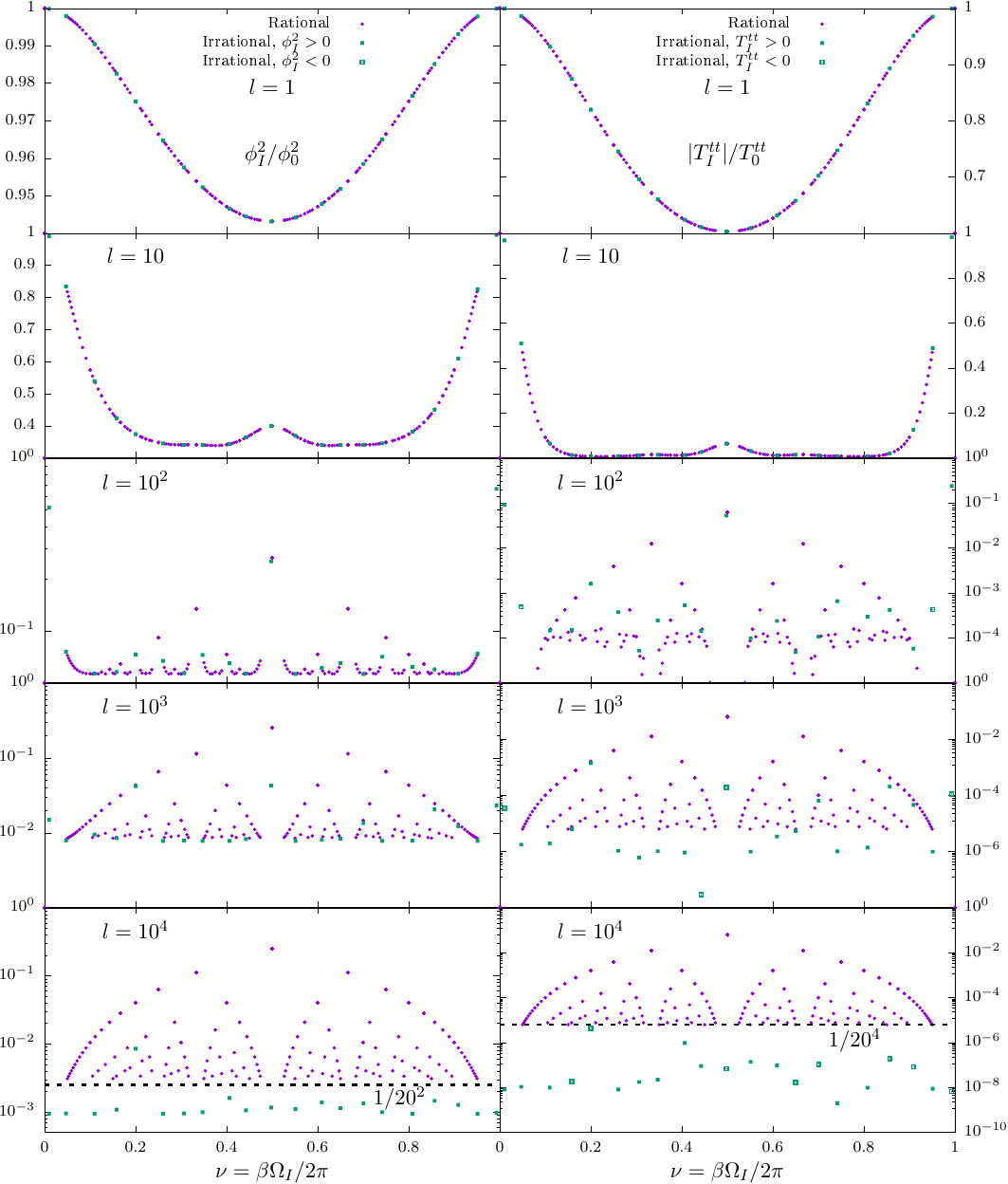}
\caption{Thermal expectation values of (left) $\phi_I^2$ normalized by $\phi_0^2 = 1/ 12\beta^2$; and (right) $|T_I^{tt}|$ normalized by $T_0^{tt} = \pi^2 / 30 \beta^4$, shown with respect to the rotation parameter $\nu = \beta \Omega_I / 2\pi$, for various values of the distance parameter $l = 2\pi \rho / \beta$ from the axis of rotation of the cylinder.
The purple circles correspond to the case when $\nu = p / q$ is a rational number (we considered all irreducible fractions with $1 \le q \le 20$). The green squares correspond to the irrational values $\nu_j$ shown in Eq.~\eqref{eq_nuj_irr}. The empty symbols indicate the case when $T_I^{tt} < 0$. The dashed line shown in the bottom panels (for $l = 10^4$) indicate the expected lower bounds (left) $1/q^2$ and (right) $1/q^4$ with $q = 20$. 
This figure should be compared with Fig.~\ref{fig_ring_thermal} for particles in the ring: As the distance $l$ to the rotation axis grows, the thermal expectation values  get fractal features similar to the fractalization of thermodynamics of scalar particles in the ring.
}
\label{fig_fractal}
\end{figure*}

For $\nu = 1/2$, we have:
\begin{align}
 \frac{\phi_I^2}{\phi^2_0} & = \frac{1}{4} + \frac{3}{2l} \tanh \frac{l}{2}\,,\nonumber\\
 \frac{T^{tt}_I}{T^{tt}_0} & = \frac{1}{16} + \frac{5}{4l^3} \left(\tanh\frac{l}{2} - 
 \frac{l/2}{\cosh^2\frac{l}{2}} + \frac{l^2 \tanh\frac{l}{2}}{\cosh^2\frac{l}{2}}\right)\,,\nonumber\\
 \frac{T_I^{\rho\rho}}{T^{tt}_0} & = \frac{1}{48} + \frac{5}{4l^3} \left(\tanh\frac{l}{2} - 
 \frac{l/2}{\cosh^2\frac{l}{2}}\right)\,,\nonumber\\
 \frac{\rho^2 T_I^{\varphi\varphi}}{T^{tt}_0} & = \frac{1}{48} - \frac{5}{2l^3} \left(\tanh\frac{l}{2} - 
 \frac{l/2}{\cosh^2\frac{l}{2}} - \frac{l^2 \tanh\frac{l}{2}}{4\cosh^2\frac{l}{2}}\right)\,,\nonumber\\
 \frac{T_I^{zz}}{T^{tt}_0} & = \frac{1}{48} + \frac{5}{2l^3} \left(\tanh\frac{l}{2} - 
 \frac{l/2}{\cosh^2\frac{l}{2}} + \frac{l^2 \tanh\frac{l}{2}}{4\cosh^2\frac{l}{2}}\right)\,,\nonumber\\
 T_I^{t \varphi} &= 0. \label{eq_fixed:OmegaI:pi}
\end{align}

The behaviour of the scalar condensate $\phi_I^2$ and energy density $T_I^{tt}$ as functions of $l$ is illustrated in panels (a) and (b) of Fig.~\ref{fig_Ttt_vs_rho}, respectively, where we consider the cases $\nu = p' / 10$ with $0 \le p' \le 5$, corresponding to irreducible fractions $p/q$ with $q = 1$, $2$, $5$ and $10$. As $l > 1$, visible differences between the curves corresponding to various values of $\nu$ can be seen. Contrary to the classical case shown in Eq.~\eqref{eq_RKT_im_Tmunu_far}, the far-field behavior of $\phi_I^2$ and $T_I^{\mu\nu}$ is dominated by quantum effects. An estimate of how these observables approach their asymptotic values $\phi^2_q$ and $T^{\mu\nu}_q$ can be obtained by considering the decay of the ``classical'' part from Eq.~\eqref{eq_RKT_im_Tmunu_far} to values of the same order of magnitude as $\phi^2_q$ and $T^{\mu\nu}_q$, which occurs at values $l \gtrsim l_q$, where
\begin{equation}
 l_q \sim \frac{q}{\nu} = \frac{q^2}{p}.
\end{equation}
The $q^2$ dependence of $l_q$ is confirmed for both $\phi_I^2$ and $T_I^{tt}$, however the $p$ dependence appears to be negligible.
The emerging fractal behaviour exhibits a stark contrast to the classical result in Eq.~\eqref{eq_RKT_im_E} derived within relativistic kinetic theory, which is also shown in Fig.~\ref{fig_Ttt_vs_rho}(b) using dashed gray lines. Sizeable deviations can be seen for the curves with smaller value of $q$, which reach the fractalized plateau at smaller values of $L$. In the $p/q = 1/10$ case, the RKT curve follows closely the QFT one, providing a good approximation also in the region where $T_I^{tt}$ becomes negative. Noting that the RKT result for $p / q = 3/10$ falls off too rapidly compared to the QFT curve leads us to conclude that the classical RKT description becomes valid only in the limit $\nu \rightarrow 0$.

As discussed above, the fractal behaviour manifests itself at large distances from the rotation axis, i.e., as $l \rightarrow \infty$. Figure~\ref{fig_fractal} illustrates the expectation values of $\phi_I^2 / \phi_0^2$ (left) and $T_I^{tt} / T^{tt}_0$ (right) with respect to $\nu$ for various values of $l$. We considered $\nu = p' / q'$ with $1 \le q' \le 20$ and $0 \le p' \le q'$, covering all irreducible fractions $p / q$ with $1 \le q \le 20$. These results are represented with purple circles. We also considered a set $\nu_j$ ($0 \le j \le n = 20$) of ``irrational'' values of $\nu$, represented using green squares, obtained as:
\begin{equation}
 \nu_j = \frac{j}{n} + \delta \nu_j, \label{eq_nuj_irr}
\end{equation}
where $-0.01< \delta \nu_j < 0.01$ is a random number.\footnote{For $j = 0$ and $j = 20$, we employed $\nu_0 = |\delta \nu_0|$ and $\nu_{20} = 1 - |\delta \nu_{20}|$, respectively, in order to ensure that $0 \le \nu_j \le 1$.}
In order to employ a logarithmic scale on the vertical axis, we represented the absolute values of our observables, with the convention that filled and empty symbols are used when the observables are positive and negative, respectively. Since $\phi_I^2 > 0$ for all values of $\nu$ and $l$, this discussion applies only to $T_I^{tt}$ (see, e.g., the $(p,q) = (1,10)$ curve in Fig.~\ref{fig_Ttt_vs_rho}). 

For $l \lesssim 1$, both $\phi_I^2$ and $T_I^{tt}$ exhibit a smooth dependence on $\nu$. As $l$ is increased, the expectation values for the case when $\nu = p/q$ is an irreducible fraction become frozen on their corresponding asymptotic values ($1/q^2$ for $\phi_I^2/\phi_0^2$ and $1/q^4$ for $T_I^{tt}/T^{tt}_0$), earlier for smaller values of $q$ than for larger values of $q$. In contrast, the expectation values corresponding to the irrational values of $\nu$ continue their decreasing trend towards $0$.

Strikingly, the thermodynamics of the scalar field in the (3+1)d cylindrically symmetric space, obtained numerically and shown in Fig.~\ref{fig_fractal}, resembles drastically the one at the ring, obtained analytically and represented in Fig.~\ref{fig_ring_thermal}, with the fractalization features becoming more pronounced as the distance $l$ to the rotation axis is increased.

\subsection{Thermodynamic limit} \label{sec_KG_thermo}

For the purpose of analyzing the large-volume limit of our system, we consider a fictitious cylinder of radius $R \equiv \beta L / 2\pi$ and of large vertical extent $L_z$, centered on the rotation axis. The volume-averaged scalar condensate and energy density can be computed by integrating Eqs.~\eqref{eq_fractal_aux}, \eqref{eq_pq_phi2} and \eqref{eq_pq_Ttt} over this cylinder and dividing by the total volume $V = \pi R^2 L_z$:
\begin{subequations}
\begin{align}
 \overline{\Phi_I^2} &= \phi^2_0(q \beta) + \frac{1}{\beta^2 L^2} \sum_{j = 1}^{q-1} \frac{1}{s_j^2} \ln \frac{\Gamma(j/q)}{|\Gamma_j|}, \label{eq_KG_thermo_phi2}\\
 \om{E}_I &= T^{tt}_0(q\beta) + \frac{1}{L^2 q^2 \beta^4} \sum_{j = 1}^{q-1} \Bigg[\left(\frac{1}{3} - \frac{1}{s_j^2}\right) {\rm Re}\psi'_j \nonumber\\
 & - \frac{{\rm Im}\psi_j}{3X_j} + \frac{1}{s_j^2} \psi'\left(\frac{j}{q}\right) \Bigg],
 \label{eq_KG_thermo_E}
\end{align}
\end{subequations}
where $\phi_0^2$ and $T^{tt}_0$ were introduced in Eq.~\eqref{eq_phi20_Ttt0}, $s_j$ was defined in Eq.~\eqref{eq_xj_sj_cj_def}, while $X_j$ corresponds to the old $x_j$ evaluated at the volume boundary:
\begin{equation}
 X_j = \frac{L s_j}{\pi q}. \label{eq_Xj_def}
\end{equation}
Furthermore, we keep the notation $\Gamma_j$ and $\psi_j$ introduced in Eq.~\eqref{eq_psij_def}, but now we understand that these functions take the argument $\frac{j}{q} + i X_j$. 

We now compute the average free energy $\om{F}_I$ from $\om{E}_I$ starting from Eq.~\eqref{eq_F_from_E}:
\begin{align}    
 \om{F}_I &= \om{F}_0 (q\beta) 
  - \frac{1}{\beta^4 L^2 q^2} \sum_{j = 1}^{q-1} \Biggl\{ \left(\frac{1}{3} - \frac{1}{s_j^2} \right) \frac{{\rm Im}\psi_j}{X_j} 
\nonumber \\
&
 + \frac{1}{s_j^2}  
 \psi' \left(\frac{j}{q}\right) - \frac{1}{3 X_j} \int_0^{X_j} \frac{d x}{x} {\rm Im}\left[\psi\left(\frac{j}{q} + i x\right)\right] \Biggr\},
 \label{eq_KG_thermo_F}
\end{align}
where $\om{F}_0(\beta) = -\pi^2 / 90\beta^4$ is the free energy of a bosonic gas in the absence of rotation.

The entropy and angular momentum given by Eq.~\eqref{eq_thermo_rel} require taking derivatives of $\om{F}_I$ with respect to $\beta$ and $\Omega_I$ at constant $\Omega_I$ and $\beta$, respectively. This is not possible at the level of the fractalized form in Eq.~\eqref{eq_KG_thermo_F}. Thus, we seek to obtain the free energy after rewriting $\om{E}_I$ for general (not necessarily rational) values of $\nu$:
\begin{equation}
 \om{E}_I = \sum_{j = 1}^\infty \frac{3 + \alpha_j^2(R) - \frac{2}{3} \sin^2(\pi j \nu)}
 {\pi^2 \beta^4 j^4 [1 + \alpha_j^2(R)]^2}.
\end{equation}
It can be checked that writing $\nu = p/q$ and $j = qQ + j'$ gives Eq.~\eqref{eq_KG_thermo_E}. Applying now Eq.~\eqref{eq_F_from_E} leads to 
\begin{multline}
 \om{F}_I = -\sum_{j = 1}^\infty \frac{1}{\pi^2 \beta^4 j^4} \left\{\frac{\alpha_j^2 + \frac{1}{3} \sin^2(\pi j \nu)}{\alpha_j^2 (\alpha_j^2 + 1)} \right.\\
 \left. - \frac{\sin^2 \pi j \nu}{3\alpha_j^3} \left[\frac{\pi}{2} - \arctan\left(\frac{1}{\alpha_j}\right)\right]\right\},
 \label{eq_KG_F_sumj}
\end{multline}
where the term $\pi/2$ appearing on the second line represents an integration constant such that $\lim_{\Omega_I \rightarrow 0} \om{F}_I = -\sum_{j = 1}^{\infty} 1/(\pi^2 \beta^4 j^4) = \om{F}_0$. Using Eq.~\eqref{eq_thermo_rel}, the average entropy $\om{S}_I$ and angular momentum $\om{M}_I$ are 
\begin{align}
 \om{S}_I &= \sum_{j = 1}^\infty \frac{1}{\pi^2 \beta^3 j^4} \Bigg\{
 \frac{2(\alpha_j^2 + 2)}{(\alpha_j^2 + 1)^2} + \frac{\sin^2(\pi j \nu) ( 1 - \alpha_j^2)}{3\alpha_j^2(\alpha_j^2 + 1)^2} \nonumber\\
 & + \frac{\pi j \nu}{\tan(\pi j \nu)} \left[ 
 \frac{2\alpha_j^2}{(\alpha_j^2 + 1)^2} + \frac{\sin^2(\pi j \nu)(3\alpha_j^2 + 1)}{3\alpha_j^2 (\alpha_j^2 + 1)^2}\right]\nonumber\\
 & 
 - \frac{\sin^2(\pi j \nu)}{3\alpha_j^3}\left(1 + \frac{\pi j \nu}{\tan(\pi j \nu)}\right) \left(\frac{\pi}{2} - \arctan \frac{1}{\alpha_j}\right) \Bigg\},\nonumber\\
 \om{M}_I &= \sum_{j = 1}^\infty \frac{\sin(2\pi j \nu)}{4\pi^2 \beta^3 j^3} \left[\frac{2L^2 / \pi^2 j^2}{(1+\alpha_j^2)^2} + \frac{1 - \alpha_j^2}{3(1+\alpha_j^2)^2} \right.\nonumber\\
 & \left. - \frac{1}{3\alpha_j^3} \left(\frac{\pi}{2} - \alpha_j - \arctan \frac{1}{\alpha_j}\right)\right].
\end{align}

Considering as before that $\nu = p / q$ and writing $j = qQ + j'$, with $0 \le Q < \infty$ and $1 \le j' \le q$, we get 
\begin{subequations}\label{eq_KG_thermo_2}
\begin{align}    
 \om{F}_I &= \om{F}_0 (q\beta) 
 - \frac{1}{\beta^4 L^2 q^2} \sum_{j = 1}^{q-1} 
 \Biggl\{ \left(\frac{1}{3} - \frac{1}{s_j^2} \right) \frac{{\rm Im} \, \psi_j}{X_j} 
 \nonumber \\
 & + \frac{1}{s_j^2} \psi' \left(\frac{j}{q}\right) 
 - \frac{1}{3 X_j} S_Q \Biggr\},\label{eq_KG_thermo_F2}\\
 \om{S}_I &= \frac{S_0(q\beta)}{q} + \frac{1}{\pi^2 q^4 j^3} \sum_{j = 1}^{q-1} \Bigg\{ \frac{2}{X_j^2} \psi'\left(\frac{j}{q}\right) - \frac{{\rm Im}\psi_j}{X_j^3} \nonumber\\
 & + \left(\frac{s_j^2}{3} - 1\right) \left({\rm Re} \, \frac{\psi'_j}{X_j^2} - \frac{c_j p \pi}{s_j X_j}{\rm Im}\psi'_j\right) \nonumber\\
 & + \frac{c_j p \pi}{s_j X_j^2} \left(\frac{s_j^2}{3} - 2\right) \left[\psi\left(\frac{j}{q}\right) - {\rm Re} \, \psi_j\right] \nonumber\\
 & - \frac{1}{3X_j} \left(\frac{\pi q}{L}\right)^2 \left(S_Q + \frac{\pi p c_j}{s_j} S_Q'\right)\Bigg\},\label{eq_KG_thermo_S2}\\
 \om{M}_I &= \sum_{j = 1}^{q-1} \frac{\sin(2\pi j \nu)}{12\pi^2 q^3 \beta^3} \Bigg\{ \frac{s_j^2 - 6}{s_j^2 X_j^2} \left[\psi\left(\frac{j}{q}\right) - {\rm Re} \, \psi_j\right]\nonumber\\
 & -\frac{s_j^2 - 3}{s_j^2 X_j} {\rm Im} \, \psi'_j - \frac{1}{X_j^3} S_Q'\Bigg\},
 \label{eq_KG_thermo_M2}
\end{align}
\end{subequations}
where $s_j$ and $c_j$ were introduced in Eq.~\eqref{eq_xj_sj_cj_def}, $\psi_j$ in Eq.~\eqref{eq_psij_def} with $x_j$ replaced by $X_j$, while $X_j$ is defined in Eq.~\eqref{eq_Xj_def}. Furthermore, the following notation was introduced:
\begin{align}
 S_Q &= \sum_{Q = 0}^\infty \frac{1}{Q+\frac{j}{q}} \left[\frac{\pi}{2} - \arctan\left(\frac{Q + \frac{j}{q}}{X_j}\right)\right], \nonumber\\
 S_Q' &= \sum_{Q = 0}^\infty \left[\frac{\pi}{2} - \arctan\left(\frac{Q + \frac{j}{q}}{X_j}\right) - \frac{X_j}{Q + \frac{j}{q}}\right].
\end{align}
Comparing Eqs.~\eqref{eq_KG_thermo_F2} and \eqref{eq_KG_thermo_F}, it can be seen that 
\begin{align}
 S_Q &= \int_0^{X_j} \frac{dx}{x} {\rm Im}\left[\psi\left(\frac{j}{q} + i x\right)\right].
 \label{eq_KG_SQ}
\end{align}
The above identity is easily established by noting that 
\begin{align}
 \frac{\partial S_Q}{\partial X_j} &= \sum_{Q = 0}^\infty \frac{1}{(Q + \frac{j}{q})^2 + X_j^2} = \frac{{\rm Im} \, \psi_j}{X_j}.
\end{align}
Integrating the above expression with respect to $X_j$ and demanding $S_Q(X_j = 0) = 0$ gives Eq.~\eqref{eq_KG_SQ}. A similar expression can be obtained for $S_Q'$. Taking the derivative with respect to $X_j$ eliminates the arctangent, such that
\begin{align}
 \frac{\partial S_Q'}{\partial X_j} &= -\sum_{Q = 0}^\infty \frac{X_j^2}{(Q+\frac{j}{q})[(Q + \frac{j}{q})^2 + X_j^2]} \nonumber\\
 &= \psi\left(\frac{j}{q}\right) - {\rm Re}\left[\psi\left(\frac{j}{q} + i X_j\right)\right].
\end{align}
Integrating the above relation with respect to $X_j$ gives
\begin{equation}
 S_Q' = X_j \psi\left(\frac{j}{q} \right) - \int_0^{X_j} dx\, {\rm Re}\left[\psi\left(\frac{j}{q} + ix\right)\right].
\end{equation}

For the case when $\nu = p / q = 1/2$, we find 
\begin{align}
 \frac{\overline{\Phi^2_I}}{\phi_0^2} &= \frac{1}{4} + \frac{6}{L^2} \ln\left(\cosh \frac{L}{2}\right),\nonumber\\
 \frac{\om{E}_I}{T_0^{tt}} &= \frac{1}{16} + \frac{5}{4L^2}\left(1 + 2 \tanh^2 \frac{L}{2} - \frac{2}{L} \tanh\frac{L}{2}\right),
\end{align}
while $\om{M}_I = 0$. 

\subsection{Slow rotation: moment of inertia and shape}
\label{sec_KG_slow}

In the case of slow rotation, 
we take the free energy in the absence of rotation, $\om{F}(0)$, as well as the first two coefficients, $K_2$ and $K_4$, by  performing a series expansion with respect to $\nu$ in Eq.~\eqref{eq_KG_F_sumj}:
\begin{align}
 \om{F}(0) &= -\sum_{j = 1}^\infty \frac{1}{\pi^2 j^4 \beta^4}, \nonumber\\
 \overline{K}_2 &= -\sum_{j = 1}^\infty \frac{2}{j^4 \pi^2 \beta^4} \left(1 + \frac{2 \pi^2 j^2}{9 L^2}\right),\nonumber\\
 \overline{K}_4 &= -\sum_{j = 1}^\infty \frac{24}{\pi^2 j^4 \beta^4} \left(1 + \frac{3\pi^2 j^2}{5L^2} + \frac{2\pi^4 j^4}{27 L^4}\right),
\end{align}
where we took into account that $\Omega_I^2 \rightarrow -\Omega^2$ for the purpose of extracting $K_{2n}$. We also employed the notation $\overline{K}_{2n} \equiv \om{F}(0) K_{2n}$.

The leading term $\om{F}(0)$ and the first coefficient $K_2$, which corresponds to the dimensionless moment of inertia, can be computed exactly:
\begin{equation}
 \om{F}(0) = -\frac{\pi^2}{90 \beta^4}, \qquad
 K_2 = 2 + \frac{20}{3L^2}\,.
 \label{eq_KG_K2}
\end{equation}
In the case of the rotational shape coefficient $K_4$, the summation of the coefficient of $L^{-4}$ cannot be performed. However, its leading-order behavior and the first $L$-dependent correction can be extracted as follows:
\begin{equation}
 K_4 = 24 + \frac{216}{L^2} + \dots.
 \label{eq_KG_K4}
\end{equation}
Comparing Eqs.~\eqref{eq_KG_K2}--\eqref{eq_KG_K4} to \eqref{eq_ring_K2n} [see also Eq.~\eqref{eq_RKT_K2n}], we see that the classical result $K_{2n} = (2n)!$ receives $L$-dependent corrections that vanish as the transverse size of the cylinder becomes infinite,  $L \rightarrow \infty$. As in the case of the $(1+1)d$ ring, shown in Fig.~\ref{fig:ring_QFT_M0} [see also Eq.~\eqref{eq:ring_QFT_K2n_an}], both $K_2(L)$ and $K_4(L)$ exceed at finite $L$ their thermodynamic limits, $K_{2n}(\infty) = (2n)!$.

In Subsect.~\ref{sec_bounded_slow}, we discuss the dimensionless moment of inertia $K_2$ and the rotational shape change coefficient $K_4$ in a cylinder of a finite radius, taking into account the proper quantization of the radial modes. We will see that the behaviour of $K_2$ qualitatively agrees with Eq.~\eq{eq_KG_K2} while the coefficient $K_4$ becomes finite and converges to the classical result~\eqref{eq_RKT_K2n} in the infinite volume limit.

\subsection{Effect of finite mass}\label{sec_KG_massive}

In this subsection, we discuss the effect of a finite mass on the fractalization of thermodynamics (seen at the level of the local observables), as well as on the slow rotation coefficients, $K_{2n}$. For this purpose, we start the analysis with the component $T_I^{tt}$, which is still given by Eq.~\eqref{eq_tev_SET_Ttt}. The only difference is that the $\omega$ integration in the functions $G^I_{abc}$ introduced in Eq.~\eqref{eq_kg:Gabc} now starts from $\omega = M$,
\begin{multline}
 G^I_{abc}(M) = \sum_{m = -\infty}^\infty \int_M^\infty \frac{d\omega}{e^{\beta \tilde{\omega}} - 1} 
 \int_{0}^\omega \frac{dk}{2\pi^2} \omega^a q^b m^c J^2_m(q\rho).
 \label{eq_kg:Gabc_massive}
\end{multline}

Focusing on the functions $G^I_{ab0} = \sum_{j=1}^\infty (-1)^j G^{j;I}_{ab0}$, as pointed out in Eq.~\eqref{eq_kg:sumj_Gabc}, and taking the same steps as described in Subsec.~\ref{sec_KG_OmegaI}, we arrive at Eq.~\eqref{eq_Gabc_aux}, which is modified to
\begin{multline}
 G_{ab0}^{j;I} = \int_{j\beta M}^\infty \frac{dx e^{-x} x^a y^{b+1}}{2\pi^2 (j\beta)^{a+b+2}} \\
 \times \int_0^{\pi/2} d\theta (\cos\theta)^{b+1} J_0(\alpha_j y \cos\theta),
 \label{eq_Gabc_aux_M}
\end{multline}
where we introduced $y = [x^2 - (j\beta M)^2]^{1/2}$ corresponding to the momentum magnitude, $p = \sqrt{\omega^2 - M^2}$.
In the case $b = 0$, which is relevant to the computation of $T_I^{tt}$, the $\theta$ integral can be performed using Eq.~\eqref{eq:int_J0_b0}:
\begin{equation}
 G_{n00}^{j;I} = \int_{j\beta M}^\infty \frac{dx \, x^{n-1} y}{2\pi^2 \alpha_j (j\beta)^{n+2}} e^{-x} \sin(\alpha_j y).
 \label{eq:unb_Gn00j_massive}
\end{equation}
% We further change the integration variable to $x \rightarrow x + j\beta M$, such that 
% \begin{multline}
%  G^j_{n00} = \frac{e^{-j \beta M}}{2\pi^2 \alpha_j (j\beta)^{n+2}}
%  \int_{0}^\infty dx e^{-x} (x+j\beta M)^{n-1}\\\times 
%  \sqrt{x(x+2j\beta M)} \sin(\alpha_j \sqrt{x(x+2j \beta M)}).
% \end{multline}

\begin{figure}
\begin{center}
\includegraphics[width=.99\linewidth]{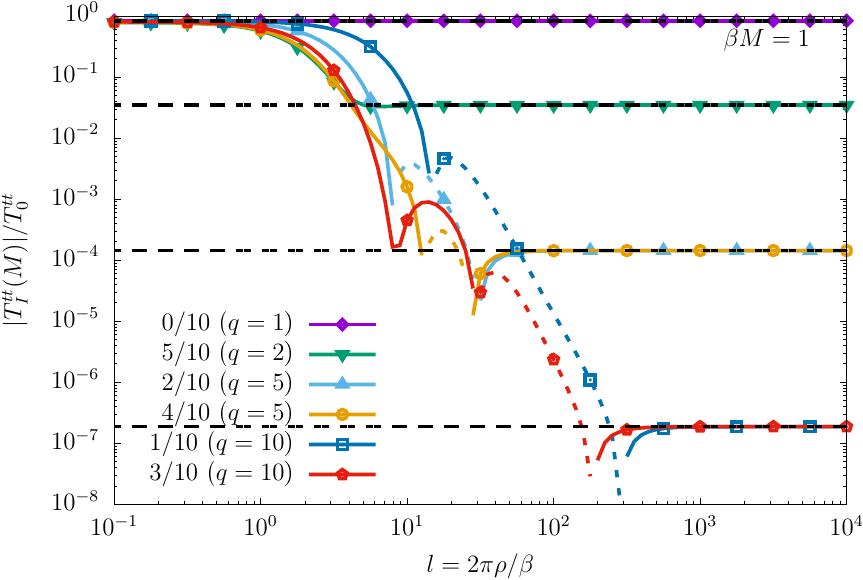}
\end{center}
\caption{Ratio $T^{tt}_I(M) / T^{tt}_0$ between the energy density at finite mass, given in Eq.~\eqref{eq:Tttj_M}, and $T^{tt}_0 = \pi^2 / 30 \beta^4$. The rotation parameter takes the values $\nu = p' / 10$, with $0 \le p' \le 5$, leading to irreducible fractions $p / q$ with $q \in \{2, 5, 10\}$. The dotted colored lines and points correspond to regions where $T_I^{tt}(M) < 0$, while the dashed black lines represent the thermodynamic limit derived in Eq.~\eqref{eq:Ttt_fractal_M}.}
\label{fig_unb_QFT_M}
\end{figure}

The integral in \eqref{eq:unb_Gn00j_massive} proves difficult for general values of the parameters. We now focus on $T_I^{tt}$, requiring $G^I_{200}$ and the function $G^{(2);I}_{000} = \rho \partial_\rho \rho \partial_\rho G^I_{000}$, for which we find:
\begin{multline}
 G^{(2);j;I}_{000} = \int_{j \beta M}^\infty \frac{dx\, e^{-x} y}{2\pi^2 j^2 \beta^2 \alpha_j x } [(1 - \alpha_j^2 y^2) \sin(\alpha_j y) \\
 - \alpha_j y \cos(\alpha_j y)].
\end{multline}
This allows $T^{tt}_{j;I}$ to be obtained using Eq.~\eqref{eq_tev_SET_Ttt} as
\begin{multline}
 T^{tt}_{j;I} = \int_{j \beta M}^\infty \frac{dx\, e^{-x} y}{2\pi^2 (j \beta)^4 \alpha_j x} \Bigg\{x^2 \sin(\alpha_j y) \\
 + \frac{\pi^2 j^2}{3 l^2} [(1-\alpha_j^2 y^2) \sin(\alpha_j y) - \alpha_j y \cos(\alpha_j y)]\Bigg\}.
 \label{eq:Tttj_M}
\end{multline}
To study the fractal properties of $T^{tt}_I$, we consider a rational rotation parameter, $\nu = p/q$, with $p<q$ being irreducible coprime numbers, and write $j = q Q + j'$, with $1 \le j' \le q$ and $0 \le Q < \infty$. Since $\alpha_j = \frac{l}{\pi j} \sin(\pi j \nu)$ vanishes when he $j' = q$, the corresponding contribution $T^{tt}_q$ becomes coordinate-independent and evaluates to
\begin{align}
 T^{tt}_q &= \sum_{Q = 1}^\infty \int_{qQ \beta M}^\infty \frac{dx\, e^{-x} y^2 x}{2\pi^2 (q Q \beta)^4} \nonumber\\
 &= \sum_{Q = 1}^\infty \frac{e^{-q Q \beta M}}{\pi^2 (qQ\beta)^4}[3 + 3 q Q \beta M + (q Q \beta M)^2] \nonumber\\
 &= \frac{1}{\pi^2 (q \beta)^4} \left[3 {\rm Li}_4(e^{-q \beta M}) + 3 q \beta M {\rm Li}_3(e^{-q \beta M}) \right.\nonumber\\
 &\left. \qquad \qquad + (q \beta M)^2 {\rm Li}_2(e^{-q \beta M})\right].
 \label{eq:Ttt_fractal_M}
\end{align}
The results of the numerical evaluation of $T_I^{tt}(M)$ for $\nu = p / q$ with $q = 10$ and $0 \le p \le 5$ are shown in Fig.~\ref{fig_unb_QFT_M}. The fractal pattern can easily be seen as $L$ is increased. The large-$L$ value coincides with the expression in Eq.~\eqref{eq:Ttt_fractal_M}, shown with horizontal dotted black lines.

We now attempt to evaluate the average free energy $\om{F}(0, M)$ in the absence of rotation, as well as the shape coefficients $K_2$ and $K_4$. For this purpose, we evaluate the average energy $\om{E}_{j;I} = \frac{2}{R^2} \int_0^R d\rho\,\rho T^{tt}_{j;I}$ starting from Eq.~\eqref{eq:Tttj_M}:
\begin{multline}
 \om{E}_{j;I} = \int_{j \beta M}^\infty \frac{dx\, e^{-x}}{\pi^2 j^4 \beta^4 A_j^2 x} \Bigg[x^2 - \frac{y s_j^2}{3A_j} \sin(A_j y) \\
 + \left(\frac{s_j^2 y^2}{3} - x^2\right) \cos(A_j y)\Bigg].
\end{multline}
where $A_j = (L / \pi j) \sin(\pi j \nu)$.
Computing the free energy via $\om{F}_I = \beta^{-1} \int d\beta\, \om{E}_I$ is cumbersome due to the the $\beta M$ dependence. Since we are interested only in the slow rotation limit, we can perform an expansion with respect to $v_R = i\nu L$,
\begin{equation}
 \om{E}_{j;I} = \om{E}_{j;0} - \om{E}_{j;2} \frac{(\nu L)^2}{2!} + \om{E}_{j;4} \frac{(\nu L)^4}{4!} + \dots,
\end{equation}
where 
\begin{align}
 \om{E}_{j;0} &= \int_{j \beta M}^\infty \frac{dx\, e^{-x} x y^2}{2\pi^2 j^4 \beta^4}, \nonumber\\
 \om{E}_{j;2} &= \int_{j \beta M}^\infty \frac{dx\, e^{-x} x y^4}{12\pi^2 j^4 \beta^4} \left(1 + \frac{8 \pi^2j^2}{3x^2 L^2}\right), \nonumber\\
 \om{E}_{j;4} &= \int_{j \beta M}^\infty \frac{dx\, e^{-x} x y^6}{30\pi^2 j^4 \beta^4} \left[1 + 
 \frac{2\pi^2 j^2}{L^2} \left(\frac{5}{y^2} + \frac{4}{x^2}\right) \right.\nonumber\\
 & \hspace{.5\columnwidth}\left. + \frac{80 \pi^4 j^4}{3L^4 x^2 y^2}\right].
\end{align}
The integral with respect to $x$ can be performed exactly for all terms displayed above. However, the $x^{-1}$ factors appearing in the $L^{-2}$ and $L^{-4}$ terms (which vanish in the thermodynamic limit) lead to results in terms of the exponential integral function, ${\rm Ei}(-j \beta M)$, which makes analytical manipulations difficult. Instead, we focus on the leading-order contributions, for which we get
\begin{align}
 \om{E}_{j;0} &= \frac{e^{-j \beta M}}{\pi^2 j^4 \beta^4} [3 + 3 j \beta M + (j \beta M)^2],\nonumber\\
 \om{E}_{j;2} &\rightarrow \frac{2e^{-j \beta M}}{3 \pi^2 j^4 \beta^4} [15 + 15 j \beta M + 6 (j \beta M)^2 + (j \beta M)^3],\nonumber\\
 \om{E}_{j;4} &\rightarrow \frac{8e^{-j \beta M}}{5 \pi^2 j^4 \beta^4} [105 + 105 j \beta M + 45 (j \beta M)^2 \nonumber\\
 & \hspace{.3\columnwidth} + 10 (j \beta M)^3 + (j \beta M)^4],
 \label{eq:Ej_M}
\end{align}
where the arrow $\rightarrow$ indicates that the thermodynamic limit $L\rightarrow \infty$ was taken.

The summation over $j$ can be performed in Eq.~\eqref{eq:Ej_M} in terms of the polylogarithm functions. However, at this stage we can already obtain the coefficients of the free energy density,
\begin{equation}
 \om{F}_I(\Omega, M) = \om{F}(0,M) - \overline{K}_2(M) \frac{(\nu L)^2}{2!} + \overline{K}_4(M) \frac{(\nu L)^4}{4!} + \dots,
\end{equation}
where $\overline{K}_{2n} \equiv \om{F}(0,M) K_{2n}$, by using the relation $\om{F}_I = \beta^{-1} \int d\beta\, \left.\om{E}_I\right|_{\beta \Omega}$:
\begin{align}
 \om{F}_j(0,M) &= -\frac{e^{-j \beta M}}{\pi^2 j^4 \beta^4}(1 + j \beta M),\nonumber\\
 \overline{K}_{2;j} &\rightarrow -\frac{2 e^{-j \beta M}}{3 \pi^2 j^4 \beta^4} [3 + 3 j \beta M + (j \beta M)^2],\nonumber\\
 \overline{K}_{4;j} &\rightarrow -\frac{8 e^{-j \beta M}}{5 \pi^2 j^4 \beta^4} [15 + 15 j\beta M + 6 (j \beta M)^2 \nonumber\\
 &\hspace{.4\columnwidth} + (j \beta M)^3].
\end{align}
Performing now the summation over $j$ via $\om{F}_I = \sum_{j = 1}^\infty \om{F}_{j;I}$, we arrive at 
\begin{align}
 \om{F}(0,M) &= -\frac{1}{\pi^2 \beta^4} [{\rm Li}_4(e^{-\beta M}) + \beta M {\rm Li}_3(e^{-\beta M})],\nonumber\\
 \overline{K}_2(M) &\rightarrow -\frac{2}{3\pi^2 \beta^4} [3{\rm Li}_4(e^{-\beta M}) + 3 \beta M {\rm Li}_3(e^{-\beta M}) \nonumber\\
 & + (\beta M)^2 {\rm Li}_2(e^{-\beta M})],\nonumber\\
 \overline{K}_4(M) &\rightarrow -\frac{8}{5\pi^2 \beta^4} [15 {\rm Li}_4(e^{-\beta M}) + 15 \beta M {\rm Li}_3(e^{-\beta M}) \nonumber\\
 & + 6 (\beta M)^2 {\rm Li}_2(e^{-\beta M}) + (\beta M)^3 {\rm Li}_1(e^{-\beta M})].
\end{align}
Evaluating the above for small values of the mass, we get the following asymptotic behaviour:
\begin{align}
 \om{F}(0,M) &\simeq \om{F}_0 \left(1 - \frac{15 \beta^2 M^2}{2\pi^2}\right),\nonumber\\
 \lim_{L \rightarrow \infty} K_2(M) &\simeq 2!\left(1 + \frac{5}{\pi^2} \beta^2 M^2\right),\nonumber\\
 \lim_{L \rightarrow \infty} K_4(M) &\simeq 4!\left(1 + \frac{6}{\pi^2} \beta^2 M^2\right).
\label{eq_K2_K4_mass_correction}
\end{align}

\section{Bounded Klein-Gordon field} \label{sec_bounded}

\subsection{Eigenspectrum of the system and observables} \label{sec_bounded_eigen}

\begin{figure*}
    \centering
    \includegraphics[width=1.9\columnwidth]{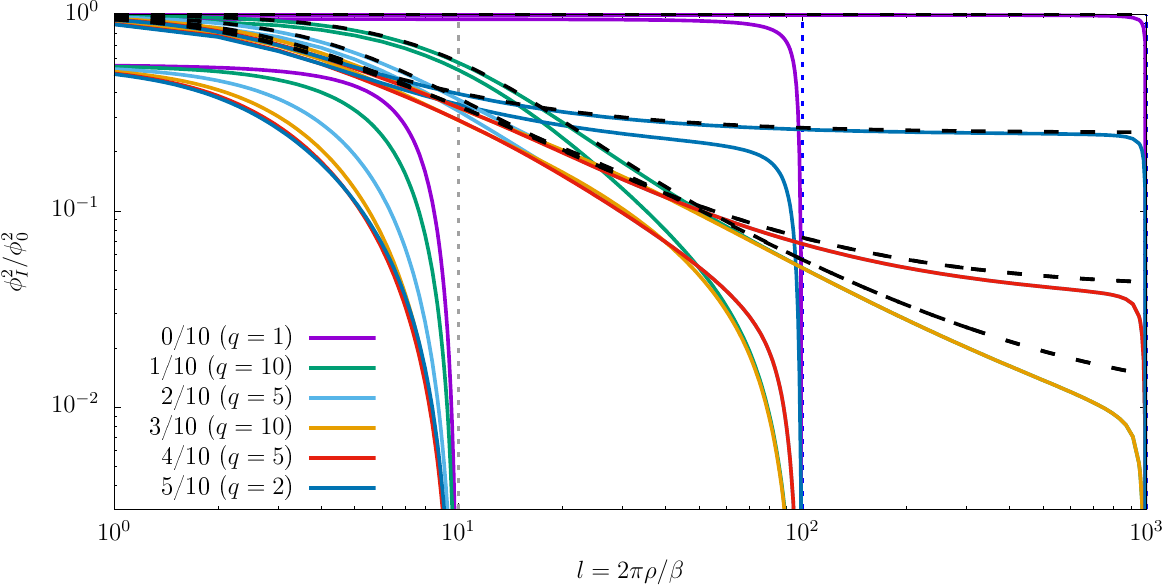} \\
    \includegraphics[width=1.9\columnwidth]{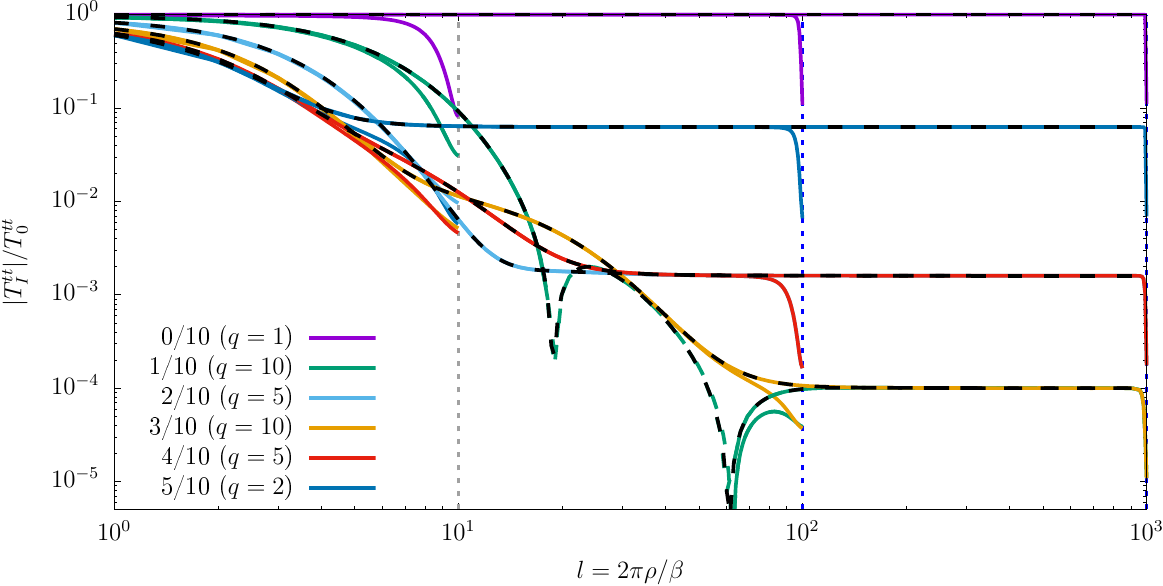}
    \caption{Same as Fig.~\ref{fig_Ttt_vs_rho} for the case when the system is enclosed within cylindrical boundaries of three different sizes, shown using vertical dotted lines, located such that $L = 2\pi R / \beta = 10$, $100$ and $1000$. The black dashed lines represent the results obtained in the unbounded case, shown in Fig.~\ref{fig_Ttt_vs_rho}.
    }
    \label{fig_b_Ttt_vs_rho}
\end{figure*}

We now enclose the system inside a cylindrical surface at a distance $R$ from the symmetry axis. Imposing Dirichlet boundary conditions on the eigenmodes $f_j$ leads to the quantization of the transverse momentum in Eq.~\eqref{eq_fj_rotation},
\begin{equation}
 J_{m_j}(q_j) = 0. \label{eq_b_quantization}
\end{equation}
The resulting normalized modes are \cite{Duffy:2002ss}
\begin{equation}
    f_{kmn} = \frac{e^{-i \omega_{mn} t + i k z + i m \varphi}}{2\pi R |J_{m+1}(q_{mn}/R)| \sqrt{\omega_{mn}}} J_m(q_{mn} \rho). 
    \label{eq_b_modes}
\end{equation}
The t.e.v.s of $\widehat{\Phi}^2$ and $\widehat{T}^{tt}$ can be expressed in the form shown in Eqs.~\eqref{eq_tev_phi2} and \eqref{eq_tev_SET_Ttt}, 
\begin{align}
 \phi^2 &= \overline{G}_{000}, & 
 T^{tt} &= \overline{G}_{200} + \frac{1}{12\rho^2} \overline{G}^{(2)}_{000},    
 \label{eq_b_tev}
\end{align}
where $\overline{G}^{(2)}_{abc} = \rho \frac{d}{d\rho} \rho \frac{d}{d\rho} \overline{G}_{abc}$ and the functions $\overline{G}_{abc}$ generalize the functions $G_{abc}$ in Eq.~\eqref{eq_kg:Gabc} to the bounded case considered here:
\begin{multline}
 \overline{G}_{abc} = \frac{1}{\pi^2 R^2} \sum_{m=-\infty}^\infty \sum_{n=1}^\infty \int_0^\infty \frac{dk / \omega_{mn}}{e^{\beta \tilde{\omega}_{mn}} - 1} \\ \times 
 \frac{J^2_m(q_{mn} \rho)}{J^2_{m+1}(q_{mn}R)} \omega_{mn}^a q_{mn}^b m^c.
 \label{eq_b_Gabc}
\end{multline}

\subsection{Scalar condensate, energy-momentum expectation values and fractalization} \label{sec_bounded_tevs}

Figure~\ref{fig_b_Ttt_vs_rho} shows the main features of $\phi_I^2$ (top panel) and $T_I^{tt}$ (lower panel) as functions of $l = 2 \pi \rho / \beta$ for several different radii $R$ chosen such that the quantity
\begin{equation}
 L = \frac{2\pi R}{\beta}
 \label{eq_L_def}
\end{equation}
takes the values $L = 10$, $100$, and $1000$. Only the case of rational (imaginary) rotation parameter is considered, with $\nu = \beta \Omega_I / 2\pi = p' / 10$ and $0 \le p' \le 5$, giving rise to all irreducible fractions $p/q \le 1/2$ with $q = 1$, $2$, $5$, and $10$. The dashed black lines represent the results obtained in the unbounded case, computed based on Eqs.~\eqref{eq_fractal_aux}, \eqref{eq_pq_phi2}, and \eqref{eq_pq_Ttt}.
As expected, the Dirichlet boundary conditions considered in this section affect the behaviour of the observables close to the boundary. Specifically, $\phi_I^2 = 0$ when $\rho = R$, since $\overline{G}^I_{000}(\rho = R)$ vanishes identically by virtue of the quantization condition $J_m(q_{mn}R) = 0$; while $T_I^{tt}$ is decreased by about a factor of $10$ compared to its bulk value. In both panels, the bounded and unbounded results stay in good agreement throughout most of the cylinder if $L$ is sufficiently large. In particular, $\phi_I^2$ exhibits a notably smaller value on the rotation axis when $L = 10$ compared to the unbounded case, while for the $L = 100$ and $1000$ cases, good agreement can be seen.

\begin{figure*}
\begin{tabular}{ccc}
\includegraphics[width=.33\linewidth]{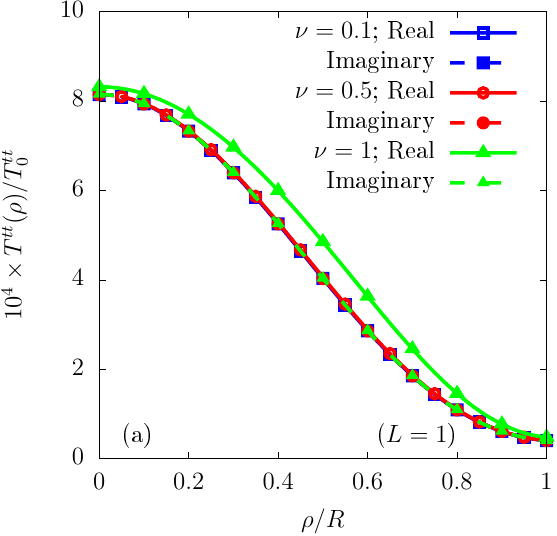} &
\includegraphics[width=.33\linewidth]{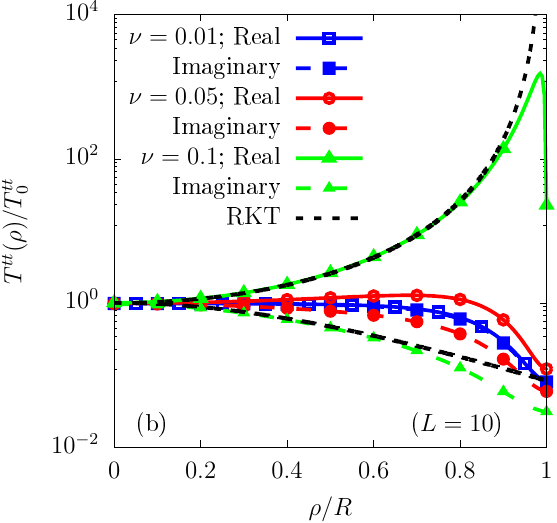} &
\includegraphics[width=.33\linewidth]{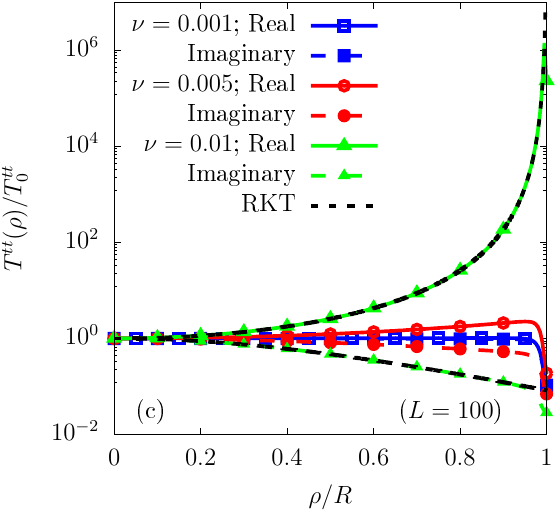} 
\end{tabular}
\caption{Profiles of $T^{tt}(\rho) / T^{tt}_0$, represented with respect to the dimensionless radial coordinate $\rho / R$ for a system enclosed within a cylindrical boundary located at $R = \beta L / 2\pi$, with $L = 1$ (a), $10$ (b), and $100$ (c). The rotation parameter satisfies $\nu L = 0.1$ (blue squares), $0.5$ (red circles), and $1$ (green triangles). Solid and dashed lines and symbols denote profiles corresponding to real ($T_{tt}$) and imaginary ($T^{tt}_I$) rotation, respectively. The black dotted lines shown in panels (b) and (c) correspond to the RKT results (see text).
\label{fig_b_comp_pr}
}
\end{figure*}

\begin{figure*}
\begin{tabular}{ccc}
\includegraphics[width=.33\linewidth]{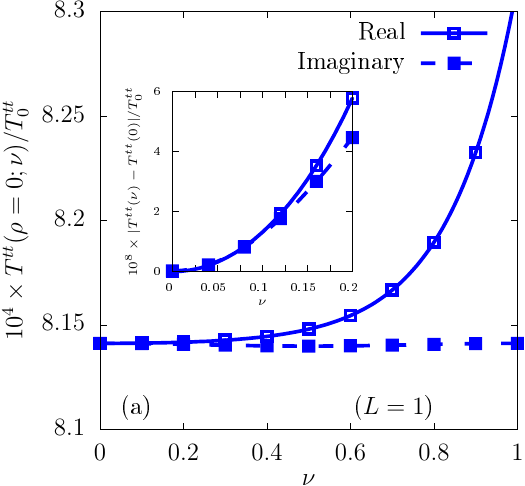} &
\includegraphics[width=.33\linewidth]{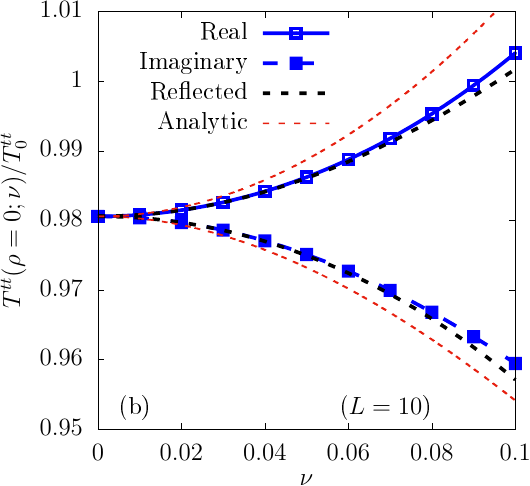} &
\includegraphics[width=.33\linewidth]{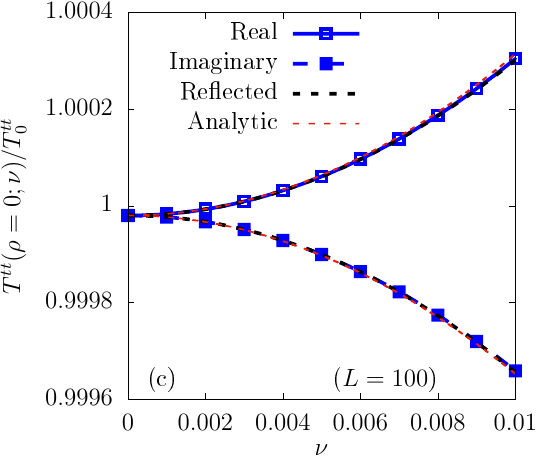} 
\end{tabular}
\caption{Value of $T^{tt}(\rho = 0; \nu)$ on the rotation axis with respect to $T^{tt}_0 = \pi^2 / 30\beta^4$ for cylindrical systems with the dimensionless radius $L = 2\pi R / \beta = 1$ (a), $10$ (b) and $100$ (c). The rotation parameter $\nu = \beta |\Omega| / 2\pi$ (shown on the $x$ axis) spans $0 \le \nu \le 1/L$. The blue solid and dashed lines with squares denote results for the case of real and imaginary rotation, respectively. The black dashed lines in panels (b) and (c) are obtained by reflecting the results corresponding to real and imaginary rotation with respect to the value $T^{tt}(\rho= 0; \nu = 0)$ obtained in the absence of rotation. The red dotted line represents the analytical expressions from the unbounded case, given in Eqs.~\eqref{eq_rho0_Ttt} and \eqref{eq_real_rho0_Ttt}, scaled by $T^{tt}(\rho=0;\nu = 0)$.
\label{fig_b_comp_c}
}
\end{figure*}

As mentioned in Sec.~\ref{sec_KG_tevs}, the boundary permits the study of a system undergoing real rigid rotation, as long as $\Omega R = \nu L \le 1$ and the light cylinder is excluded from the system. It is thus interesting to compare expectation values computed for imaginary and real rotation, $\nu_I$ and $\nu_R$. To keep the comparison meaningful, both $\nu_I$ and $\nu_R$ are restricted to be lower than or equal to $1 / L$. Fig.~\ref{fig_b_comp_pr} shows the radial profile $T^{tt}(\rho)$ for three cylinders, with L = $1$ (a), $10$ (b) and $100$ (c), in the case of slow ($\nu L = 0.1$, blue), medium ($\nu L = 0.5$, red) and fast ($\nu L = 1$, green) rotation. At small $L = 2 \pi R / \beta$, the boundary effects dominate over thermal ones and $T^{tt}$ decreases monotonically from the rotation axis towards the boundary. Furthermore, $T^{tt}(\rho = 0)$ is strongly suppressed (by four orders of magnitude at $L = 1$) compared to its value for a boson gas at rest, $T^{tt}_0 = \pi^2 / 30 \beta^4$. The effect of imaginary rotation is negligible, while in the case of real rotation, $T^{tt}(\rho)$ increases slightly at $\nu L = 1$. At $L \gtrsim 10$, the bulk of the system is dominated by thermal effects. Panels (b) and (c) also show the RKT result for~$T^{tt}$:
\begin{equation}
 T^{tt}_{\rm cl} = \frac{\pi^2 \gamma^4}{90\beta^4} (4\gamma^2 - 1), \quad 
 T^{tt}_{{\rm cl}; I} = \frac{\pi^2 \gamma_I^4}{90\beta^4} (4\gamma_I^2 - 1), 
\end{equation}
where $\gamma^2 = 1 / (1 - \nu_R^2 l^2)$ and $\gamma_I^2 = 1 / (1 + \nu_I^2 l^2)$. In the case when $L = 100$, the QFT results deviate from the RKT ones only in a small vicinity of the boundary. Such good agreement is also a consequence of the fact that at large $L$, $\nu$ is constrained to be small. As discussed in Sec.~\ref{sec_KG_fractal}, RKT is expected to agree with QFT for small values of $\nu$ and sufficiently far from the boundary (see also Fig.~\ref{fig_Ttt_vs_rho}.

Next, the value of $T^{tt}(\rho = 0; \nu)$ on the rotation axis for both real and imaginary rotation is shown in Fig.~\ref{fig_b_comp_c} for (a) $L = 1$, (b) $L = 10$ and (c) $L = 100$. As before, in the case $L = 1$, the value of $T^{tt}$ is suppressed by over three orders of magnitude. Here, the rotation parameter covers an entire period of the system undergoing imaginary rotation. In the case of real rotation, no such periodicity arises, contrary to the expectation based on the result in Eq.~\eqref{eq_real_rho0_Ttt} for the unbounded case. The inset in panel (a) shows the effect of rotation on $T^{tt}(\rho=0;\nu)$ for smaller values of $\nu$. It can be seen that for $\nu \lesssim 0.1$, the quantity $|T^{tt}(\nu) - T^{tt}(0)|$ has the same behavior for both real and imaginary rotation. 
At $L = 10$ and $100$, $T^{tt}(\rho = 0; \nu = 0)$ approaches $T^{tt}_0 = \pi^2 / 30\beta^4$. The maximum value of $\nu$ is greatly reduced. It can be seen that for such a small interval of $\nu$, the result corresponding to imaginary rotation is approximately equal to that obtained for real rotation, mirrored with respect to the value $T^{tt}(\rho=0;\nu = 0)$ obtained in the absence of rotation, as shown with black dotted lines. Furthermore, the red dotted lines indicate the analytical predictions in Eqs.~\eqref{eq_rho0_Ttt} and \eqref{eq_real_rho0_Ttt}, scaled by the value $T^{tt}(\rho = 0, \nu = 0) / T^{tt}_0$ on the rotation axis in the absence of rotation, corresponding to the given value of $L$. The agreement with the analytical predictions is better at larger values of $L$, which may also be due to the smaller range allowed for $\nu$.

\begin{figure*}
    \centering
    \includegraphics[width=1.9\columnwidth]{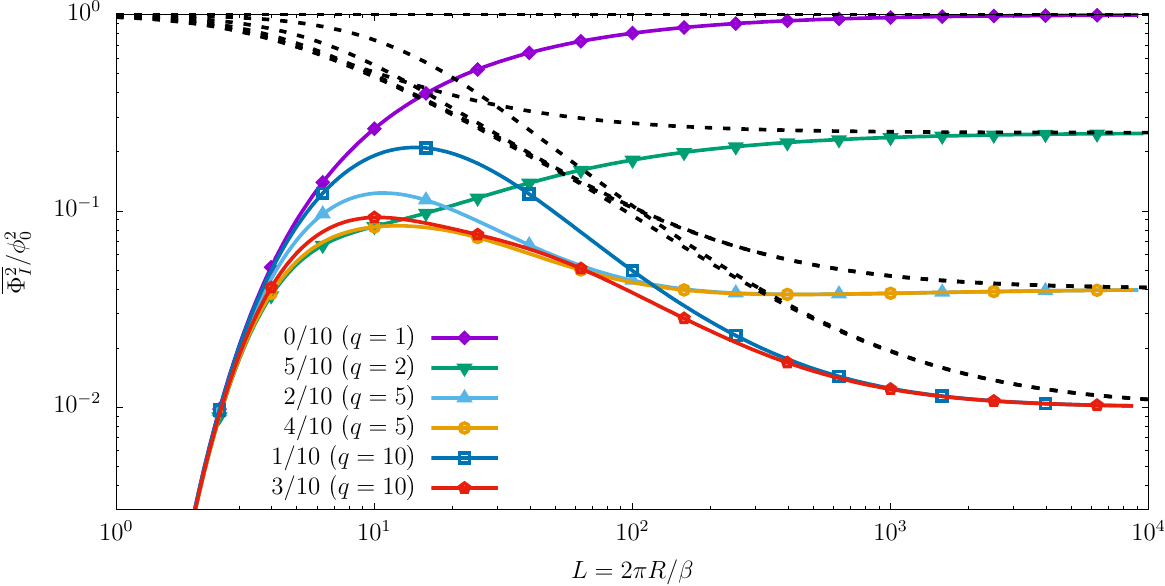} \\
    \includegraphics[width=1.9\columnwidth]{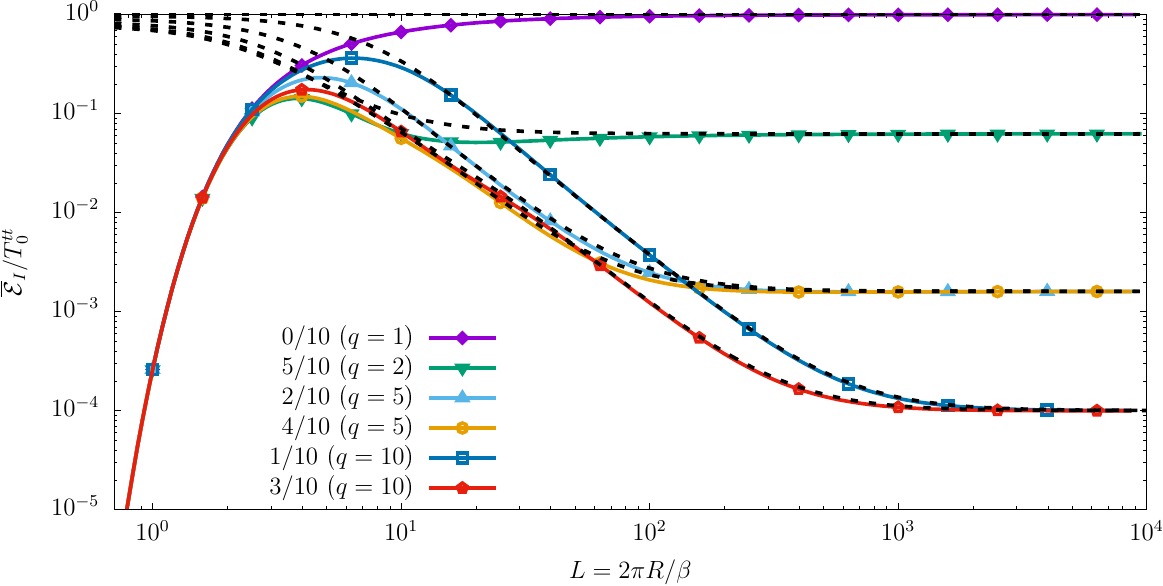}
    \caption{Integrated quantities $\overline{\Phi^2_I} / \phi_0^2$ (top) and $\om{E}_I / T^{tt}_0$ for the case when the system is enclosed within a cylindrical boundary located at $R = \beta L / 2\pi$, with $L$ shown on the horizontal axis. The colored lines with points show the results for rational (imaginary) rotation parameter $\nu = p' / q'$ shown in the legend ($q' = 10$ and $0 \le p' \le 5$), corresponding to the irreducible fraction $p / q$, with $q$ shown between the parentheses. The black dotted lines represent the results obtained in the unbounded case, shown in Fig.~\ref{fig_Ttt_vs_rho}.
    }
    \label{fig_b_vs_R}
\end{figure*}

\begin{figure}
\begin{center}
\includegraphics[width=.99\linewidth]{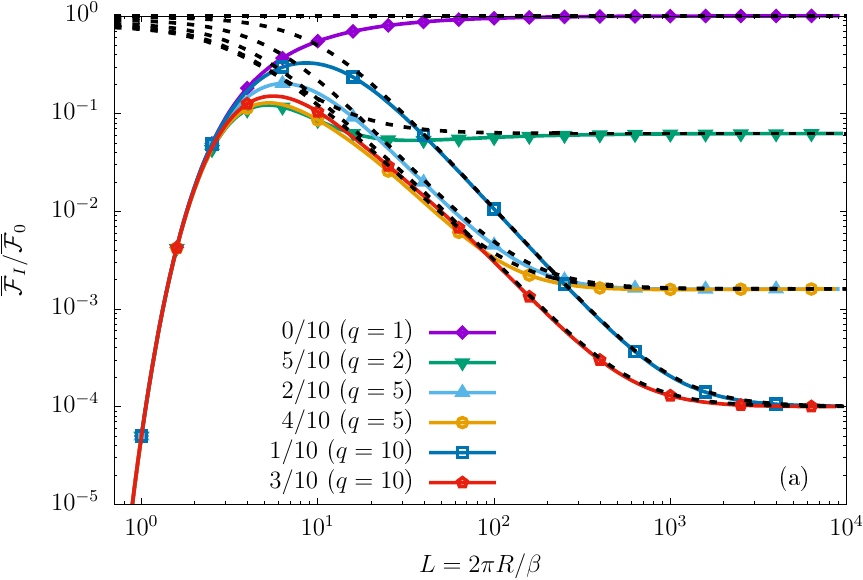} \\
\includegraphics[width=.99\linewidth]{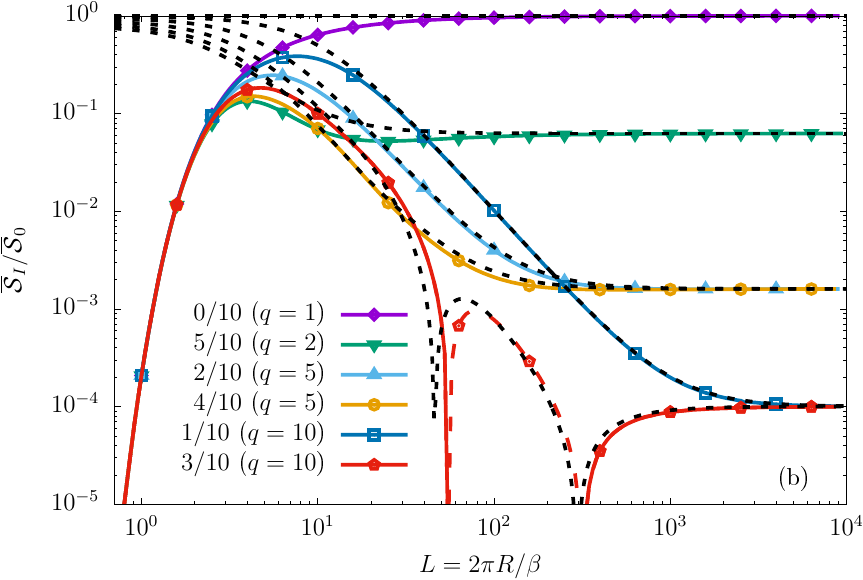} \\
\includegraphics[width=.99\linewidth]{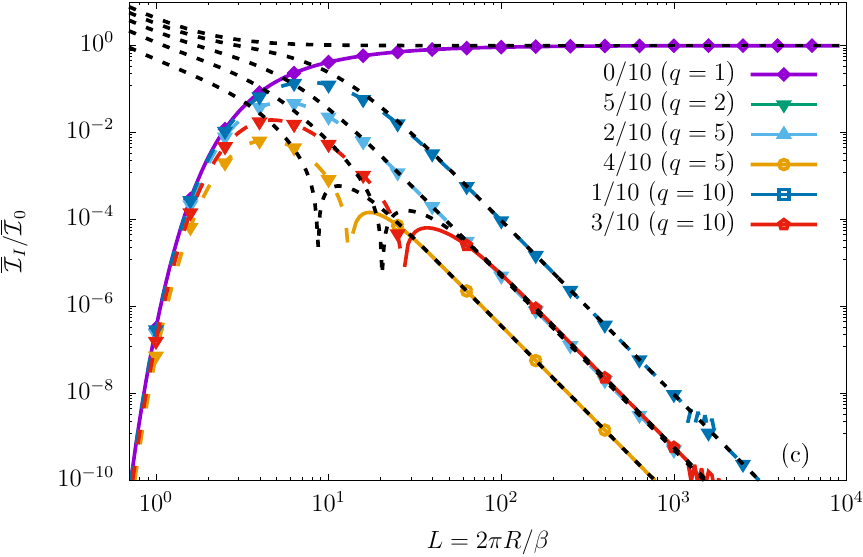} 
\end{center}
    \caption{Same as Fig.~\ref{fig_b_vs_R} for (a) $\om{F}_I / \om{F}_0$, (b) $\om{S}_I / \om{S}_0$ and (c) $\om{I}_I / \om{I}_0$, where the quantities $\om{F}_0$, $\om{S}_0$, and $\om{I}_0$ corresponding to a non-rotating massless boson gas are given in Eq.~\eqref{eq:b_refq}.
    }
    \label{fig_b_vs_R_small}
\end{figure}

We now consider the thermodynamic system as a whole and discuss the volume-averaged free energy $\om{F} = \mathcal{F} / V$, computed via the equivalent of Eq.~\eqref{eq_RKT_Fcl_cosh}:
\begin{subequations}\label{eq_b_all}
\begin{align}
 \om{F}(\Omega,R) &= \frac{1}{\pi^2 R^2 \beta} \sum_{m = -\infty}^\infty \sum_{n = 1}^\infty \int_0^\infty dk \ln(1 - e^{-\beta \tilde{\omega}_{mn}}) \nonumber\\
 &= -\frac{4}{L^2 \beta^2} \sum_{m = -\infty}^\infty \sum_{n = 1}^\infty \int_0^\infty \frac{dk\, k^2 / \omega_{mn}}{e^{\beta\tilde{\omega}_{mn}} - 1}.
 \label{eq_b_F}
\end{align}
Applying Eqs.~\eqref{eq_P_from_F} and \eqref{eq_thermo_rel} leads to the following expressions for the radial pressure $\mathcal{P}_R$, average entropy $\om{S}$, and average angular momentum $\om{M}$:
\begin{align}
 \mathcal{P}_R &= \frac{2}{L^2 \beta^2} \sum_{m = -\infty}^\infty \sum_{n = 1}^\infty \int_0^\infty \frac{dk}{e^{\beta\tilde{\omega}_{mn}} - 1} \left(\omega_{mn} - \frac{k^2}{\omega_{mn}}\right), \label{eq_b_PR}\\
 \om{S} &= \frac{4}{L^2 \beta} \sum_{m = -\infty}^\infty \sum_{n = 1}^\infty \int_0^\infty \frac{dk}{e^{\beta\tilde{\omega}_{mn}} - 1} \left(\tilde{\omega}_{mn} + \frac{k^2}{\omega_{mn}}\right), \label{eq_b_S}\\
 \om{M} &= \frac{4}{L^2\beta^2} \sum_{m = -\infty}^\infty \sum_{n = 1}^\infty \int_0^\infty \frac{dk\,m}{e^{\beta\tilde{\omega}_{mn}} - 1}, \label{eq_b_M}
\end{align}
while $\mathcal{P}_z = -\om{F}$. The average energy $\om{E}$ and scalar condensate $\overline{\Phi^2}$ can be obtained by taking the volume average of the expressions in Eq.~\eqref{eq_b_tev}:
\begin{align}
 \om{E} &= \frac{1}{\pi^2 R^2} \sum_{m = -\infty}^\infty \sum_{n = 1}^\infty \int_0^\infty \frac{dk\, \omega_{mn}}{e^{\beta\tilde{\omega}_{mn}} - 1}, \label{eq_b_E} \\
 \overline{\Phi^2} &= \frac{1}{\pi^2 R^2} \sum_{m = -\infty}^\infty \sum_{n = 1}^\infty \int_0^\infty \frac{dk / \omega_{mn}}{e^{\beta\tilde{\omega}_{mn}} - 1}.
 \label{eq_b_phi2}
\end{align}
\end{subequations}
In deriving the above expressions, we employed the integration formula
\begin{multline}
    \int_0^R d\rho\, \rho J_m^2(q\rho) = \frac{R^2}{2}[J^2_m(qR) + J^2_{m+1}(qR)] \\
    - \frac{m R}{q} J_m(qR) J_{m+1}(qR),
    \label{eq_b_intJ}
\end{multline}
together with the Dirichlet boundary conditions $J_m(q_{mn} R) = 0$.
Comparing Eqs.~\eqref{eq_b_E}, \eqref{eq_b_F} and \eqref{eq_b_PR}, it is easy to see that $\om{E} = \mathcal{P} / 3$ with $\mathcal{P} = \frac{2}{3} \mathcal{P}_R + \frac{1}{3} \mathcal{P}_z$ being the isotropic pressure. Moreover, the relations in Eq.~\eqref{eq_b_all} are compatible with the Euler relation \eqref{eq_Euler}.

The relations in Eq.~\eqref{eq_b_all} are valid for both real and imaginary rotation. In the latter case, $\om{M}$ becomes 
\begin{equation}
    \om{M}_I = \frac{8}{L^2 \beta^2}\sum_{m = 1}^\infty \sum_{n = 1}^\infty \int_0^\infty \frac{dk\, m\, e^{\beta \omega} \sin (\beta\Omega_I m)}{e^{2\beta\omega} - 2e^{\beta\omega} \cos(\beta\Omega_I m) + 1}.
\end{equation}
Thus, $\om{M}_I$ vanishes in the imaginary rotation case when $\nu = 1/2$, as was the case also in the unbounded system [see Eq.~\eqref{eq_KG_thermo_M2}].

The results for $\overline{\Phi_I^2}$ and $\om{E}_I$ are shown for the case of imaginary rotation in the top and bottom panels of Fig.~\ref{fig_b_vs_R}, while $\om{F}_I / \om{F}_0$, $\om{S}_I / \om{S}_0$, and $\om{I}_I / \om{I}_0$ are shown in Fig.~\ref{fig_b_vs_R_small}, where $\om{I} = \om{M} / \Omega$ ($\om{I}_I = \om{M}_I / \Omega_I$) and 
\begin{align}
 \om{F}_0 &= -\frac{\pi^2}{90\beta^4}, &
 \om{S}_0 &= \frac{2\pi^2}{45 \beta^4}, & 
 \om{I}_0 &= \frac{L^2}{180\beta^2}. \label{eq:b_refq}
\end{align}
As before, we set $\nu = p' / 10$ with $0 \le p' \le 5$, leading to irreducible fractions $p / q$ with $q \in \{1, 2, 5, 10\}$. The horizontal axis shows $L = 2\pi R / \beta$, where $R$ represents the radius of the bounding cylinder. The dotted black lines represent the same quantities computed for the unbounded system using Eqs.~\eqref{eq_KG_thermo_phi2} and \eqref{eq_KG_thermo_E} for the same value of $L$. For $L \lesssim 10$, the boundary effects lead to strong quenching of all five observables, such that they tend to $0$ as $L \rightarrow 0$ in the bounded case. This is contrary to the unbounded case, where the $L \rightarrow 0$ limit is finite. As already seen in Fig.~\ref{fig_b_Ttt_vs_rho}, with increasing $L$, the boundary effects become localized around a small vicinity of the boundary and the bounded and unbounded results approach each other. While for $\overline{\Phi^2_I}$, visible discrepancies remain even for $L \gtrsim 10^3$, in the case of $\om{E}_I$, $\om{F}_I$, $\om{S}_I$, and $\om{M}_I$, the results obtained in the bounded case start following the ones corresponding to the unbounded case already when $L \gtrsim 10$. As in the previous sections, the fractal structure reveals itself at large values of $L$.

\subsection{Slow rotation: moment of inertia and shape}
\label{sec_bounded_slow}

\begin{figure}
 \centering    
 \includegraphics[width=0.99\columnwidth,clip=true]{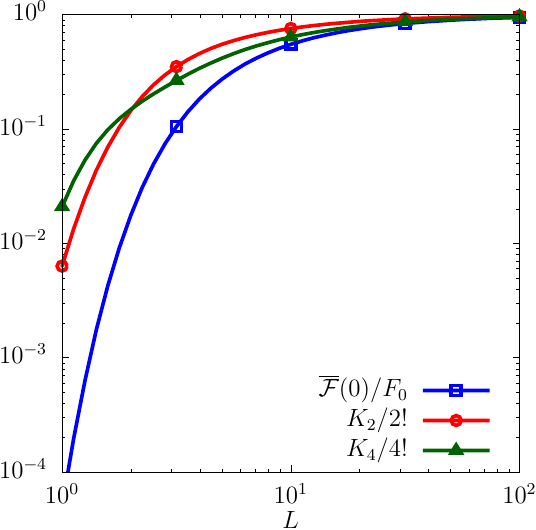}
 \caption{Ratios $\om{F}(0) / \om{F}_0$, $K_2 / 2!$ and $K_4 /4!$ computed at various values of the normalized transverse size of the system, $L = 2\pi R/\beta$ for the bounded system discussed in Sec.~\ref{sec_bounded}.}
 \label{fig_b_Kn}
\end{figure}

Finally, we discuss the expansion in Eq.~\eqref{eq_Kn_def} of the free energy density $\om{F}$ in the case of slow rotation. In particular, we focus on the free energy in the absence of rotation, $\om{F}(0)$, as well as the first two coefficients, $K_2$ (related to the moment of inertia) and $K_4$ (related to rotation-induced deformability), which we evaluate using:
\begin{subequations}\label{eq_b_Kn}
\begin{align}
 \om{F}(0) &= -\frac{4}{L^2 \beta^2} \sum_{m = -\infty}^\infty \sum_{n = 1}^\infty \int_0^\infty \frac{dk\, k^2 / \omega_{mn}}{e^{\beta\omega_{mn}} - 1}, \label{eq_b_F0}\\
 \overline{K}_2 &= -\frac{8 \pi^2}{\beta^3 L^4} \sum_{m = 1}^\infty \sum_{n = 1}^\infty \int_0^\infty \frac{m^2\,dk}{\sinh^2(\beta \omega_{m,n} / 2)}, \label{eq_b_K2}\\
 \overline{K}_4 &= -\frac{16 \pi^4}{L^6 \beta^3} \sum_{m = 1}^\infty \sum_{n = 1}^\infty \int_0^\infty \frac{dk\, m^4[2 + \cosh(\beta \omega_{mn})]}{\sinh^4(\beta \omega_{mn} / 2)}, \label{eq_b_K4}
\end{align}
\end{subequations}
where $\overline{K}_{2n} = K_{2n} \om{F}(0)$. The coefficients $\om{F}(0)$, $K_2$ and $K_4$ are studied as functions of the transverse size of the system $L$ with respect to their unbounded counterparts.

The ratios $\om{F}(0) / \om{F}_0$, $K_2 / 2!$ and $K_4/ 4!$ are represented in Fig.~\ref{fig_b_Kn}, where the denominators of these expressions are the classical expectations given in Eq.~\eqref{eq_RKT_K2n}. As we already noted in Subsect.~\ref{sec_slow_rotation_shape_0}, a strong quenching due to the boundary can be seen at small values of $L$, which is however less pronounced for the shape coefficient $K_4$ compared to the moment of inertia coefficient $K_2$ and the average free energy $\om{F}_0$. At $L = 100$, these coefficients already reach their expected asymptotic values $K_{2n} = (2n)!$, Eq.~\eq{eq_RKT_K2n}. Surprisingly, both $K_2(L)$ and $K_4(L)$ evaluated at finite $L$ are smaller than their asymptotic values, $K_{2n}(\infty) = (2n)!$. This property is in contrast to the behaviour seen for the $(1+1)$d ring [see Fig.~\ref{fig:ring_QFT_M0} and Eq.~\eqref{eq:ring_QFT_K2n_an}] and in the unbounded case [see Eqs.~\eqref{eq_KG_K2}--\eqref{eq_KG_K4}]. It remains an open question whether this feature remains robust when other boundary conditions (e.g., von Neumann) are considered.

In a cylinder of a finite radius $R$, a large-size $L \to \infty$ limit corresponds also to the high-temperature limit, $T \to \infty$ since $L = 2 \pi R/\beta \equiv 2 \pi R T$. Figure~\ref{fig_b_Kn} shows that the dimensionless moment of inertia $K_2$ approaches the asymptotic value $K_2 = 2$ from below, indicating that the moment of inertia should decrease as temperature decreases. This effect is related to the presence of an effective energy gap between the states with zero, $m=0$, and non-zero, $m \neq 0$, orbital momenta due to the finite size of the system. Therefore, at lower temperatures, the system mostly resides in the $m=0$ state and the rotational modes, which contribute to the moment of inertia~\eq{eq_b_K2}, are not excited. Since the latter modes do not participate in rotation at low $T$, the moment of inertia of the system decreases as the system gets colder. This effect should evidently also occur for $K_4$ and higher coefficients. Interestingly, the same qualitative behaviour for the moment of inertia $K_2$ is also observed in the first-principle simulations of gluon plasma in the high-temperature phase of Yang-Mills theory: as the temperature increases, the moment of inertia approaches the high-temperature value
$K_{2} = 2$, Eq.~\eq{eq_RKT_K2n}, from below~\cite{Braguta:2023yjn}.

\begin{figure}
 \centering    
 \begin{tabular}{c} 
  \includegraphics[width=0.99\columnwidth,clip=true]{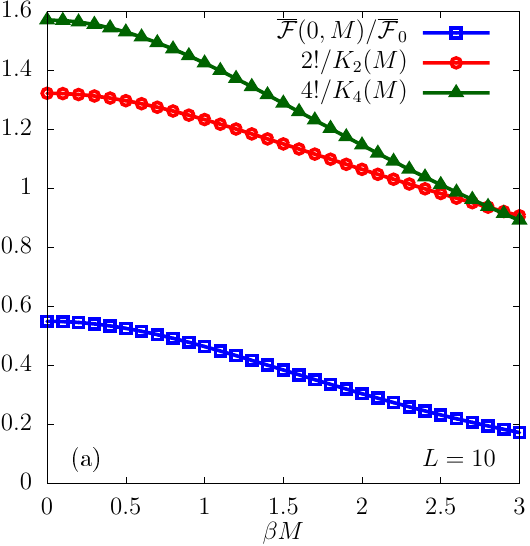} \\
  \includegraphics[width=0.99\columnwidth,clip=true]{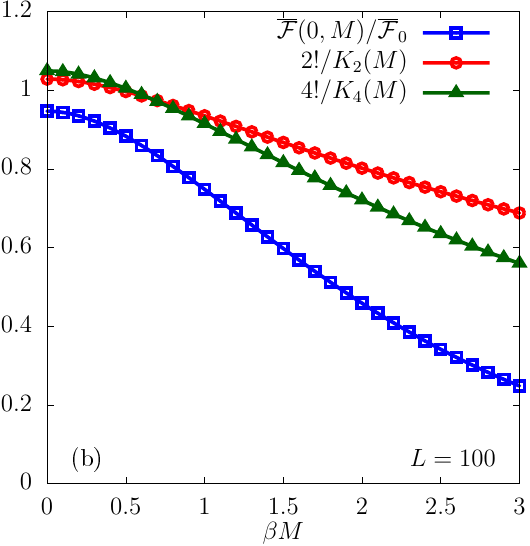}
 \end{tabular}
 \caption{Ratios $\om{F}(0) / \om{F}_0$, as well as the inverse ratios, $2!/K_2$ and $4!/K_4 $, computed at various values of the mass $M$, for (a) $L = 10$ and (b) $L = 100$.}
 \label{fig_b_Kn_M}
\end{figure}

The above discussion indicates that the restriction in size of the quantum system reduces the $K_{2n}$ coefficients. We now consider the behavior of the $K_{2n}$ coefficients for particles of finite mass $M$. As indicated in Sections~\ref{sec_ring_RKT}, \ref{sec_ring_massive}, \ref{sec:RKT:massive}, and \ref{sec_KG_massive}, the $K_{2n}$ coefficients increase as $M$ increases. In Fig.~\ref{fig_b_Kn_M}, we demonstrate that the same qualitative behavior is achieved for the spatially bounded system.

%%%%%%%%%%%%%

\section{Conclusions}

In the present work, we studied the thermodynamic properties of a massless scalar field subjected to rigid rotation and an inter-relation of the real rotation with its imaginary analogue. The latter concept -- a rigid rotation with an imaginary frequency~\cite{Chernodub:2020qah,Chen:2022smf} -- has a practical interest since rotating systems cannot be implemented in Euclidean path-integral formalism, suitable, for example, for numerical first-principle calculations on the lattice~\cite{Yamamoto:2013zwa,Braguta:2020biu,Braguta:2021jgn,Braguta:2023yjn,Chernodub:2022veq}. In this sense, rotation shares the deficiency suffered by finite-density systems, namely the sign problem~\cite{Yamamoto:2013zwa}, and needs to be implemented in Euclidean spacetime via its imaginary version supplemented with the subsequent analytical continuation to real angular frequencies~\cite{Braguta:2020biu,Braguta:2021jgn,Braguta:2023yjn}.

Using the $1+1$-dimensional toy model of a scalar field under rigid rotation on a ring, we explicitly demonstrated that the analytical no-go theorem~\cite{Chernodub:2022qlz}, which describes the impossibility of continuation of the imaginary-angular-frequency thermodynamics to the real angular frequencies is related to the development of the fractality of thermodynamics for the former. The result applies in the thermodynamic limit. Within this model, thermodynamic functions such as pressure and energy density can be expressed analytically via the Dedekind $\eta$ function~\eq{eq_Dedekind}. The latter function tends to the fractal, non-analytical Thomae function~\eq{eq_Thomae} as its argument approaches the real axis \cite{nechaev2017number}, which corresponds to the infinite-volume (thermodynamic) limit.

In the case of the (3+1)-dimensional Minkowski space, we first considered a classical description of scalar particles under rotation using relativistic kinetic theory. In the absence of boundaries, rigid rotation with a real rotation parameter leads to a violation of causality and subsequent divergence of all observables on the light cylinder. Imaginary rotation can be described by a real distribution function only as an average of clockwise and counterclockwise rotations, leading to a seemingly non-equilibrium state. As expected, observables such as the energy-momentum tensor $T^{\mu\nu}$ decrease as the distance to the rotation axis is increased, at a faster rate for faster rotation.

Under the quantum field theoretical treatment, rigid rotation with a real rotation parameter leads to the divergence of $T^{\mu\nu}$ at each point of the space-time, also inside the light cylinder. Under imaginary rotation, $T^{\mu\nu}$ evaluated on the rotation axis can be expressed via Bernoulli polynomials~\eq{eq_sumj}. Away from the rotation axis, we were able to demonstrate the fractalization of both the field fluctuations $\phi^2$ and of $T^{\mu\nu}$ in the case of imaginary rotation, in complete analogy to the $1+1$-dimensional toy model discussed above. In all cases, our results exhibit a periodicity with respect to the imaginary rotation parameter which is not present in the classical, kinetic analysis. For this reason, we found agreement with kinetic theory only in the limit of slow rotation and only in the vicinity of the rotation axis (before fractalization sets in).

For comparison with the case of real rotation, we took results obtained using a perturbative calculation for slow rotation (with respect to a stationary background) and found on the rotation axis an analytical result~\eq{eq_real_rho0} which can be related to the one obtained for imaginary rotation in an essentially non-analytical way~\eq{eq_no_analytical}. 
We conclude that even on the axis of rotation -- which appears to be static in the rotating system -- the non-analyticity is strong and unavoidable.  

We demonstrated the same analytic--non-analytical transition using numerical calculations for the thermal scalar field system enclosed in a cylinder, undergoing rotation with imaginary angular frequencies. As the radius of the cylinder grows, the pressure becomes a non-analytical function of temperature expressed, again, via the Thomae function (demonstrated in Figs.~\ref{fig_Ttt_vs_rho}--\ref{fig_b_vs_R_small}). In this limit, the boundary effects become less important and the results obtained in the unbounded case provide a good approximation for our observables inside the cylindrical boundary. For values of the rotation parameter respecting the causality constraints (i.e., when the light cylinder is outside the boundary), the fractalization features do not appear in the case of imaginary rotation. For sufficiently slow rotation and high temperature, our numerically-obtained results are compatible with the flip $\Omega_I^2 \rightarrow -\Omega^2$ from imaginary to real rotation, signaling the restoration of analytical continuation in this limit.

The exotic fractalization properties discussed above are related to a ninionic deformation~\eq{eq_ninionic} of statistical distributions at imaginary angular frequencies~\cite{Chernodub:2022qlz}. For this reason, we conclude that the results obtained in the infinite-volume system subjected to imaginary rotation cannot be analytically related to the properties of the physically rotating system (with a real angular frequency).  
However, we explicitly demonstrated both for the analytically-treatable case on the ring and numerically-accessible case of rotating cylinder that the imaginary rotation in a spatially bounded system in Euclidean space can be continued to the real-frequency domain in Minkowski spacetime provided that the causality is respected for the latter spacetime.

We have also shown that for real-frequency rotation, the dimensionless moment of inertia $K_2$, normalized per one degree of freedom, is equal to two, $K_2 = 2$, in the thermodynamic limit of large radius $R$ of the cylinder~\eq{eq_K2}. The quantity $K_2$ determines the correction to the free energy~\eq{eq_Kn_def}, %or, equivalently, to the pressure of a non-interacting gas of scalar bosons, 
\begin{align}
 %P(\Omega) = P(0) \biggl(1 + \frac{1}{2} K_2 R^2 \Omega^2  + \dots \biggr)\,,
 \mathcal{F}(\Omega) = \mathcal{F}(0)\biggl(1 + \frac{1}{2} K_2 R^2 \Omega^2  + \dots \biggr)\,,
\end{align} 
due to small nonzero angular frequency, $\Omega \to 0$. This result matches well the first-principle result of Ref.~\cite{Braguta:2023yjn} on the behavior of gluons in high-temperature limit of Yang-Mills theory. We have also shown that in the thermodynamic limit, the generic expansion of the pressure (free energy) in higher orders in angular frequency~\eq{eq_Kn_def} is characterized by exactly-calculable coefficients~$K_{2n}$ with $n=1,2, \dots$, given by identical expressions ~\eq{eq_ring_K2n} for the ring and \eq{eq_RKT_K2n} for the cylinder, which also includes the shape coefficient $K_4 = 4!$ responsible for an $\Omega^4$ correction to the pressure.

Below we summarize our main findings:
\begin{enumerate}
\item We demonstrated the no-go theorem~\cite{Chernodub:2022qlz} regarding the impossibility of continuation of the imaginary-angular-frequency thermodynamics to the real angular frequencies using an analytical $1+1$-dimensional toy model, revealing the development of the fractality of thermodynamics for the former.
\item Since fractalization does not show up in the classical, kinetic theory treatment of imaginary rotation, we conclude that this is a purely quantum effect.
\item We found similar fractalization on the unbounded Minkowski space for imaginary rotation, as well as in the thermodynamic (infinite volume) limit of the bounded system.
\item For the case of a causal boundary that excludes the light cylinder, we were able to restore the analytical continuation from imaginary to real rotation in the limit of slow rotation and large temperatures.
\item We attributed the exotic fractalization properties to a ninionic deformation~\eq{eq_ninionic} of statistical distributions at imaginary angular frequencies~\cite{Chernodub:2022qlz}. 
\item These statements are applicable both for massless and massive fields, with the fractal features become pronounced in the thermodynamic limit for massive particles. While an increasing mass of bosons enhances the moment of inertia $K_2$ and the shape coefficient $K_4$ of the rotating gas, this effect vanishes in the high-temperature limit. 

\end{enumerate}

The results obtained in this paper shed light on the implications of the effects of imaginary rotation obtained in the context of first-principle lattice simulations and constitute the basis for the analysis of more complicated systems, e.g., free fermions or the chiral phase transition in the effective QCD models such as the Nambu--Jona-Lasinio model or nonlinear sigma models. Extending the present analysis to the case of Dirac fermions is a logical avenue of future research \cite{Palermo:2021hlf}.

\begin{acknowledgments}
V.E.A.~gratefully acknowledges the support through a grant of the Ministry of Research, Innovation and Digitization, CNCS - UEFISCDI, project number PN-III-P1-1.1-TE-2021-1707, within PNCDI III.
The authors gratefully acknowledge support by the European Union - NextGenerationEU through grant No. 760079/23.05.2023, funded by the Romanian Ministry of Research, Innovation, and Digitalization through Romania’s National Recovery and Resilience Plan, call no. PNRR-III-C9-2022-I8, during the revision and extension of this manuscript.
\end{acknowledgments}

\bibliography{plasma}

\begin{thebibliography}{60}%
\makeatletter
\providecommand \@ifxundefined [1]{%
 \@ifx{#1\undefined}
}%
\providecommand \@ifnum [1]{%
 \ifnum #1\expandafter \@firstoftwo
 \else \expandafter \@secondoftwo
 \fi
}%
\providecommand \@ifx [1]{%
 \ifx #1\expandafter \@firstoftwo
 \else \expandafter \@secondoftwo
 \fi
}%
\providecommand \natexlab [1]{#1}%
\providecommand \enquote  [1]{``#1''}%
\providecommand \bibnamefont  [1]{#1}%
\providecommand \bibfnamefont [1]{#1}%
\providecommand \citenamefont [1]{#1}%
\providecommand \href@noop [0]{\@secondoftwo}%
\providecommand \href [0]{\begingroup \@sanitize@url \@href}%
\providecommand \@href[1]{\@@startlink{#1}\@@href}%
\providecommand \@@href[1]{\endgroup#1\@@endlink}%
\providecommand \@sanitize@url [0]{\catcode `\\12\catcode `\$12\catcode
  `\&12\catcode `\#12\catcode `\^12\catcode `\_12\catcode `\%12\relax}%
\providecommand \@@startlink[1]{}%
\providecommand \@@endlink[0]{}%
\providecommand \url  [0]{\begingroup\@sanitize@url \@url }%
\providecommand \@url [1]{\endgroup\@href {#1}{\urlprefix }}%
\providecommand \urlprefix  [0]{URL }%
\providecommand \Eprint [0]{\href }%
\providecommand \doibase [0]{http://dx.doi.org/}%
\providecommand \selectlanguage [0]{\@gobble}%
\providecommand \bibinfo  [0]{\@secondoftwo}%
\providecommand \bibfield  [0]{\@secondoftwo}%
\providecommand \translation [1]{[#1]}%
\providecommand \BibitemOpen [0]{}%
\providecommand \bibitemStop [0]{}%
\providecommand \bibitemNoStop [0]{.\EOS\space}%
\providecommand \EOS [0]{\spacefactor3000\relax}%
\providecommand \BibitemShut  [1]{\csname bibitem#1\endcsname}%
\let\auto@bib@innerbib\@empty
%</preamble>
\bibitem [{\citenamefont {Beams}(1968)}]{Beams1968}%
  \BibitemOpen
  \bibfield  {author} {\bibinfo {author} {\bibfnamefont {J.~W.}\ \bibnamefont
  {Beams}},\ }\bibfield  {title} {\enquote {\bibinfo {title} {Potentials on
  rotor surfaces},}\ }\href {\doibase 10.1103/physrevlett.21.1093} {\bibfield
  {journal} {\bibinfo  {journal} {Physical Review Letters}\ }\textbf {\bibinfo
  {volume} {21}},\ \bibinfo {pages} {1093--1096} (\bibinfo {year}
  {1968})}\BibitemShut {NoStop}%
\bibitem [{\citenamefont {Barnett}(1915)}]{Barnett1915}%
  \BibitemOpen
  \bibfield  {author} {\bibinfo {author} {\bibfnamefont {S.~J.}\ \bibnamefont
  {Barnett}},\ }\bibfield  {title} {\enquote {\bibinfo {title} {Magnetization
  by rotation},}\ }\href {\doibase 10.1103/physrev.6.239} {\bibfield  {journal}
  {\bibinfo  {journal} {Physical Review}\ }\textbf {\bibinfo {volume} {6}},\
  \bibinfo {pages} {239--270} (\bibinfo {year} {1915})}\BibitemShut {NoStop}%
\bibitem [{\citenamefont {Einstein}\ and\ \citenamefont
  {De~Haas}(1915)}]{Einstein1915}%
  \BibitemOpen
  \bibfield  {author} {\bibinfo {author} {\bibfnamefont {A}~\bibnamefont
  {Einstein}}\ and\ \bibinfo {author} {\bibfnamefont {WJ}~\bibnamefont
  {De~Haas}},\ }\bibfield  {title} {\enquote {\bibinfo {title} {Experimental
  proof of the existence of amp{\`e}re’s molecular currents},}\ }in\
  \href@noop {} {\emph {\bibinfo {booktitle} {Proc. KNAW}}},\ Vol.\ \bibinfo
  {volume} {181}\ (\bibinfo {year} {1915})\ p.\ \bibinfo {pages}
  {696}\BibitemShut {NoStop}%
\bibitem [{\citenamefont {Arabgol}\ and\ \citenamefont
  {Sleator}(2019)}]{Arabgol2019}%
  \BibitemOpen
  \bibfield  {author} {\bibinfo {author} {\bibfnamefont {Mohsen}\ \bibnamefont
  {Arabgol}}\ and\ \bibinfo {author} {\bibfnamefont {Tycho}\ \bibnamefont
  {Sleator}},\ }\bibfield  {title} {\enquote {\bibinfo {title} {Observation of
  the nuclear barnett effect},}\ }\href {\doibase
  10.1103/physrevlett.122.177202} {\bibfield  {journal} {\bibinfo  {journal}
  {Physical Review Letters}\ }\textbf {\bibinfo {volume} {122}} (\bibinfo
  {year} {2019}),\ 10.1103/physrevlett.122.177202}\BibitemShut {NoStop}%
\bibitem [{\citenamefont {Adamczyk}\ \emph {et~al.}(2017)\citenamefont
  {Adamczyk} \emph {et~al.}}]{STAR:2017ckg}%
  \BibitemOpen
  \bibfield  {author} {\bibinfo {author} {\bibfnamefont {L.}~\bibnamefont
  {Adamczyk}} \emph {et~al.} (\bibinfo {collaboration} {STAR}),\ }\bibfield
  {title} {\enquote {\bibinfo {title} {{Global $\Lambda$ hyperon polarization
  in nuclear collisions: evidence for the most vortical fluid}},}\ }\href
  {\doibase 10.1038/nature23004} {\bibfield  {journal} {\bibinfo  {journal}
  {Nature}\ }\textbf {\bibinfo {volume} {548}},\ \bibinfo {pages} {62--65}
  (\bibinfo {year} {2017})},\ \Eprint {http://arxiv.org/abs/1701.06657}
  {arXiv:1701.06657 [nucl-ex]} \BibitemShut {NoStop}%
\bibitem [{\citenamefont {Deng}\ and\ \citenamefont
  {Huang}(2016)}]{Deng:2016gyh}%
  \BibitemOpen
  \bibfield  {author} {\bibinfo {author} {\bibfnamefont {Wei-Tian}\
  \bibnamefont {Deng}}\ and\ \bibinfo {author} {\bibfnamefont {Xu-Guang}\
  \bibnamefont {Huang}},\ }\bibfield  {title} {\enquote {\bibinfo {title}
  {{Vorticity in Heavy-Ion Collisions}},}\ }\href {\doibase
  10.1103/PhysRevC.93.064907} {\bibfield  {journal} {\bibinfo  {journal} {Phys.
  Rev. C}\ }\textbf {\bibinfo {volume} {93}},\ \bibinfo {pages} {064907}
  (\bibinfo {year} {2016})},\ \Eprint {http://arxiv.org/abs/1603.06117}
  {arXiv:1603.06117 [nucl-th]} \BibitemShut {NoStop}%
\bibitem [{\citenamefont {Jiang}\ \emph {et~al.}(2016)\citenamefont {Jiang},
  \citenamefont {Lin},\ and\ \citenamefont {Liao}}]{Jiang:2016woz}%
  \BibitemOpen
  \bibfield  {author} {\bibinfo {author} {\bibfnamefont {Yin}\ \bibnamefont
  {Jiang}}, \bibinfo {author} {\bibfnamefont {Zi-Wei}\ \bibnamefont {Lin}}, \
  and\ \bibinfo {author} {\bibfnamefont {Jinfeng}\ \bibnamefont {Liao}},\
  }\bibfield  {title} {\enquote {\bibinfo {title} {{Rotating quark-gluon plasma
  in relativistic heavy ion collisions}},}\ }\href {\doibase
  10.1103/PhysRevC.94.044910} {\bibfield  {journal} {\bibinfo  {journal} {Phys.
  Rev. C}\ }\textbf {\bibinfo {volume} {94}},\ \bibinfo {pages} {044910}
  (\bibinfo {year} {2016})},\ \bibinfo {note} {[Erratum: Phys.Rev.C 95, 049904
  (2017)]},\ \Eprint {http://arxiv.org/abs/1602.06580} {arXiv:1602.06580
  [hep-ph]} \BibitemShut {NoStop}%
\bibitem [{\citenamefont {Becattini}\ and\ \citenamefont
  {Lisa}(2020)}]{Becattini:2020ngo}%
  \BibitemOpen
  \bibfield  {author} {\bibinfo {author} {\bibfnamefont {Francesco}\
  \bibnamefont {Becattini}}\ and\ \bibinfo {author} {\bibfnamefont
  {Michael~A.}\ \bibnamefont {Lisa}},\ }\bibfield  {title} {\enquote {\bibinfo
  {title} {{Polarization and Vorticity in the Quark\textendash{}Gluon
  Plasma}},}\ }\href {\doibase 10.1146/annurev-nucl-021920-095245} {\bibfield
  {journal} {\bibinfo  {journal} {Ann. Rev. Nucl. Part. Sci.}\ }\textbf
  {\bibinfo {volume} {70}},\ \bibinfo {pages} {395--423} (\bibinfo {year}
  {2020})},\ \Eprint {http://arxiv.org/abs/2003.03640} {arXiv:2003.03640
  [nucl-ex]} \BibitemShut {NoStop}%
\bibitem [{\citenamefont {Huang}\ \emph {et~al.}(2020)\citenamefont {Huang},
  \citenamefont {Liao}, \citenamefont {Wang},\ and\ \citenamefont
  {Xia}}]{Huang:2020dtn}%
  \BibitemOpen
  \bibfield  {author} {\bibinfo {author} {\bibfnamefont {Xu-Guang}\
  \bibnamefont {Huang}}, \bibinfo {author} {\bibfnamefont {Jinfeng}\
  \bibnamefont {Liao}}, \bibinfo {author} {\bibfnamefont {Qun}\ \bibnamefont
  {Wang}}, \ and\ \bibinfo {author} {\bibfnamefont {Xiao-Liang}\ \bibnamefont
  {Xia}},\ }\bibfield  {title} {\enquote {\bibinfo {title} {{Vorticity and Spin
  Polarization in Heavy Ion Collisions: Transport Models}},}\ }\href {\doibase
  10.1007/978-3-030-71427-7\_9} {\  (\bibinfo {year} {2020}),\
  10.1007/978-3-030-71427-7\_9},\ \Eprint {http://arxiv.org/abs/2010.08937}
  {arXiv:2010.08937 [nucl-th]} \BibitemShut {NoStop}%
\bibitem [{\citenamefont {Chen}\ \emph {et~al.}(2016)\citenamefont {Chen},
  \citenamefont {Fukushima}, \citenamefont {Huang},\ and\ \citenamefont
  {Mameda}}]{Chen:2015hfc}%
  \BibitemOpen
  \bibfield  {author} {\bibinfo {author} {\bibfnamefont {Hao-Lei}\ \bibnamefont
  {Chen}}, \bibinfo {author} {\bibfnamefont {Kenji}\ \bibnamefont {Fukushima}},
  \bibinfo {author} {\bibfnamefont {Xu-Guang}\ \bibnamefont {Huang}}, \ and\
  \bibinfo {author} {\bibfnamefont {Kazuya}\ \bibnamefont {Mameda}},\
  }\bibfield  {title} {\enquote {\bibinfo {title} {{Analogy between rotation
  and density for Dirac fermions in a magnetic field}},}\ }\href {\doibase
  10.1103/PhysRevD.93.104052} {\bibfield  {journal} {\bibinfo  {journal} {Phys.
  Rev. D}\ }\textbf {\bibinfo {volume} {93}},\ \bibinfo {pages} {104052}
  (\bibinfo {year} {2016})},\ \Eprint {http://arxiv.org/abs/1512.08974}
  {arXiv:1512.08974 [hep-ph]} \BibitemShut {NoStop}%
\bibitem [{\citenamefont {Jiang}\ and\ \citenamefont
  {Liao}(2016)}]{Jiang:2016wvv}%
  \BibitemOpen
  \bibfield  {author} {\bibinfo {author} {\bibfnamefont {Yin}\ \bibnamefont
  {Jiang}}\ and\ \bibinfo {author} {\bibfnamefont {Jinfeng}\ \bibnamefont
  {Liao}},\ }\bibfield  {title} {\enquote {\bibinfo {title} {{Pairing Phase
  Transitions of Matter under Rotation}},}\ }\href {\doibase
  10.1103/PhysRevLett.117.192302} {\bibfield  {journal} {\bibinfo  {journal}
  {Phys. Rev. Lett.}\ }\textbf {\bibinfo {volume} {117}},\ \bibinfo {pages}
  {192302} (\bibinfo {year} {2016})},\ \Eprint
  {http://arxiv.org/abs/1606.03808} {arXiv:1606.03808 [hep-ph]} \BibitemShut
  {NoStop}%
\bibitem [{\citenamefont {Chernodub}\ and\ \citenamefont
  {Gongyo}(2017{\natexlab{a}})}]{Chernodub:2016kxh}%
  \BibitemOpen
  \bibfield  {author} {\bibinfo {author} {\bibfnamefont {M.~N.}\ \bibnamefont
  {Chernodub}}\ and\ \bibinfo {author} {\bibfnamefont {Shinya}\ \bibnamefont
  {Gongyo}},\ }\bibfield  {title} {\enquote {\bibinfo {title} {{Interacting
  fermions in rotation: chiral symmetry restoration, moment of inertia and
  thermodynamics}},}\ }\href {\doibase 10.1007/JHEP01(2017)136} {\bibfield
  {journal} {\bibinfo  {journal} {JHEP}\ }\textbf {\bibinfo {volume} {01}},\
  \bibinfo {pages} {136} (\bibinfo {year} {2017}{\natexlab{a}})},\ \Eprint
  {http://arxiv.org/abs/1611.02598} {arXiv:1611.02598 [hep-th]} \BibitemShut
  {NoStop}%
\bibitem [{\citenamefont {Chernodub}\ and\ \citenamefont
  {Gongyo}(2017{\natexlab{b}})}]{Chernodub:2017ref}%
  \BibitemOpen
  \bibfield  {author} {\bibinfo {author} {\bibfnamefont {M.~N.}\ \bibnamefont
  {Chernodub}}\ and\ \bibinfo {author} {\bibfnamefont {Shinya}\ \bibnamefont
  {Gongyo}},\ }\bibfield  {title} {\enquote {\bibinfo {title} {{Effects of
  rotation and boundaries on chiral symmetry breaking of relativistic
  fermions}},}\ }\href {\doibase 10.1103/PhysRevD.95.096006} {\bibfield
  {journal} {\bibinfo  {journal} {Phys. Rev. D}\ }\textbf {\bibinfo {volume}
  {95}},\ \bibinfo {pages} {096006} (\bibinfo {year} {2017}{\natexlab{b}})},\
  \Eprint {http://arxiv.org/abs/1702.08266} {arXiv:1702.08266 [hep-th]}
  \BibitemShut {NoStop}%
\bibitem [{\citenamefont {Wang}\ \emph {et~al.}(2019)\citenamefont {Wang},
  \citenamefont {Wei}, \citenamefont {Li},\ and\ \citenamefont
  {Huang}}]{Wang:2018sur}%
  \BibitemOpen
  \bibfield  {author} {\bibinfo {author} {\bibfnamefont {Xinyang}\ \bibnamefont
  {Wang}}, \bibinfo {author} {\bibfnamefont {Minghua}\ \bibnamefont {Wei}},
  \bibinfo {author} {\bibfnamefont {Zhibin}\ \bibnamefont {Li}}, \ and\
  \bibinfo {author} {\bibfnamefont {Mei}\ \bibnamefont {Huang}},\ }\bibfield
  {title} {\enquote {\bibinfo {title} {{Quark matter under rotation in the NJL
  model with vector interaction}},}\ }\href {\doibase
  10.1103/PhysRevD.99.016018} {\bibfield  {journal} {\bibinfo  {journal} {Phys.
  Rev. D}\ }\textbf {\bibinfo {volume} {99}},\ \bibinfo {pages} {016018}
  (\bibinfo {year} {2019})},\ \Eprint {http://arxiv.org/abs/1808.01931}
  {arXiv:1808.01931 [hep-ph]} \BibitemShut {NoStop}%
\bibitem [{\citenamefont {Zhang}\ \emph {et~al.}(2020)\citenamefont {Zhang},
  \citenamefont {Shi}, \citenamefont {He}, \citenamefont {Luo},\ and\
  \citenamefont {Zong}}]{Zhang:2020hha}%
  \BibitemOpen
  \bibfield  {author} {\bibinfo {author} {\bibfnamefont {Zheng}\ \bibnamefont
  {Zhang}}, \bibinfo {author} {\bibfnamefont {Chao}\ \bibnamefont {Shi}},
  \bibinfo {author} {\bibfnamefont {Xiao-Tao}\ \bibnamefont {He}}, \bibinfo
  {author} {\bibfnamefont {Xiaofeng}\ \bibnamefont {Luo}}, \ and\ \bibinfo
  {author} {\bibfnamefont {Hong-Shi}\ \bibnamefont {Zong}},\ }\bibfield
  {title} {\enquote {\bibinfo {title} {{Chiral phase transition inside a
  rotating cylinder within the Nambu\textendash{}Jona-Lasinio model}},}\ }\href
  {\doibase 10.1103/PhysRevD.102.114023} {\bibfield  {journal} {\bibinfo
  {journal} {Phys. Rev. D}\ }\textbf {\bibinfo {volume} {102}},\ \bibinfo
  {pages} {114023} (\bibinfo {year} {2020})},\ \Eprint
  {http://arxiv.org/abs/2012.01017} {arXiv:2012.01017 [hep-ph]} \BibitemShut
  {NoStop}%
\bibitem [{\citenamefont {Sadooghi}\ \emph {et~al.}(2021)\citenamefont
  {Sadooghi}, \citenamefont {Tabatabaee~Mehr},\ and\ \citenamefont
  {Taghinavaz}}]{Sadooghi:2021upd}%
  \BibitemOpen
  \bibfield  {author} {\bibinfo {author} {\bibfnamefont {N.}~\bibnamefont
  {Sadooghi}}, \bibinfo {author} {\bibfnamefont {S.~M.~A.}\ \bibnamefont
  {Tabatabaee~Mehr}}, \ and\ \bibinfo {author} {\bibfnamefont {F.}~\bibnamefont
  {Taghinavaz}},\ }\bibfield  {title} {\enquote {\bibinfo {title} {{Inverse
  magnetorotational catalysis and the phase diagram of a rotating hot and
  magnetized quark matter}},}\ }\href {\doibase 10.1103/PhysRevD.104.116022}
  {\bibfield  {journal} {\bibinfo  {journal} {Phys. Rev. D}\ }\textbf {\bibinfo
  {volume} {104}},\ \bibinfo {pages} {116022} (\bibinfo {year} {2021})},\
  \Eprint {http://arxiv.org/abs/2108.12760} {arXiv:2108.12760 [hep-ph]}
  \BibitemShut {NoStop}%
\bibitem [{\citenamefont {Braguta}\ \emph {et~al.}(2020)\citenamefont
  {Braguta}, \citenamefont {Kotov}, \citenamefont {Kuznedelev},\ and\
  \citenamefont {Roenko}}]{Braguta:2020biu}%
  \BibitemOpen
  \bibfield  {author} {\bibinfo {author} {\bibfnamefont {V.~V.}\ \bibnamefont
  {Braguta}}, \bibinfo {author} {\bibfnamefont {A.~Yu.}\ \bibnamefont {Kotov}},
  \bibinfo {author} {\bibfnamefont {D.~D.}\ \bibnamefont {Kuznedelev}}, \ and\
  \bibinfo {author} {\bibfnamefont {A.~A.}\ \bibnamefont {Roenko}},\ }\bibfield
   {title} {\enquote {\bibinfo {title} {{Study of the Confinement/Deconfinement
  Phase Transition in Rotating Lattice SU(3) Gluodynamics}},}\ }\href {\doibase
  10.31857/S1234567820130029} {\bibfield  {journal} {\bibinfo  {journal} {Pisma
  Zh. Eksp. Teor. Fiz.}\ }\textbf {\bibinfo {volume} {112}},\ \bibinfo {pages}
  {9--16} (\bibinfo {year} {2020})}\BibitemShut {NoStop}%
\bibitem [{\citenamefont {Chen}\ \emph {et~al.}(2021)\citenamefont {Chen},
  \citenamefont {Zhang}, \citenamefont {Li}, \citenamefont {Hou},\ and\
  \citenamefont {Huang}}]{Chen:2020ath}%
  \BibitemOpen
  \bibfield  {author} {\bibinfo {author} {\bibfnamefont {Xun}\ \bibnamefont
  {Chen}}, \bibinfo {author} {\bibfnamefont {Lin}\ \bibnamefont {Zhang}},
  \bibinfo {author} {\bibfnamefont {Danning}\ \bibnamefont {Li}}, \bibinfo
  {author} {\bibfnamefont {Defu}\ \bibnamefont {Hou}}, \ and\ \bibinfo {author}
  {\bibfnamefont {Mei}\ \bibnamefont {Huang}},\ }\bibfield  {title} {\enquote
  {\bibinfo {title} {{Gluodynamics and deconfinement phase transition under
  rotation from holography}},}\ }\href {\doibase 10.1007/JHEP07(2021)132}
  {\bibfield  {journal} {\bibinfo  {journal} {JHEP}\ }\textbf {\bibinfo
  {volume} {07}},\ \bibinfo {pages} {132} (\bibinfo {year} {2021})},\ \Eprint
  {http://arxiv.org/abs/2010.14478} {arXiv:2010.14478 [hep-ph]} \BibitemShut
  {NoStop}%
\bibitem [{\citenamefont {Chernodub}(2021{\natexlab{a}})}]{Chernodub:2020qah}%
  \BibitemOpen
  \bibfield  {author} {\bibinfo {author} {\bibfnamefont {M.~N.}\ \bibnamefont
  {Chernodub}},\ }\bibfield  {title} {\enquote {\bibinfo {title}
  {{Inhomogeneous confining-deconfining phases in rotating plasmas}},}\ }\href
  {\doibase 10.1103/PhysRevD.103.054027} {\bibfield  {journal} {\bibinfo
  {journal} {Phys. Rev. D}\ }\textbf {\bibinfo {volume} {103}},\ \bibinfo
  {pages} {054027} (\bibinfo {year} {2021}{\natexlab{a}})},\ \Eprint
  {http://arxiv.org/abs/2012.04924} {arXiv:2012.04924 [hep-ph]} \BibitemShut
  {NoStop}%
\bibitem [{\citenamefont {Fujimoto}\ \emph {et~al.}(2021)\citenamefont
  {Fujimoto}, \citenamefont {Fukushima},\ and\ \citenamefont
  {Hidaka}}]{Fujimoto:2021xix}%
  \BibitemOpen
  \bibfield  {author} {\bibinfo {author} {\bibfnamefont {Yuki}\ \bibnamefont
  {Fujimoto}}, \bibinfo {author} {\bibfnamefont {Kenji}\ \bibnamefont
  {Fukushima}}, \ and\ \bibinfo {author} {\bibfnamefont {Yoshimasa}\
  \bibnamefont {Hidaka}},\ }\bibfield  {title} {\enquote {\bibinfo {title}
  {{Deconfining Phase Boundary of Rapidly Rotating Hot and Dense Matter and
  Analysis of Moment of Inertia}},}\ }\href {\doibase
  10.1016/j.physletb.2021.136184} {\bibfield  {journal} {\bibinfo  {journal}
  {Phys. Lett. B}\ }\textbf {\bibinfo {volume} {816}},\ \bibinfo {pages}
  {136184} (\bibinfo {year} {2021})},\ \Eprint
  {http://arxiv.org/abs/2101.09173} {arXiv:2101.09173 [hep-ph]} \BibitemShut
  {NoStop}%
\bibitem [{\citenamefont {Braguta}\ \emph {et~al.}(2021)\citenamefont
  {Braguta}, \citenamefont {Kotov}, \citenamefont {Kuznedelev},\ and\
  \citenamefont {Roenko}}]{Braguta:2021jgn}%
  \BibitemOpen
  \bibfield  {author} {\bibinfo {author} {\bibfnamefont {V.~V.}\ \bibnamefont
  {Braguta}}, \bibinfo {author} {\bibfnamefont {A.~Yu.}\ \bibnamefont {Kotov}},
  \bibinfo {author} {\bibfnamefont {D.~D.}\ \bibnamefont {Kuznedelev}}, \ and\
  \bibinfo {author} {\bibfnamefont {A.~A.}\ \bibnamefont {Roenko}},\ }\bibfield
   {title} {\enquote {\bibinfo {title} {{Influence of relativistic rotation on
  the confinement-deconfinement transition in gluodynamics}},}\ }\href
  {\doibase 10.1103/PhysRevD.103.094515} {\bibfield  {journal} {\bibinfo
  {journal} {Phys. Rev. D}\ }\textbf {\bibinfo {volume} {103}},\ \bibinfo
  {pages} {094515} (\bibinfo {year} {2021})},\ \Eprint
  {http://arxiv.org/abs/2102.05084} {arXiv:2102.05084 [hep-lat]} \BibitemShut
  {NoStop}%
\bibitem [{\citenamefont {Golubtsova}\ \emph {et~al.}(2022)\citenamefont
  {Golubtsova}, \citenamefont {Gourgoulhon},\ and\ \citenamefont
  {Usova}}]{Golubtsova:2021agl}%
  \BibitemOpen
  \bibfield  {author} {\bibinfo {author} {\bibfnamefont {Anastasia~A.}\
  \bibnamefont {Golubtsova}}, \bibinfo {author} {\bibfnamefont {Eric}\
  \bibnamefont {Gourgoulhon}}, \ and\ \bibinfo {author} {\bibfnamefont
  {Marina~K.}\ \bibnamefont {Usova}},\ }\bibfield  {title} {\enquote {\bibinfo
  {title} {{Heavy quarks in rotating plasma via holography}},}\ }\href
  {\doibase 10.1016/j.nuclphysb.2022.115786} {\bibfield  {journal} {\bibinfo
  {journal} {Nucl. Phys. B}\ }\textbf {\bibinfo {volume} {979}},\ \bibinfo
  {pages} {115786} (\bibinfo {year} {2022})},\ \Eprint
  {http://arxiv.org/abs/2107.11672} {arXiv:2107.11672 [hep-th]} \BibitemShut
  {NoStop}%
\bibitem [{\citenamefont {Chen}\ \emph {et~al.}(2022)\citenamefont {Chen},
  \citenamefont {Fukushima},\ and\ \citenamefont {Shimada}}]{Chen:2022smf}%
  \BibitemOpen
  \bibfield  {author} {\bibinfo {author} {\bibfnamefont {Shi}\ \bibnamefont
  {Chen}}, \bibinfo {author} {\bibfnamefont {Kenji}\ \bibnamefont {Fukushima}},
  \ and\ \bibinfo {author} {\bibfnamefont {Yusuke}\ \bibnamefont {Shimada}},\
  }\bibfield  {title} {\enquote {\bibinfo {title} {{Perturbative Confinement in
  Thermal Yang-Mills Theories Induced by Imaginary Angular Velocity}},}\ }\href
  {\doibase 10.1103/PhysRevLett.129.242002} {\bibfield  {journal} {\bibinfo
  {journal} {Phys. Rev. Lett.}\ }\textbf {\bibinfo {volume} {129}},\ \bibinfo
  {pages} {242002} (\bibinfo {year} {2022})},\ \Eprint
  {http://arxiv.org/abs/2207.12665} {arXiv:2207.12665 [hep-ph]} \BibitemShut
  {NoStop}%
\bibitem [{\citenamefont {Golubtsova}\ and\ \citenamefont
  {Tsegelnik}(2022)}]{Golubtsova2022}%
  \BibitemOpen
  \bibfield  {author} {\bibinfo {author} {\bibfnamefont {Anastasia~A.}\
  \bibnamefont {Golubtsova}}\ and\ \bibinfo {author} {\bibfnamefont
  {Nikita~S.}\ \bibnamefont {Tsegelnik}},\ }\bibfield  {title} {\enquote
  {\bibinfo {title} {{Probing the holographic model of $\mathcal{N} =4$ SYM
  rotating quark-gluon plasma}},}\ }\href@noop {} {\  (\bibinfo {year}
  {2022})},\ \Eprint {http://arxiv.org/abs/2211.11722} {arXiv:2211.11722
  [hep-th]} \BibitemShut {NoStop}%
\bibitem [{\citenamefont {Zhao}\ \emph {et~al.}(2022)\citenamefont {Zhao},
  \citenamefont {He}, \citenamefont {Hou}, \citenamefont {Li},\ and\
  \citenamefont {Li}}]{Zhao:2022uxc}%
  \BibitemOpen
  \bibfield  {author} {\bibinfo {author} {\bibfnamefont {Yan-Qing}\
  \bibnamefont {Zhao}}, \bibinfo {author} {\bibfnamefont {Song}\ \bibnamefont
  {He}}, \bibinfo {author} {\bibfnamefont {Defu}\ \bibnamefont {Hou}}, \bibinfo
  {author} {\bibfnamefont {Li}~\bibnamefont {Li}}, \ and\ \bibinfo {author}
  {\bibfnamefont {Zhibin}\ \bibnamefont {Li}},\ }\bibfield  {title} {\enquote
  {\bibinfo {title} {{Phase diagram of holographic thermal dense QCD matter
  with rotation}},}\ }\href@noop {} {\  (\bibinfo {year} {2022})},\ \Eprint
  {http://arxiv.org/abs/2212.14662} {arXiv:2212.14662 [hep-ph]} \BibitemShut
  {NoStop}%
\bibitem [{\citenamefont {Chernodub}\ \emph {et~al.}(2022)\citenamefont
  {Chernodub}, \citenamefont {Goy},\ and\ \citenamefont
  {Molochkov}}]{Chernodub:2022veq}%
  \BibitemOpen
  \bibfield  {author} {\bibinfo {author} {\bibfnamefont {M.~N.}\ \bibnamefont
  {Chernodub}}, \bibinfo {author} {\bibfnamefont {V.~A.}\ \bibnamefont {Goy}},
  \ and\ \bibinfo {author} {\bibfnamefont {A.~V.}\ \bibnamefont {Molochkov}},\
  }\bibfield  {title} {\enquote {\bibinfo {title} {{Inhomogeneity of rotating
  gluon plasma and Tolman-Ehrenfest law in imaginary time: lattice results for
  fast imaginary rotation}},}\ }\href@noop {} {\  (\bibinfo {year} {2022})},\
  \Eprint {http://arxiv.org/abs/2209.15534} {arXiv:2209.15534 [hep-lat]}
  \BibitemShut {NoStop}%
\bibitem [{\citenamefont {Braguta}\ \emph {et~al.}(2023)\citenamefont
  {Braguta}, \citenamefont {Chernodub}, \citenamefont {Roenko},\ and\
  \citenamefont {Sychev}}]{Braguta:2023yjn}%
  \BibitemOpen
  \bibfield  {author} {\bibinfo {author} {\bibfnamefont {Victor~V.}\
  \bibnamefont {Braguta}}, \bibinfo {author} {\bibfnamefont {Maxim~N.}\
  \bibnamefont {Chernodub}}, \bibinfo {author} {\bibfnamefont {Artem~A.}\
  \bibnamefont {Roenko}}, \ and\ \bibinfo {author} {\bibfnamefont {Dmitrii~A.}\
  \bibnamefont {Sychev}},\ }\bibfield  {title} {\enquote {\bibinfo {title}
  {{Negative moment of inertia and rotational instability of gluon plasma}},}\
  }\href@noop {} {\  (\bibinfo {year} {2023})},\ \Eprint
  {http://arxiv.org/abs/2303.03147} {arXiv:2303.03147 [hep-lat]} \BibitemShut
  {NoStop}%
\bibitem [{\citenamefont {Ambru\c{s}}\ and\ \citenamefont
  {Winstanley}(2014)}]{Ambrus:2014uqa}%
  \BibitemOpen
  \bibfield  {author} {\bibinfo {author} {\bibfnamefont {Victor~E.}\
  \bibnamefont {Ambru\c{s}}}\ and\ \bibinfo {author} {\bibfnamefont
  {Elizabeth}\ \bibnamefont {Winstanley}},\ }\bibfield  {title} {\enquote
  {\bibinfo {title} {{Rotating quantum states}},}\ }\href {\doibase
  10.1016/j.physletb.2014.05.031} {\bibfield  {journal} {\bibinfo  {journal}
  {Phys. Lett. B}\ }\textbf {\bibinfo {volume} {734}},\ \bibinfo {pages}
  {296--301} (\bibinfo {year} {2014})},\ \Eprint
  {http://arxiv.org/abs/1401.6388} {arXiv:1401.6388 [hep-th]} \BibitemShut
  {NoStop}%
\bibitem [{\citenamefont {Ambrus}\ and\ \citenamefont
  {Winstanley}(2016)}]{Ambrus:2015lfr}%
  \BibitemOpen
  \bibfield  {author} {\bibinfo {author} {\bibfnamefont {Victor~E.}\
  \bibnamefont {Ambrus}}\ and\ \bibinfo {author} {\bibfnamefont {Elizabeth}\
  \bibnamefont {Winstanley}},\ }\bibfield  {title} {\enquote {\bibinfo {title}
  {{Rotating fermions inside a cylindrical boundary}},}\ }\href {\doibase
  10.1103/PhysRevD.93.104014} {\bibfield  {journal} {\bibinfo  {journal} {Phys.
  Rev. D}\ }\textbf {\bibinfo {volume} {93}},\ \bibinfo {pages} {104014}
  (\bibinfo {year} {2016})},\ \Eprint {http://arxiv.org/abs/1512.05239}
  {arXiv:1512.05239 [hep-th]} \BibitemShut {NoStop}%
\bibitem [{\citenamefont {Yamamoto}\ and\ \citenamefont
  {Hirono}(2013)}]{Yamamoto:2013zwa}%
  \BibitemOpen
  \bibfield  {author} {\bibinfo {author} {\bibfnamefont {Arata}\ \bibnamefont
  {Yamamoto}}\ and\ \bibinfo {author} {\bibfnamefont {Yuji}\ \bibnamefont
  {Hirono}},\ }\bibfield  {title} {\enquote {\bibinfo {title} {{Lattice QCD in
  rotating frames}},}\ }\href {\doibase 10.1103/PhysRevLett.111.081601}
  {\bibfield  {journal} {\bibinfo  {journal} {Phys. Rev. Lett.}\ }\textbf
  {\bibinfo {volume} {111}},\ \bibinfo {pages} {081601} (\bibinfo {year}
  {2013})},\ \Eprint {http://arxiv.org/abs/1303.6292} {arXiv:1303.6292
  [hep-lat]} \BibitemShut {NoStop}%
\bibitem [{\citenamefont {Chernodub}(2022{\natexlab{a}})}]{Chernodub:2022wsw}%
  \BibitemOpen
  \bibfield  {author} {\bibinfo {author} {\bibfnamefont {M.~N.}\ \bibnamefont
  {Chernodub}},\ }\bibfield  {title} {\enquote {\bibinfo {title} {{Instantons
  in rotating finite-temperature Yang-Mills gas}},}\ }\href@noop {} {\
  (\bibinfo {year} {2022}{\natexlab{a}})},\ \Eprint
  {http://arxiv.org/abs/2208.04808} {arXiv:2208.04808 [hep-th]} \BibitemShut
  {NoStop}%
\bibitem [{\citenamefont {Chernodub}(2022{\natexlab{b}})}]{Chernodub:2022qlz}%
  \BibitemOpen
  \bibfield  {author} {\bibinfo {author} {\bibfnamefont {M.~N.}\ \bibnamefont
  {Chernodub}},\ }\bibfield  {title} {\enquote {\bibinfo {title} {{Fractal
  thermodynamics and ninionic statistics of coherent rotational states:
  realization via imaginary angular rotation in imaginary time formalism}},}\
  }\href@noop {} {\  (\bibinfo {year} {2022}{\natexlab{b}})},\ \Eprint
  {http://arxiv.org/abs/2210.05651} {arXiv:2210.05651 [quant-ph]} \BibitemShut
  {NoStop}%
\bibitem [{\citenamefont {Duffy}\ and\ \citenamefont
  {Ottewill}(2003)}]{Duffy:2002ss}%
  \BibitemOpen
  \bibfield  {author} {\bibinfo {author} {\bibfnamefont {Gavin}\ \bibnamefont
  {Duffy}}\ and\ \bibinfo {author} {\bibfnamefont {Adrian~C.}\ \bibnamefont
  {Ottewill}},\ }\bibfield  {title} {\enquote {\bibinfo {title} {{The Rotating
  quantum thermal distribution}},}\ }\href {\doibase
  10.1103/PhysRevD.67.044002} {\bibfield  {journal} {\bibinfo  {journal} {Phys.
  Rev. D}\ }\textbf {\bibinfo {volume} {67}},\ \bibinfo {pages} {044002}
  (\bibinfo {year} {2003})},\ \Eprint {http://arxiv.org/abs/hep-th/0211096}
  {arXiv:hep-th/0211096} \BibitemShut {NoStop}%
\bibitem [{\citenamefont {Romeo}\ and\ \citenamefont
  {Saharian}(2002)}]{Romeo:2000wt}%
  \BibitemOpen
  \bibfield  {author} {\bibinfo {author} {\bibfnamefont {August}\ \bibnamefont
  {Romeo}}\ and\ \bibinfo {author} {\bibfnamefont {Aram~A.}\ \bibnamefont
  {Saharian}},\ }\bibfield  {title} {\enquote {\bibinfo {title} {{Casimir
  effect for scalar fields under Robin boundary conditions on plates}},}\
  }\href {\doibase 10.1088/0305-4470/35/5/312} {\bibfield  {journal} {\bibinfo
  {journal} {J. Phys. A}\ }\textbf {\bibinfo {volume} {35}},\ \bibinfo {pages}
  {1297--1320} (\bibinfo {year} {2002})},\ \Eprint
  {http://arxiv.org/abs/hep-th/0007242} {arXiv:hep-th/0007242} \BibitemShut
  {NoStop}%
\bibitem [{\citenamefont {Romeo}\ and\ \citenamefont
  {Saharian}(2001)}]{Romeo:2001dd}%
  \BibitemOpen
  \bibfield  {author} {\bibinfo {author} {\bibfnamefont {August}\ \bibnamefont
  {Romeo}}\ and\ \bibinfo {author} {\bibfnamefont {Aram~A.}\ \bibnamefont
  {Saharian}},\ }\bibfield  {title} {\enquote {\bibinfo {title} {{Vacuum
  densities and zero point energy for fields obeying Robin conditions on
  cylindrical surfaces}},}\ }\href {\doibase 10.1103/PhysRevD.63.105019}
  {\bibfield  {journal} {\bibinfo  {journal} {Phys. Rev. D}\ }\textbf {\bibinfo
  {volume} {63}},\ \bibinfo {pages} {105019} (\bibinfo {year} {2001})},\
  \Eprint {http://arxiv.org/abs/hep-th/0101155} {arXiv:hep-th/0101155}
  \BibitemShut {NoStop}%
\bibitem [{\citenamefont {Ambru\c{s}}(2019)}]{Ambrus:2019vkx}%
  \BibitemOpen
  \bibfield  {author} {\bibinfo {author} {\bibfnamefont {Victor~E.}\
  \bibnamefont {Ambru\c{s}}},\ }\bibfield  {title} {\enquote {\bibinfo {title}
  {{Rigidly-rotating quantum thermal states in bounded systems}},}\ }in\ \href
  {\doibase 10.1142/9789811258251_0211} {\emph {\bibinfo {booktitle} {{15th
  Marcel Grossmann Meeting on Recent Developments in Theoretical and
  Experimental General Relativity, Astrophysics, and Relativistic Field
  Theories}}}}\ (\bibinfo {year} {2019})\ \Eprint
  {http://arxiv.org/abs/1904.01123} {arXiv:1904.01123 [hep-th]} \BibitemShut
  {NoStop}%
\bibitem [{\citenamefont {Kapusta}\ and\ \citenamefont
  {Gale}(2006)}]{Kapusta2006}%
  \BibitemOpen
  \bibfield  {author} {\bibinfo {author} {\bibfnamefont {Joseph~I}\
  \bibnamefont {Kapusta}}\ and\ \bibinfo {author} {\bibfnamefont {Charles}\
  \bibnamefont {Gale}},\ }\href@noop {} {\emph {\bibinfo {title}
  {Finite-temperature field theory: {Principles} and applications}}}\ (\bibinfo
   {publisher} {Cambridge University Press},\ \bibinfo {year}
  {2006})\BibitemShut {NoStop}%
\bibitem [{\citenamefont {Rothe}(2012)}]{Rothe2012}%
  \BibitemOpen
  \bibfield  {author} {\bibinfo {author} {\bibfnamefont {Heinz~J}\ \bibnamefont
  {Rothe}},\ }\href@noop {} {\emph {\bibinfo {title} {Lattice gauge theories:
  an introduction}}}\ (\bibinfo  {publisher} {World Scientific Publishing
  Company},\ \bibinfo {year} {2012})\BibitemShut {NoStop}%
\bibitem [{\citenamefont {de~Forcrand}(2009)}]{deForcrand:2009zkb}%
  \BibitemOpen
  \bibfield  {author} {\bibinfo {author} {\bibfnamefont {Philippe}\
  \bibnamefont {de~Forcrand}},\ }\bibfield  {title} {\enquote {\bibinfo {title}
  {{Simulating QCD at finite density}},}\ }\href {\doibase 10.22323/1.091.0010}
  {\bibfield  {journal} {\bibinfo  {journal} {PoS}\ }\textbf {\bibinfo {volume}
  {LAT2009}},\ \bibinfo {pages} {010} (\bibinfo {year} {2009})},\ \Eprint
  {http://arxiv.org/abs/1005.0539} {arXiv:1005.0539 [hep-lat]} \BibitemShut
  {NoStop}%
\bibitem [{\citenamefont {Chernodub}(2021{\natexlab{b}})}]{Chernodub:2021fpm}%
  \BibitemOpen
  \bibfield  {author} {\bibinfo {author} {\bibfnamefont {M.~N.}\ \bibnamefont
  {Chernodub}},\ }\bibfield  {title} {\enquote {\bibinfo {title} {{Rotational
  diode: Clockwise/counterclockwise asymmetry in conducting and mechanical
  properties of rotating (semi)conductors}},}\ }\href {\doibase
  10.3390/sym13091569} {\bibfield  {journal} {\bibinfo  {journal} {Symmetry}\
  }\textbf {\bibinfo {volume} {13}},\ \bibinfo {pages} {1569} (\bibinfo {year}
  {2021}{\natexlab{b}})},\ \Eprint {http://arxiv.org/abs/2104.05032}
  {arXiv:2104.05032 [cond-mat.mes-hall]} \BibitemShut {NoStop}%
\bibitem [{\citenamefont {Flack}\ \emph {et~al.}(2023)\citenamefont {Flack},
  \citenamefont {Gorsky},\ and\ \citenamefont {Nechaev}}]{Flack:2023uop}%
  \BibitemOpen
  \bibfield  {author} {\bibinfo {author} {\bibfnamefont {Ana}\ \bibnamefont
  {Flack}}, \bibinfo {author} {\bibfnamefont {Alexander}\ \bibnamefont
  {Gorsky}}, \ and\ \bibinfo {author} {\bibfnamefont {Sergei}\ \bibnamefont
  {Nechaev}},\ }\bibfield  {title} {\enquote {\bibinfo {title} {{Generalized
  Devil's staircase and RG flows}},}\ }\href@noop {} {\  (\bibinfo {year}
  {2023})},\ \Eprint {http://arxiv.org/abs/2304.07640} {arXiv:2304.07640
  [cond-mat.stat-mech]} \BibitemShut {NoStop}%
\bibitem [{\citenamefont {Devaney}(1999)}]{Devaney1999}%
  \BibitemOpen
  \bibfield  {author} {\bibinfo {author} {\bibfnamefont {Robert~L.}\
  \bibnamefont {Devaney}},\ }\bibfield  {title} {\enquote {\bibinfo {title}
  {The mandelbrot set, the farey tree, and the fibonacci sequence},}\ }\href
  {\doibase 10.1080/00029890.1999.12005046} {\bibfield  {journal} {\bibinfo
  {journal} {The American Mathematical Monthly}\ }\textbf {\bibinfo {volume}
  {106}},\ \bibinfo {pages} {289--302} (\bibinfo {year} {1999})}\BibitemShut
  {NoStop}%
\bibitem [{\citenamefont {Nechaev}\ and\ \citenamefont
  {Polovnikov}(2017)}]{nechaev2017number}%
  \BibitemOpen
  \bibfield  {author} {\bibinfo {author} {\bibfnamefont {S}~\bibnamefont
  {Nechaev}}\ and\ \bibinfo {author} {\bibfnamefont {K}~\bibnamefont
  {Polovnikov}},\ }\bibfield  {title} {\enquote {\bibinfo {title}
  {Number-theoretic aspects of 1d localization:" popcorn function" with
  {Lifshitz} tails and its continuous approximation by the dedekind
  $\backslash\eta$},}\ }\href@noop {} {\bibfield  {journal} {\bibinfo
  {journal} {arXiv preprint arXiv:1702.06757}\ } (\bibinfo {year}
  {2017})}\BibitemShut {NoStop}%
\bibitem [{\citenamefont {Landau}\ and\ \citenamefont {Lifshitz}(1996)}]{LL5}%
  \BibitemOpen
  \bibfield  {author} {\bibinfo {author} {\bibfnamefont {L~D}\ \bibnamefont
  {Landau}}\ and\ \bibinfo {author} {\bibfnamefont {E~M}\ \bibnamefont
  {Lifshitz}},\ }\href@noop {} {\emph {\bibinfo {title} {Statistical
  Physics}}},\ \bibinfo {edition} {3rd}\ ed.\ (\bibinfo  {publisher}
  {Butterworth-Heinemann},\ \bibinfo {address} {Oxford, England},\ \bibinfo
  {year} {1996})\BibitemShut {NoStop}%
\bibitem [{\citenamefont {Milton}(2001)}]{Milton2001}%
  \BibitemOpen
  \bibfield  {author} {\bibinfo {author} {\bibfnamefont {K~A}\ \bibnamefont
  {Milton}},\ }\href {\doibase 10.1142/4505} {\emph {\bibinfo {title} {The
  Casimir Effect}}}\ (\bibinfo  {publisher} {{WORLD} {SCIENTIFIC}},\ \bibinfo
  {year} {2001})\BibitemShut {NoStop}%
\bibitem [{\citenamefont {de~Groot}\ \emph {et~al.}(1980)\citenamefont
  {de~Groot}, \citenamefont {van Leeuwen},\ and\ \citenamefont {van
  Weert}}]{Groot.1980}%
  \BibitemOpen
  \bibfield  {author} {\bibinfo {author} {\bibfnamefont {S.~R.}\ \bibnamefont
  {de~Groot}}, \bibinfo {author} {\bibfnamefont {W.~A.}\ \bibnamefont {van
  Leeuwen}}, \ and\ \bibinfo {author} {\bibfnamefont {C.~G.}\ \bibnamefont {van
  Weert}},\ }\href@noop {} {\emph {\bibinfo {title} {Relativistic kinetic
  theory: Principles and applications}}}\ (\bibinfo  {publisher}
  {{North-Holland Publ. Comp}},\ \bibinfo {address} {Amsterdam},\ \bibinfo
  {year} {1980})\BibitemShut {NoStop}%
\bibitem [{\citenamefont {Cercignani}\ and\ \citenamefont
  {Kremer}(2002)}]{Cercignani.2002}%
  \BibitemOpen
  \bibfield  {author} {\bibinfo {author} {\bibfnamefont {C.}~\bibnamefont
  {Cercignani}}\ and\ \bibinfo {author} {\bibfnamefont {G.~M.}\ \bibnamefont
  {Kremer}},\ }\href@noop {} {\emph {\bibinfo {title} {The Relativistic
  Boltzmann Equation: Theory and Applications}}}\ (\bibinfo  {publisher}
  {Springer},\ \bibinfo {year} {2002})\BibitemShut {NoStop}%
\bibitem [{\citenamefont {Rezzolla}\ and\ \citenamefont
  {Zanotti}(2013)}]{Rezzolla.2013}%
  \BibitemOpen
  \bibfield  {author} {\bibinfo {author} {\bibfnamefont {L.}~\bibnamefont
  {Rezzolla}}\ and\ \bibinfo {author} {\bibfnamefont {O.}~\bibnamefont
  {Zanotti}},\ }\href@noop {} {\emph {\bibinfo {title} {Relativistic
  Hydrodynamics}}}\ (\bibinfo  {publisher} {Oxford University Press},\ \bibinfo
  {address} {Oxford, United Kingdom},\ \bibinfo {year} {2013})\BibitemShut
  {NoStop}%
\bibitem [{\citenamefont {Denicol}\ \emph {et~al.}(2012)\citenamefont
  {Denicol}, \citenamefont {Niemi}, \citenamefont {Molnar},\ and\ \citenamefont
  {Rischke}}]{Denicol.2012}%
  \BibitemOpen
  \bibfield  {author} {\bibinfo {author} {\bibfnamefont {G.~S.}\ \bibnamefont
  {Denicol}}, \bibinfo {author} {\bibfnamefont {H.}~\bibnamefont {Niemi}},
  \bibinfo {author} {\bibfnamefont {E.}~\bibnamefont {Molnar}}, \ and\ \bibinfo
  {author} {\bibfnamefont {D.~H.}\ \bibnamefont {Rischke}},\ }\bibfield
  {title} {\enquote {\bibinfo {title} {Derivation of transient relativistic
  fluid dynamics from the boltzmann equation},}\ }\href {\doibase
  10.1103/PhysRevD.85.114047} {\bibfield  {journal} {\bibinfo  {journal} {Phys.
  Rev. D}\ }\textbf {\bibinfo {volume} {85}},\ \bibinfo {pages} {114047}
  (\bibinfo {year} {2012})}\BibitemShut {NoStop}%
\bibitem [{\citenamefont {Tolman}\ and\ \citenamefont
  {Ehrenfest}(1930)}]{Tolman:1930ona}%
  \BibitemOpen
  \bibfield  {author} {\bibinfo {author} {\bibfnamefont {Richard}\ \bibnamefont
  {Tolman}}\ and\ \bibinfo {author} {\bibfnamefont {Paul}\ \bibnamefont
  {Ehrenfest}},\ }\bibfield  {title} {\enquote {\bibinfo {title} {{Temperature
  Equilibrium in a Static Gravitational Field}},}\ }\href {\doibase
  10.1103/PhysRev.36.1791} {\bibfield  {journal} {\bibinfo  {journal} {Phys.
  Rev.}\ }\textbf {\bibinfo {volume} {36}},\ \bibinfo {pages} {1791--1798}
  (\bibinfo {year} {1930})}\BibitemShut {NoStop}%
\bibitem [{\citenamefont {Tolman}(1930)}]{Tolman:1930zza}%
  \BibitemOpen
  \bibfield  {author} {\bibinfo {author} {\bibfnamefont {Richard~C.}\
  \bibnamefont {Tolman}},\ }\bibfield  {title} {\enquote {\bibinfo {title} {{On
  the Weight of Heat and Thermal Equilibrium in General Relativity}},}\ }\href
  {\doibase 10.1103/PhysRev.35.904} {\bibfield  {journal} {\bibinfo  {journal}
  {Phys. Rev.}\ }\textbf {\bibinfo {volume} {35}},\ \bibinfo {pages} {904--924}
  (\bibinfo {year} {1930})}\BibitemShut {NoStop}%
\bibitem [{\citenamefont {Christensen}(1976)}]{Christensen:1976vb}%
  \BibitemOpen
  \bibfield  {author} {\bibinfo {author} {\bibfnamefont {S.~M.}\ \bibnamefont
  {Christensen}},\ }\bibfield  {title} {\enquote {\bibinfo {title} {{Vacuum
  Expectation Value of the Stress Tensor in an Arbitrary Curved Background: The
  Covariant Point Separation Method}},}\ }\href {\doibase
  10.1103/PhysRevD.14.2490} {\bibfield  {journal} {\bibinfo  {journal} {Phys.
  Rev. D}\ }\textbf {\bibinfo {volume} {14}},\ \bibinfo {pages} {2490--2501}
  (\bibinfo {year} {1976})}\BibitemShut {NoStop}%
\bibitem [{\citenamefont {Davies}\ \emph {et~al.}(1996)\citenamefont {Davies},
  \citenamefont {Dray},\ and\ \citenamefont {Manogue}}]{Davies:1996ks}%
  \BibitemOpen
  \bibfield  {author} {\bibinfo {author} {\bibfnamefont {Paul C.~W.}\
  \bibnamefont {Davies}}, \bibinfo {author} {\bibfnamefont {Tevian}\
  \bibnamefont {Dray}}, \ and\ \bibinfo {author} {\bibfnamefont {Corinne~A.}\
  \bibnamefont {Manogue}},\ }\bibfield  {title} {\enquote {\bibinfo {title}
  {{The Rotating quantum vacuum}},}\ }\href {\doibase 10.1103/PhysRevD.53.4382}
  {\bibfield  {journal} {\bibinfo  {journal} {Phys. Rev. D}\ }\textbf {\bibinfo
  {volume} {53}},\ \bibinfo {pages} {4382--4387} (\bibinfo {year} {1996})},\
  \Eprint {http://arxiv.org/abs/gr-qc/9601034} {arXiv:gr-qc/9601034}
  \BibitemShut {NoStop}%
\bibitem [{\citenamefont {Gradshteyn}\ and\ \citenamefont
  {Ryzhik}(2015)}]{gradshteyn2015table}%
  \BibitemOpen
  \bibfield  {author} {\bibinfo {author} {\bibfnamefont {I.~S.}\ \bibnamefont
  {Gradshteyn}}\ and\ \bibinfo {author} {\bibfnamefont {I.~M.}\ \bibnamefont
  {Ryzhik}},\ }\href@noop {} {\emph {\bibinfo {title} {Table of integrals,
  series, and products}}},\ \bibinfo {edition} {eighth edition}\ ed.\ (\bibinfo
   {publisher} {Academic Press},\ \bibinfo {year} {2015})\BibitemShut {NoStop}%
\bibitem [{\citenamefont {Olver}\ \emph {et~al.}(2010)\citenamefont {Olver},
  \citenamefont {Lozier}, \citenamefont {Boisvert},\ and\ \citenamefont
  {Clark}}]{olver10}%
  \BibitemOpen
  \bibfield  {author} {\bibinfo {author} {\bibfnamefont {F.~W.~J}\ \bibnamefont
  {Olver}}, \bibinfo {author} {\bibfnamefont {D.~W.}\ \bibnamefont {Lozier}},
  \bibinfo {author} {\bibfnamefont {R.~F.}\ \bibnamefont {Boisvert}}, \ and\
  \bibinfo {author} {\bibfnamefont {C.~W.}\ \bibnamefont {Clark}},\ }\href@noop
  {} {\emph {\bibinfo {title} {{NIST} handbook of mathematical functions}}}\
  (\bibinfo  {publisher} {Cambridge University Press},\ \bibinfo {address} {New
  York, NY},\ \bibinfo {year} {2010})\BibitemShut {NoStop}%
\bibitem [{\citenamefont {Ambrus}(2014)}]{Ambrus:2014itg}%
  \BibitemOpen
  \bibfield  {author} {\bibinfo {author} {\bibfnamefont {Victor~Eugen}\
  \bibnamefont {Ambrus}},\ }\emph {\bibinfo {title} {{Dirac fermions on
  rotating space-times}}},\ \href@noop {} {Ph.D. thesis},\ \bibinfo  {school}
  {Sheffield U.} (\bibinfo {year} {2014})\BibitemShut {NoStop}%
\bibitem [{\citenamefont {Becattini}\ and\ \citenamefont
  {Grossi}(2015)}]{Becattini:2015nva}%
  \BibitemOpen
  \bibfield  {author} {\bibinfo {author} {\bibfnamefont {F.}~\bibnamefont
  {Becattini}}\ and\ \bibinfo {author} {\bibfnamefont {E.}~\bibnamefont
  {Grossi}},\ }\bibfield  {title} {\enquote {\bibinfo {title} {{Quantum
  corrections to the stress-energy tensor in thermodynamic equilibrium with
  acceleration}},}\ }\href {\doibase 10.1103/PhysRevD.92.045037} {\bibfield
  {journal} {\bibinfo  {journal} {Phys. Rev. D}\ }\textbf {\bibinfo {volume}
  {92}},\ \bibinfo {pages} {045037} (\bibinfo {year} {2015})},\ \Eprint
  {http://arxiv.org/abs/1505.07760} {arXiv:1505.07760 [gr-qc]} \BibitemShut
  {NoStop}%
\bibitem [{\citenamefont {Ambrus}(2017)}]{Ambrus:2017opa}%
  \BibitemOpen
  \bibfield  {author} {\bibinfo {author} {\bibfnamefont {Victor~E.}\
  \bibnamefont {Ambrus}},\ }\bibfield  {title} {\enquote {\bibinfo {title}
  {{Quantum non-equilibrium effects in rigidly-rotating thermal states}},}\
  }\href {\doibase 10.1016/j.physletb.2017.05.038} {\bibfield  {journal}
  {\bibinfo  {journal} {Phys. Lett. B}\ }\textbf {\bibinfo {volume} {771}},\
  \bibinfo {pages} {151--156} (\bibinfo {year} {2017})},\ \Eprint
  {http://arxiv.org/abs/1704.02933} {arXiv:1704.02933 [hep-th]} \BibitemShut
  {NoStop}%
\bibitem [{\citenamefont {Becattini}\ \emph {et~al.}(2021)\citenamefont
  {Becattini}, \citenamefont {Buzzegoli},\ and\ \citenamefont
  {Palermo}}]{Becattini:2020qol}%
  \BibitemOpen
  \bibfield  {author} {\bibinfo {author} {\bibfnamefont {F.}~\bibnamefont
  {Becattini}}, \bibinfo {author} {\bibfnamefont {M.}~\bibnamefont
  {Buzzegoli}}, \ and\ \bibinfo {author} {\bibfnamefont {A.}~\bibnamefont
  {Palermo}},\ }\bibfield  {title} {\enquote {\bibinfo {title} {{Exact
  equilibrium distributions in statistical quantum field theory with rotation
  and acceleration: scalar field}},}\ }\href {\doibase 10.1007/JHEP02(2021)101}
  {\bibfield  {journal} {\bibinfo  {journal} {JHEP}\ }\textbf {\bibinfo
  {volume} {02}},\ \bibinfo {pages} {101} (\bibinfo {year} {2021})},\ \Eprint
  {http://arxiv.org/abs/2007.08249} {arXiv:2007.08249 [hep-th]} \BibitemShut
  {NoStop}%
\bibitem [{\citenamefont {Palermo}\ \emph {et~al.}(2021)\citenamefont
  {Palermo}, \citenamefont {Buzzegoli},\ and\ \citenamefont
  {Becattini}}]{Palermo:2021hlf}%
  \BibitemOpen
  \bibfield  {author} {\bibinfo {author} {\bibfnamefont {Andrea}\ \bibnamefont
  {Palermo}}, \bibinfo {author} {\bibfnamefont {Matteo}\ \bibnamefont
  {Buzzegoli}}, \ and\ \bibinfo {author} {\bibfnamefont {Francesco}\
  \bibnamefont {Becattini}},\ }\bibfield  {title} {\enquote {\bibinfo {title}
  {{Exact equilibrium distributions in statistical quantum field theory with
  rotation and acceleration: Dirac field}},}\ }\href {\doibase
  10.1007/JHEP10(2021)077} {\bibfield  {journal} {\bibinfo  {journal} {JHEP}\
  }\textbf {\bibinfo {volume} {10}},\ \bibinfo {pages} {077} (\bibinfo {year}
  {2021})},\ \Eprint {http://arxiv.org/abs/2106.08340} {arXiv:2106.08340
  [hep-th]} \BibitemShut {NoStop}%
\end{thebibliography}%

\end{document}